\documentclass[onecolumn,authoryear]{els-mrw} 

\usepackage{amsmath,amssymb,amsfonts,amsthm,makeidx,graphicx}
\usepackage{txfonts}
\usepackage{helvet}
\usepackage{hyperref}

\newcommand{\oversim}[2]{\protect{\mbox{\lower0.5ex\vbox{%
   \baselineskip=0pt\lineskip=0.2ex
   \ialign{$\mathsurround=0pt #1\hfil##\hfil$\crcr#2\crcr\sim\crcr}}}}} 
\newcommand{\simgreat}{\mbox{$\,\mathrel{\mathpalette\oversim>}\,$}} 
\newcommand{\simless} {\mbox{$\,\mathrel{\mathpalette\oversim<}\,$}} 
\begin{document}

\chapter{The initial mass function of stars}
\label{chap:chIMF}

\author[1,2]{Pavel Kroupa}%
\author[3,4]{Eda Gjergo}%
\author[5]{Tereza Jerabkova}%
\author[3,4]{Zhiqiang Yan}%

\address[1]{\orgname{University of Bonn}, \orgdiv{Helmholtz-Institut
    fuer Strahlen und Kernphysik},
  \orgaddress{Nussallee 14-16, 53115 Bonn, Germany}}
\address[2]{\orgname{Charles University}, \orgdiv{Astronomical
    Institute}, \orgaddress{V Holesovickach 2, 18000 Prague, Czech
    Republic}}
\address[3]{\orgname{Nanjing University}, \orgdiv{School of
    Astronomy and Space Science}, \orgaddress{Nanjing 210093, People’s Republic of China}}
\address[4]{\orgname{Nanjing University}, \orgdiv{Key Laboratory of
    Modern Astronomy and Astrophysics},
  \orgaddress{Ministry of Education, Nanjing 210093, People’s Republic of China}}
\address[5]{\orgname{European Southern Observatory}, \orgaddress{Karl-Schwarzschild-Strasse 2,
  85748 Garching, Germany}}

\maketitle

\begin{glossary}[Glossary]

\term{Bottom/top-heavy/light IMF} With an IMF normalised to unity at $m=1\,M_\odot$ and relative to the canonical IMF: 
   
    {\it Bottom-heavy}: An IMF with a larger number of low-mass ($<1\,M_\odot$) stars.

    {\it Bottom-light IMF}: An IMF with a smaller number of low-mass ($<1\,M_\odot$) stars. 

    {\it Top-heavy}: An IMF with a larger number of high-mass ($>1\,M_\odot$) stars.
    
    {\it Top-light IMF}: An IMF with a smaller number of high-mass ($>1\,M_\odot$) stars.\\

\term{Brown dwarf (BD)} A substellar object with insufficient mass to ignite the hydrogen fusion that powers stars, but massive enough to trigger deuterium burning. \\
    
\term{Chemical enrichment} The progressive enrichment of gas in a galaxy by elements heavier than H and He, driven primarily by  stellar nucleosynthesis.\\ 
    
\term{Dark star formation} A dwarf galaxy with a very small true SFR ($\psi < 10^{-3}\,M_\odot/$yr) can appear as an H$\alpha$-dark galaxy because its gwIMF is significantly top-light.\\   

\term{Early/late-type star} {\it Early-type stars} are hot and massive white-to-blue stars (O-, B- and A-type stars) with high luminosities and short lifespans. Their mass exceeds $m=1.4\,M_\odot$.

    {\it Late-type stars} are less hot and less massive red-to-yellow stars (F-, G-, K-, and M-type stars) with low luminosities and long lifespans. Their mass does not exceed $m=1.4\,M_\odot$.\\

\term{Embedded cluster} A spatially (about $<1\,$pc) and temporally (about $<1\,$Myr) correlated dense region, or clump, in a molecular cloud that undergoes gravitational collapse and forms stars. Embedded clusters are the fundamental building blocks of galaxies, each containing a simple stellar population.\\

\term{Galaxy}  A gravitationally bound stellar population with a half-mass two-body relaxation time (i.e. energy equipartition time scale) longer than a Hubble time. A galaxy can contain multiple stellar populations and gas from which new stars can be forming. 

    {\it Late-type galaxy}: A rotationally-supported, thin-disk star-forming galaxy characterized by a mix of old and young stellar populations. Its morphological types include spiral and irregular (dwarf) galaxies. More than 90~per cent of all galaxies are of this type.

    {\it Elliptical (E) galaxy}: Unlike spiral galaxies, which are supported against gravitational collapse by angular momentum, elliptical galaxies are primarily supported by the random motion of stars, making them pressure-supported systems. E galaxies experience negligible star formation, they are passively-evolving and their stellar populations are generally very old. Only a few~per cent of all galaxies are of this type. 

    {\it Early-type galaxies (ETGs)} include elliptical (E) and lenticular (S0) galaxies. All ETGs are bulge-dominated and contain mostly old stars with little to no gas. S0 have a prominent smooth and featureless (no arms) thickened disk, while E galaxies have none. E galaxies are dominated by random motions, while S0 galaxies show rotation in their disk component.

    {\it Ultra-compact dwarf galaxies (UCDs)}: Very rare extremely massive stellar systems ten to a hundred times more massive than GCs and GC-like radii that increase with their mass similar to those of elliptical galaxies. Despite appearing linked to star clusters, UCDs are referred to as galaxies by them having half-mass two-body relaxation times longer than a Hubble time.
    
    {\it Ultra-faint dwarf galaxies (UFDs)}:  Small and faint dwarf galaxies whose stellar luminosities are typically below $10^4 L_{\odot}$, where $L_{\odot}$ is the solar luminosity. While their stellar masses are comparable to those of massive open star clusters, their sizes are $\simgreat$~ten times larger.\\

    \term{H$\alpha$ emission in the context of the IMF} Emission originating from the recombination of hydrogen in ionised gas. The gas ionisation is driven by nearby young, massive stars. H$\alpha$ emission is used as a SFR tracer for a galaxy, but the conversion is strongly sensitive to the shape of the gwIMF.\\

    \term{IGIMF Theory} The integrated-galaxy initial-mass-function theory for calculating the gwIMF by adding the stellar IMFs in all freshly formed embedded clusters.\\
 
\term{Initial mass function (IMF)} A function that represents the number distribution of 
the initial stellar masses in a complete population of stars. All IMFs have units of number of stars per solar mass [\#/$M_{\odot}$].

    The {\it stellar IMF} constitutes the IMF of stars born in a single embedded cluster in one molecular cloud clump. It is the distribution of all initial stellar masses in a simple stellar population. 

    The {\it composite IMF} is the initial mass distribution of a population of stars born in a larger region encompassing more than one embedded cluster, for example, that of a molecular cloud or of multiple molecular clouds. i.e., it is the sum of more than one stellar IMF. It can be calculated using the LIGIMF theory.

    The {\it canonical IMF} is the composite IMF derived from star counts in the local Solar neighborhood and is therewith a bench mark distribution function by being an average constituting the mixture of embedded clusters that gave rise to the Solar neighbourhood stellar ensemble. This distribution aligns with star formation processes in most environments within spiral galaxies. 

    {\it The canonical stellar IMF} is the stellar IMF with a shape equal to that of the canonical IMF.

    {\it Galaxy-wide IMF (gwIMF)}: The {\it galaxy-wide IMF} is the sum of all stellar IMFs in a galaxy. It is the composite IMF of a whole galaxy. This quantity can vary over time.

    {\it Present-day stellar mass function (PDMF)}: The distribution of stellar masses in a population of stars that are currently alive.\\

\term{Interstellar medium (ISM)} The matter dispersed in between stars in a galaxy. It is mainly composed of gas and dust. It can form stars after cooling sufficiently (to below $100^{\rm o}\,$K) under sufficiently high density to form molecular clouds. \\

\term{Isotopologues} Molecules that differ only in their isotopic composition with at least one atom having a different number of neutrons.  \\ 

\term{LIGIMF theory} The local IGIMF, also known as the composite IMF, is obtained by adding the stellar IMFs in all embedded clusters found within a given region in a galaxy (e.g. in one molecular cloud).\\
    
\term{Luminosity function of stars (LF)} The number distribution of stellar luminosities among main sequence stars.\\
    
\term{Main sequence stars} Stars in a stable phase of their evolution (being "alive"), during which they undergo hydrogen burning in their cores.\\
    
\term{Mass of an embedded cluster ($M_{\rm ecl}$)} The total stellar mass of all stars formed within an embedded cluster.  Just like the IMF, this quantity emerges over a time interval of about a million years. It is therefore not directly observable. \\

\term{Molecular cloud clump} A dense region in a molecular cloud that is gravitationally unstable and eventually collapses to form an embedded cluster. Typically, it has a scale of less than 1 pc.\\
    
\term{Optimal sampling} Deterministic sampling of a distribution function without Poisson scatter upon arbitrary binning of the sampled quantity.\\
    
\term{Pre-stellar molecular cloud core} A collapsing gas core, progenitor to a proto-star or proto-stellar binary. It has a size of $<0.01-0.1\,$pc.\\
    
\term{Proto-star} An accreting pre-stellar core that will end up as a star or a tight binary.\\
    
\term{Simple-stellar population} A stellar population of an embedded star cluster formed within a few free-fall times of the molecular cloud clump. Simple stellar populations are composed of stars with the same ages and chemical abundances.\\ 

\term{Star cluster} A gravitationally bound simple stellar population with a half-mass two-body relaxation time (i.e. energy equipartition time scale) shorter than a Hubble time. 

    {\it Open star cluster}: Open star clusters typically have half-mass radii of a few~pc and are typically found on near-circular orbits in the disks of galaxies. They are thought to be remnants of embedded star clusters after these have blown out their residual gas from which they formed. 

    {\it Globular star cluster (GC)}: A densely-packed, mostly spherical simple but partially chemically self-enriched stellar population typically of great age.  A present-day GC may contain a few hundred thousand to a few million of ancient, late-type stars and have a half-mass radius of a few~pc. GCs are massive versions of open star clusters with life times surpassing a Hubble time by virtue of their large masses. GCs form in galaxies with very large SFRs ($\psi > {\rm hundreds}\, M_\odot/$yr) and are typically the fossils of the violent onset of star formation in the early Universe.  \\
    
\term{Star-formation history (SFH)} The temporal evolution of the star formation rate (SFR, $\psi$, in units of $M_\odot/$yr), i.e., the SFH is $\psi(t)$. It is a record of a galaxy or a system's star-formation rate over its lifetime.\\

\end{glossary}

\begin{glossary}[Nomenclature]
  \begin{tabular}{@{}lp{34pc}@{}}
    BD & Brown dwarf\\
    GC & Globular star cluster\\
    ETG & Early-type galaxy\\
    gwIMF & Galaxy-wide IMF\\
    IGIMF & Integrated-galaxy initial-mass-function theory for calculating the gwIMF \\
    IMF & Initial mass function of stars
    \\
    ISM & Inter-stellar medium \\
    LIGIMF & Local IGIMF encompassing a region in a galaxy \\
    LF & Luminosity function of main sequence stars\\
    $M_{\rm ecl}$ & Cumulative mass of all stars formed within an embedded cluster, i.e. a molecular cloud clump\\
    ONC & Orion Nebula Cluster \\
   PDMF & Present-day stellar mass function of a population of stars \\
   SFE & Star formation efficiency, $\epsilon$, of a molecular cloud clump (i.e., embedded star cluster) \\
    SFH & Star-formation history \\
    SFR & Star-formation rate, $\psi$, of a galaxy\\
   SMBH & Super-massive black hole\\
   stellar IMF & Initial mass function of stars born in one molecular cloud clump, i.e. in one embedded cluster and at the same time
    \\
    UCD & Ultra-compact dwarf galaxy \\
    UFD & Ultra-faint dwarf galaxy \\
   VLMS & Very low mass star (of mass $m<0.2\,M_\odot$)\\
  \end{tabular}
\end{glossary}

\newpage

\begin{abstract}[Abstract]

 The initial mass function of stars (IMF) is one of the most important functions in astrophysics because it is key to reconstructing the cosmological matter cycle, understanding the formation of super-massive black holes, and deciphering the light from high-redshift stellar populations. 
 The IMF's dependency on the physical conditions of the star-forming gas and its connection to the galaxy-wide IMF connects the molecular clump scale to the cosmological scale. 

Significant advancements have been made in extracting the IMF from observational data. This process requires a thorough understanding of stellar evolution, the time-dependent stellar multiplicity, the stellar-dynamical evolution of dense stellar populations, and the structures, star formation histories, and chemical enrichment histories of galaxies. The IMF in galaxies, referred to as the galaxy-wide IMF (gwIMF), and the IMF in individual star-forming regions (the stellar IMF) need not be the same, although the former must be related to the latter. 
 
Observational surveys inform on whether star-forming regions provide evidence for the stellar IMF being a probability density distribution function. They may also indicate 
star formation to optimally follow an IMF shaped by the physical conditions of the star-forming gas. Both theoretical and observational evidence suggest a relationship between the initial mass function of brown dwarfs and that of stars. Late-type stars may arise from feedback-regulated fragmentation of molecular cloud filaments, which build up embedded clusters. In contrast, early-type stars form under more violent accretion and feedback-regulated conditions near the centers of these clusters.

The integration over all star-forming molecular cloud clumps and their stellar IMFs in a galaxy via the IGIMF theory yields its gwIMF which sensitively depends on the physical properties of the molecular cloud clumps and the range of their masses that depends on the SFR of the galaxy.

\end{abstract}

\begin{BoxTypeA}[chIMF:box:objectives]{Key points/Objectives box}
\noindent 
\begin{itemize}

\item Explain the current knowledge on the shape of the stellar initial mass function (IMF) as an outcome of star formation in molecular cloud clumps. Investigate the dependence of the variation of the IMF on physical conditions.

\item Discuss the nature of the stellar IMF as a probability-density distribution function or an optimally sampled distribution function. 

\item Explore the relation between this stellar IMF and the galaxy-wide IMF (gwIMF) and their implied variations.

\item Consider the astrophysical and cosmological implications of the systematically varying stellar IMF and gwIMF.

\end{itemize}

\end{BoxTypeA}

\section{Introduction} \label{chIMF:sec:introd} The advent of quantum mechanics during the first half of the 20th century allowed the understanding of the energy source that enables stars to shine for billions of years, namely hydrogen fusion, which eventually leads to the synthesis of heavier elements. This breakthrough made it possible to explain both stellar luminosities and to derive stellar lifetimes.  Heavier stars of about 100 solar masses ($M_{\odot}$) live for a few million years (Myr), while less massive stars of around 1~$M_{\odot}$ may live for billions of years (Gyr).  High mass and low mass stars are routinely observed, implying that star~formation is ongoing, and that the stellar mass distribution in galaxies changes over time.  Furthermore, this process needs theories of gas accretion onto galaxies, gas cooling, and fragmentation to sustain the observed ongoing formation of stars.

With such questions in mind, \cite{Salpeter1955} bridged quantum physics into cosmology by assessing the ``\emph{relative probability for the creation of stars of mass} $m$'' \citep{KJ2019}.  Salpeter used the available distribution of stellar luminosities, which had been obtained by astronomers from star counts in the Solar neighborhood, and applied corrections for stellar death to estimate this relative stellar creation probability, which today is referred to as the initial mass function of stars (IMF).  The observed ensemble of stars, after correcting the stellar luminosities for stellar evolution (e.g., main sequence brightening), yields the present-day mass function of stars (PDMF).  The calculation of the IMF from the PDMF requires several assumptions, including an understanding of the star-formation rate (SFR), the evolution of the SFR over time (SFH), the age of the Galaxy’s disk and its stellar-age-dependent thickness, and how stellar orbits diffuse over time in the galaxy.  These assumptions are important for connecting the local counts, which are mostly composed of faint, long-lived stars in a small volume (less than $100\,$pc) around the Sun, to the kpc-distant counts of rare, bright, short-lived massive stars.  Salpeter initially estimated the age of the Galactic disk to be $6\,$Gyr. This value is about half of the presently-known age of approximately $12\,$Gyr. Despite this discrepancy, the star counts and structure of the Galactic disk known at the time led to the formulation of the ``\emph{Salpeter-power-law IMF}''.

Salpeter assumed the IMF to be an invariant probability density distribution function. He then considered field stars with masses $0.4 < m/M_\odot < 10$ and found the IMF to be a power-law whose index is $\alpha_{\rm S} \approx 2.35$, where the number of stars ${\rm d}N$ with masses $m$ in the range to $m+{\rm d}m$ is: \begin{equation}
    {\rm d}N = \xi(m)\,{\rm d}m\, ,
\end{equation}
where $\xi(m)\propto m^{-\alpha_{\rm S}}$ is the Salpeter IMF. A thorough discussion of the above assumptions and of the biases affecting star counts has been compiled by \cite{Scalo1986, Kroupa+2013, Hopkins2018}.

The following chapter documents the present-day knowledge of the IMF based on Solar-neighbourhood star counts and resolved stellar populations (Sec.~\ref{chIMF:sec:local}).  Theoretical ideas on the IMF and the sub-stellar problem are touched upon in Sec.~\ref{chIMF:sec:origin}.  Over the past decade, evidence has accumulated that indicates the IMF is dependent on the density and metallicity of the star-forming gas cloud, i.e. on its ability to cool (Sec.~\ref{chIMF:sec:varIMF}). The fundamental question whether the IMF is a probability density distribution function or an optimal, self-regulated function of stellar mass is raised in Sec.~\ref{chIMF:sec:sampling}. Sec.~\ref{chIMF:sec:gwIMF} discusses the relation between the stellar IMF and the composite IMF such as the galaxy-wide IMF (gwIMF) and thus the redshift-dependent evolution of initial stellar-populations with an outline of the IGIMF theory.  The observational evidence is addressed in Sec.~\ref{chIMF:sec:galobs} of the connection between the IMF and star-forming galaxies, early-type galaxies (ETGs), ultra-compact dwarf galaxies (UCDs) and globular star clusters (GCs), and supermassive black holes (SMBHs). The existence and reality of the distribution functions that define a new stellar population is addressed in Sec.~\ref{chIMF:sec:existence}.  Available computer programs to set-up and simulate stellar populations and galaxies are listed in Sec.~\ref{chIMF:sec:codes}. Sec.~\ref{chIMF:sec:concs} contains the conclusions and an outlook.

\paragraph{Note:} 
The \emph{stellar IMF}, $\xi(m)$, is defined as the number distribution of initial stellar masses for all stars born in one molecular cloud clump, i.e., in one embedded star cluster, representing a simple stellar population.
Note that $m$ is always in units of the Solar mass, $M_\odot$, unless stated otherwise and that if $m$ is the argument in a function (e.g. as in  eq.~\ref{chIMF:eq:canIMF1}, \ref{chIMF:eq:lognormal} or ${\rm log}_{10}m$) then implicitly it is assumed that $m=m/M_\odot$.
The \emph{composite IMF} is defined as the number distribution of initial stellar masses of multiple simple stellar populations, i.e., it is the sum of several stellar IMFs forming at the same time. The composite IMF can be a galaxy-wide IMF, gwIMF. It is equal in shape to the stellar IMF if the physical properties of the star-forming gas are sufficiently similar for all simple stellar populations included in the composite IMF, or if the stellar IMF is an invariant probability density distribution function.

Different notations have been adopted in the literature concerning the IMF: The function $\xi(m)$ can be rewritten in terms of the logarithmic mass scale as $\xi_{\rm L}(m) = \left(m\, {\rm ln}10\right)\,\xi(m)$ such that
\begin{equation}
{\rm d}N = \xi(m)\,{\rm d}m = \xi_{\rm L}(m)\, {\rm dlog}_{\rm 10}m\, ,
\end{equation}
where ${\rm log}_{10}m = {\rm ln}m/({\rm ln}10)$. Given $\xi(m) \propto m^{-\alpha}$ and $\xi_{\rm L}(m) \propto m^\Gamma \propto m^{-x}$, we have $\Gamma = -x = 1-\alpha$.

\section{The canonical stellar IMF and its functional form}
\label{chIMF:sec:local}

Currently, it is possible in the solar neighbourhood to obtain volume-complete star counts of the faintest M~dwarf stars to a distance of not much more than 5~pc, and of G~dwarfs out to distances of about 20~pc. These distances are known through trigonometric parallax measurements, and enable to derive their stellar luminosity function (LF). 
Independent and larger ensembles of late-type stars,
based on photometric-parallax measurements, reach to larger distances.  
Late-type stars do not evolve much over a Hubble time, but it is necessary to account for two parts: the main-sequence brightening at their massive-end ($m>0.8\,M_\odot$), and the pre-main-sequence long contraction times below a few tenths of a Solar-mass. It is necessary to apply statistical corrections for incompleteness as introduced by \cite{Kroupa+1991}, because approximately half of all late-type stars in the Solar-neighbourhood are binary
systems (e.g. Fig.~\ref{chIMF:fig:binaries} below). Some systems are even higher-order multiples, this being particularly relevant for the deep
photometric-parallax ensembles. In a survey of 100 late-type Galactic field ``stars'', typically about 40 would be binaries and 5~triples with perhaps a quadruple being there too. The observer would thus miss 53~faint stars if none of the multiples were to be resolved. 
While most IMF constraints have been
based on the trigonometric-parallax-based sample
(e.g. \citealt{Kirkpatrick+2023} using Gaia data), the combination
with photometric-parallax-based star~counts provides a more robust assessment of the shape of the stellar IMF. That is, as long as all biases  are taken into account. For example, parallax measurements are affected by the Lutz-Kelker bias, while photometric-parallax surveys are affected by the Malmquist bias (\citealt{Kroupa+1993} and references therein).

Of central importance when constructing a stellar IMF from a star-count survey after a suitably-complete ensemble of main-sequence stellar masses has been obtained is the following: Either individual stellar masses
need to be calculated from their luminosities, or the stellar
luminosity function, $\Psi_{\rm X}(M_{\rm X})$, needs to be
transformed into $\xi(m)$, with

\begin{equation}
{\rm d}N=\Psi_{\rm X}\,{\rm d}M_{\rm X}
\end{equation}

\noindent being the number of stars in
the absolute magnitude interval $M_{\rm X}$ to
$M_{\rm X}+{\rm d}M_{\rm X}$ and X~being a photometric pass band
(e.g. X$=$V for the V-band). The stellar IMF, $\xi$, and luminosity function, $\Psi$, are related by

\begin{equation}
\xi(m) = - \left( {\rm d}m/{\rm d}M_{\rm X} \right)^{-1}\,\Psi_{\rm
  X}(M_{\rm X}).  
\end{equation}
  
\begin{figure}[t]
\centering
\includegraphics[width=.6\textwidth]{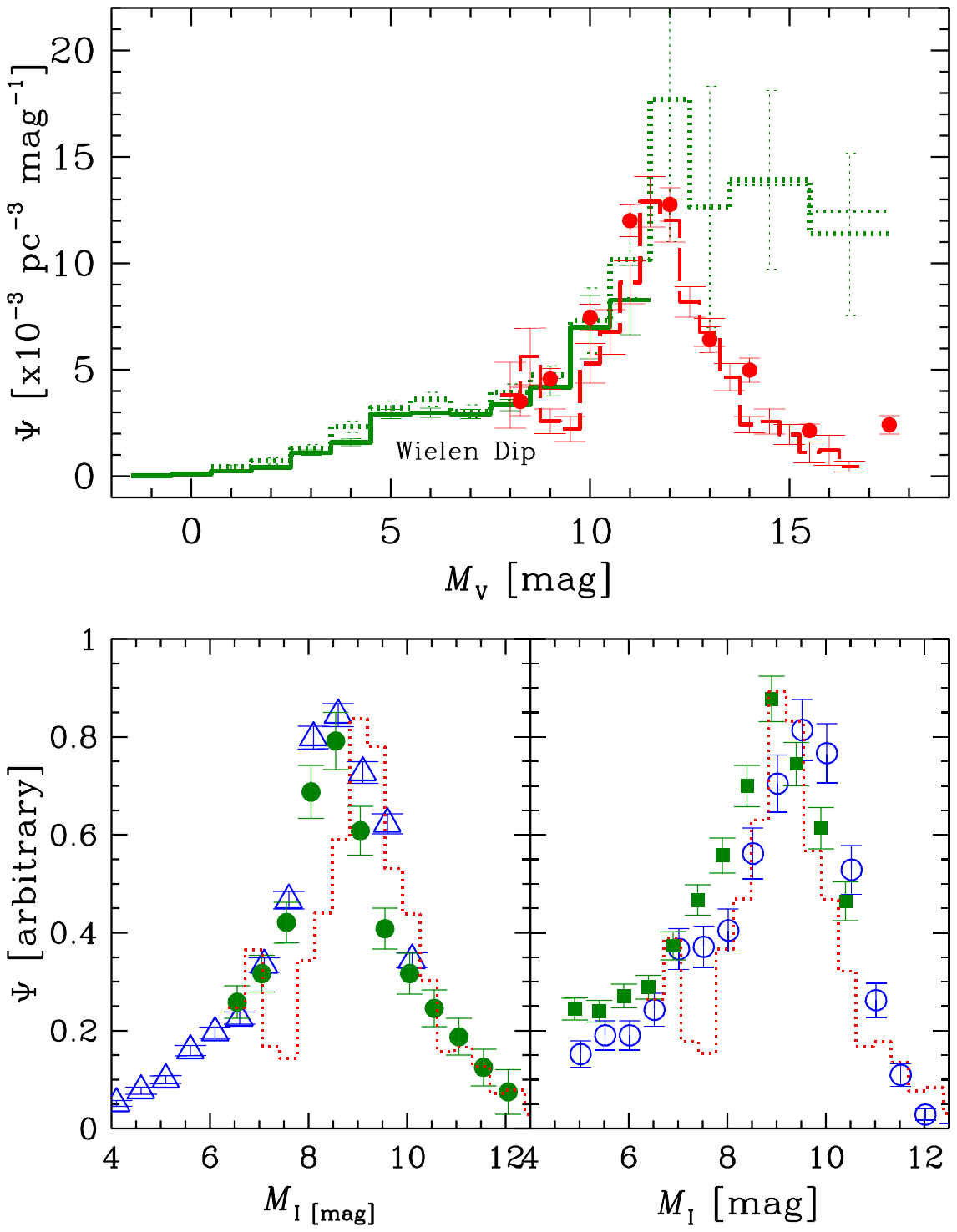}
\caption{
Stellar luminosity function of late-type stars, $\Psi(M_{\rm X})$, as a function of their photometric magnitudes in V- (\emph{top panel}) and I-bands (\emph{bottom panels}). Note that all the stellar luminosities have sharp maxima at $\approx 11.5$ for $M_{\rm V}$, and at $\approx 8.5$ for $M_{\rm I}$. These peaks match in location, amplitude, and width the red curves in the lower panels of Fig.~\ref{chIMF:fig:MLR}. 
\emph{(Top panel)}: In green are the histograms for the ensemble of stars in the solar neighborhood, employing a trigonometric-parallax based method. The \emph{solid green line} includes the complete sample up to about 20\,pc, while the \emph{dotted green lines} include the complete sample up to about 5\,pc from the Sun. The \emph{red histogram data with a dashed line and solid circles} show the star samples based on photometric parallax distance estimates in pencil beam sky surveys out to distances of~100\,pc to~300\,pc.
Marked on the graph is the ``\emph{Wielen Dip},'' which originates from structure in the mass-luminosity relation. The H-ion provides an increasing opacity source in the outer stellar envelope with decreasing stellar masses. For higher V-band magnitudes ($M_{\rm V}>11$), faint stellar companions are not counted in the deep surveys. This causes a difference in star counts obtained via trigonometric vs. photometric parallax, which manifests as different stellar volume densities, $\Psi$.
\emph{(Left bottom panel)}: The LFs of simple stellar populations. Here we see the stellar luminosity,  $\Psi$, as a function of the I-band magnitude, $M_I$, for two globular clusters: M15 \emph{(blue triangles)} and NGC~6397 \emph{(green solid circles)}.
\emph{(Right bottom panel)}: Similarly to the left bottom panel, but for the young Pleiades star
cluster \emph{(blue open circles)} and the globular cluster 47~Tuc \emph{(green solid squares)}.
In both of the bottom panels, the red histogram matches the red histogram from the upper panel, scaled to a peak value matching that of the LFs in the clusters.
Note that although the upper panels show the LFs of the composite Solar-neighbourhood population, 
correcting the star counts for Lutz-Kelker and Malmquist bias makes the LFs equivalent to those of an average simple stellar population. This simple stellar population represents a population with an average age of about 5\,Gyr with near-Solar metallicity. 
For more details, see figures~9 and~10 in \cite{Kroupa+2013}.}
\label{chIMF:fig:LF}
\end{figure}

\noindent In this step of constructing $\xi(m)$, a point of
fundamental importance is the use of the correct
stellar-luminosity--mass, $l(m)$, relation because the short-lived H$^-$ ion becomes an important opacity source in stars with  $m< 0.9 \,M_\odot$, 
the formation of the H$_2$ molecule affects the opacity and mean molecular wight for stars $m<0.4\,M_\odot$, stars with
$m < 0.3\,M_\odot$ are fully convective, while more massive stars have
a radiative core (e.g. \citealt{MansfieldKroupa2023}), and the core of a star with $m<0.1\,M_\odot$ is supported significantly by degenerate electron pressure. This leads to
inflections, noted by \cite{Kroupa+1990},  in the $l(m)$ relation that causes the "Wielen dip" near $m\approx 0.7\,M_\odot$ and a very pronounced
sharp metallicity-dependent and age-dependent maximum in the stellar luminosity
function at $M_{\rm V}\approx 11.5, M_{\rm I}\approx 8.5 \; (m\approx 0.3\,M_\odot)$ seen in all simple and composite stellar populations (Fig.~\ref{chIMF:fig:LF}). 

\begin{figure}[t]
\centering
\includegraphics[width=.4\textwidth]{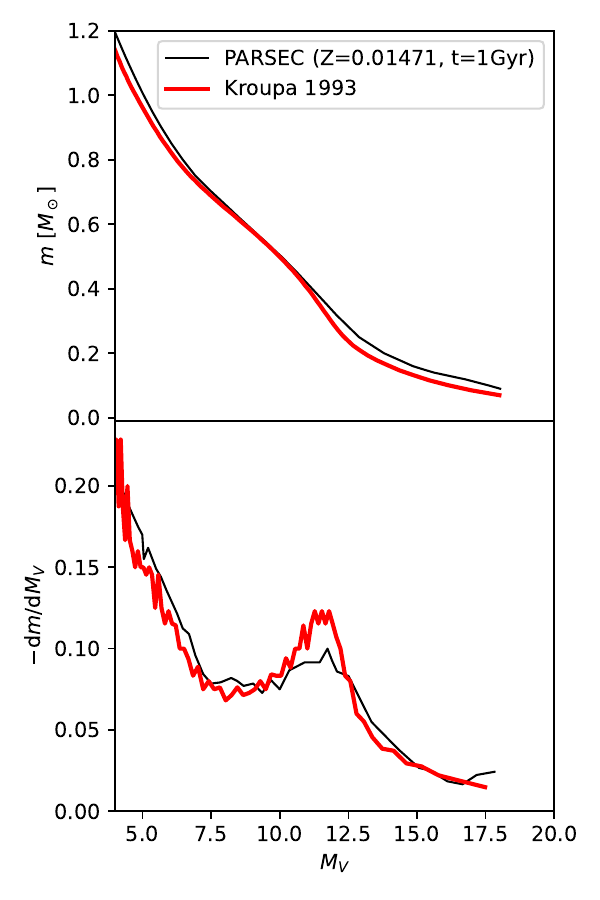}
\includegraphics[width=.4\textwidth]{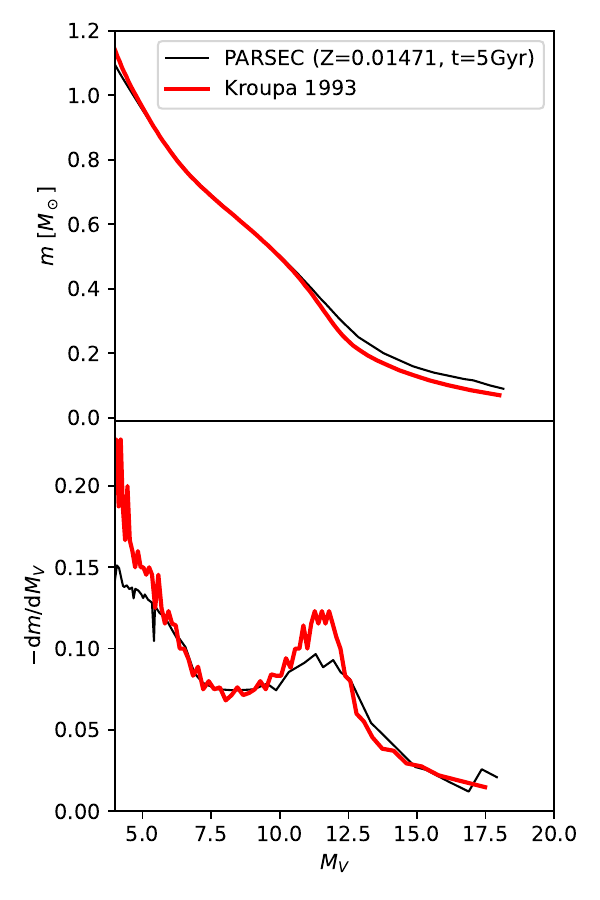}
\caption{Visualisation of the problems encountered when transforming
  stellar luminosities to stellar masses.  The stellar
  mass--luminosity relation ($m(M_{\rm V})$, upper panels) and the
  derivative ($-{\rm d}m/{\rm d}M_{\rm V}$), lower panels) in the
  V-photometric pass band for a theoretical (black curves, for details see
  footnote~\ref{chIMF_ftn_MLR}) population
  of stars of age~1Gyr (left panels) and~5~Gyr (right panels) are shown. The red
  lines are for an empirically gauged $m(M_{\rm V})$ relation valid for a
  population of Solar-neighbourhood stars of an average age of
  about~5\,Gyr and of solar metallicity. 
In the upper panels, main-sequence
  brightening of the theoretical (black curves) stars with
  $m>0.75\,M_\odot$ is evident in the 5\,Gyr panel.  If an observer
  chooses the theoretical (black) model then the observed amplitude of
  the stellar luminosity function at $M_{\rm V}\approx 11.5$ needs to
  be compensated by values of the power-law indices of the stellar IMF
  of $\alpha_1$ and $\alpha_2$ that are approximately twice larger
  than when using the red $m(M_{\rm V})$ relation, given that the
  theoretical (black) $m(M_{\rm V})$ relation has an approximately two
  times smaller amplitude in $-{\rm d}m/{\rm d}M_{\rm V}$ than the red
  relation.The stellar IMF derived from star counts will thus be more
  bottom-heavy in the approximate stellar mass range
  $0.3 < m/M_\odot < 0.7$ for the theoretical $m(M_{\rm V})$ relation than
  the empirically gauged one. }
\label{chIMF:fig:MLR}
\end{figure}

Errors can be introduced into
$\xi(m)$ when an incorrect mass-luminosity relation is applied. This is demonstrated
in Fig.~\ref{chIMF:fig:MLR}, which displays the relation between the stellar mass $m$ and the respective magnitude $M_V$\footnote{\label{chIMF_ftn_MLR}The red lines in
  Fig.~\ref{chIMF:fig:MLR} are the $m(M_{\rm V})$ relation gauged from
  Galactic-field data by \cite{Kroupa+1993} subject to strong
  constraints given by the stellar-mass and luminosity data from the
  orbital solutions of binary stars and the amplitude and width of the
  bias-corrected stellar luminosity function,
  $\Psi_{\rm V}(M_{\rm V})$.  The black lines are stellar isochrones
  generated by CMD~3.7
  (http://stev.oapd.inaf.it/cgi-bin/cmd\_3.7), based on PARSEC release
  v1.2S, for Solar metallicity stars ($Z=0.01471$) with an age of
  $1\,$Gyr or $5\,$Gyr (main-sequence stars) for the UBVRIJHK
  photometric system \citep{MA2006}
    using the YBC version of bolometric corrections \citep{Chen+2019}
   and assuming a dust composition of 60\% Silicate and 40\% AlOx for
    M stars, 85\% AMC and 15\% SiC for C stars
    \citep{Groenewegen2006}, and no interstellar extinction.}.
Purely theoretical mass-luminosity relations  are likely to result in incorrect estimates of the stellar IMF. The mass-luminosity relation is also affected by stellar rotation. The theoretical relations generally do not reproduce the mass-luminosity relation sufficiently well \citep{KroupaTout1997}. It is therefore essential to gauge carefully the mass-luminosity relation in order to calculate correctly the shape of the stellar IMF.
Fig.~\ref{chIMF:fig:LF} demonstrates that the radiative/convection boundary in late type stars near $M_{\rm V}=11.5$ (Fig.~\ref{chIMF:fig:MLR}) is evident in a very pronounced and sharp maximum in the stellar LFs of different simple stellar populations. This maximum needs to be mapped correctly to the IMF, $\xi(m)$. Failing to do so affects the value of the deduced IMF power-law index $\alpha_{1,2}$, as explained in Fig.~\ref{chIMF:fig:MLR}. This issue is of great importance when the power-law index of the IMF is estimated from star counts of stars with masses in the range $0.3-0.7\,M_\odot$, as is the case for distant GCs and ultra-faint dwarf galaxies (UFDs), for example.

\begin{BoxTypeA}[chIMF:box:MLR]{The derivative of the stellar-luminosity--mass relation}

\noindent If purely theoretical stellar models are used, the amplitude of the derivative, $dm/dM_{\rm X}$, of the stellar-luminisoty--mass relation will map into an incorrect value of the power-law index of the stellar IMF. 

\end{BoxTypeA}

After correcting the star counts statistically for unresolved binary
systems \citep{Kroupa+1991}, i.e. counting all individual stars, the combined solution to
the nearby trigonometric-parallax-based and photometric-parallax-based
star counts \citep{Kroupa+1993} provides a benchmark stellar population, referred to as
the canonical IMF \citep{Kroupa+2013}. It is representative of the
late-type stellar population in the Solar vicinity accumulated from
many star-formation events over a wide range of
metallicities (with
$m_{\rm H}\approx 0.08\,M_\odot$ being the
slightly-metallicity-dependent hydrogen-burning mass limit below which
the objects are referred to as brown dwarfs, BDs,
\citealt{Burrows+2011}). Adapting the field star-count analysis by
\cite{Scalo1986} for early-type stars one obtains the {\it canonical
  field-star IMF}. Combining instead star-count data of early-type stars in young star
clusters and associations in the Galaxy, the Large and Small
Magellanic Clouds with stellar-evolution models by \cite{Massey2003}
and stellar-dynamical models of initially binary-rich star clusters \citep{Kroupa2001} one 
obtains the {\it canonical stellar IMF}. The most-massive star in a population, $m_{\rm max}$, is discussed in
Sec.~\ref{chIMF:sec:origin}. We thus have:

\begin{BoxTypeA}[chIMF:box:fieldIMF]{The canonical field-star and stellar IMF}
  
  \noindent With continuity ($k_{\rm i}$) and normalisation ($k_\xi$) constants,
  the canonical field-star and the canonical stellar IMF can be
  writen as

  \begin{equation}
   \xi(m) = k_\xi \, k_{\rm i} \, m^{-\alpha_{\rm i}},
  \label{chIMF:eq:canIMF1}
  \end{equation}

  \noindent where 
  \begin{align}
      \alpha_{1, {\rm can}} &= 1.3 \pm 0.3
  &(m_{{\rm s}0} \simless m/M_\odot \le m_{{\rm s}1})\\
  \alpha_{2, {\rm can}} &= 2.3\pm0.3 &(m_{{\rm s}1} < m/M_\odot \le m_{{\rm s}2}) \, ,
  \label{chIMF:eq:canIMF2}
  \end{align}
with $m_{{\rm s}0}=m_{\rm H}=0.08\,M_\odot$, 
$m_{{\rm s}1}=0.5\,M_\odot$ and $m_{{\rm s}2}=1.0\,M_\odot$.\\

  \noindent For the \emph{canonical stellar IMF}:
  \begin{equation}
      \alpha_{3, {\rm can}}=2.3 \pm 0.36 \quad
  (m_{{\rm s}2} < m/M_\odot \le m_{\rm max}).
  \label{chIMF:eq:canIMF3}
  \end{equation}

  \noindent and for the \emph{canonical field-star IMF}
  \begin{equation}
   \alpha_{3, {\rm field}}= 2.7 \pm0.4 \quad (m_{{\rm s}2} < m/M_\odot),   
   \label{chIMF:eq:canIMF4}
  \end{equation}

\end{BoxTypeA}

\noindent The more recent trigonometric-parallax-based star-count analysis by \cite{Kirkpatrick+2023} confirms these values but suggests that $\alpha_{1, {\rm Kirk}} = 0.25$ for $m< 0.25\,M_\odot$ which is similar to $\alpha_0=0.3\pm0.4$ for $0.01\simless m/M\odot \simless 0.15$ in \cite{Kroupa+2013}, where for very low-mass stars and brown dwarfs, $\xi_{\rm BD}(m) = k_\xi \, k_{\rm BD} \, \left(m/0.07\right)^{-\alpha_0}$ with $k_{\rm BD}\approx 1/3$ (for a discussion of the sub-stellar/brown dwarf regime see Sec.~\ref{chIMF:sec:BDs}).

With $\alpha_{3, {\rm field}} > \alpha_{3, {\rm can}}$, the canonical
field-star IMF contains a smaller number of massive stars per
late-type star than the canonical stellar IMF, the former thus being
{\it top-light} relative to the latter which represents stellar
populations born in one embedded cluster, i.e. in one molecular cloud
clump. This difference appears to be real and it is confirmed by
\cite{RybizkiJust2015, Mor+2017, Mor+2018, Sollima2019}. This difference can be
understood in terms of the IGIMF theory (Sec.~\ref{chIMF:sec:caseC}).

A log-normal form for the canonical IMF has been suggested by \cite{MS1979} and other researchers (e.g. \citealt{Chabrier2003}) instead of a multi-power-law form.  To be consistent with the star-counts, the log-normal form needs a power-law extension for $m\simgreat 1\,M_\odot$ which is as above ($\alpha_{3, {\rm can}}$) such that this IMF is very similar to the above two-part-power-law form of the canonical stellar IMF (fig.~24 in \citealt{Kroupa+2013}). The {\it log-normal canonical IMF} can be written \begin{equation}
  \xi_\mathrm{star} (m) = k_\xi\left\{
  \begin{array}{l@{\quad\quad\quad\quad\quad,\quad}l@{\quad}}
{1\over m}\,{\rm exp}\left[ -{ ({\rm log}_{10}m - {\rm log}_{10}m_{\rm c})^2 \over 2\, \sigma_{{\rm lm}}^2} \right]
&0.1 \simless m/M_\odot \le 1.0,\\
A \, m^{-2.3\pm0.36} &1.0 < m/M_\odot \le 150\\
  \end{array}\right.
\label{chIMF:eq:lognormal}
\end{equation}

\noindent
In eq.~\ref{chIMF:eq:lognormal} continuity (but not differentiability) is assured at $1\,M_\odot$ and $\xi_{\rm BD}(m)$ is as above.

A least-squares fit to the two-part power-law form (eq.~\ref{chIMF:eq:canIMF1}--\ref{chIMF:eq:canIMF3}), whereby $\int_{0.07}^{150}m\,\xi(m)\,dm=1\,M_\odot$ for both, yields \citep{Kroupa+2013} $m_c=0.055\,M_\odot$ and $\sigma_{lm}=0.75$ with $A=0.2440$ for continuity at~$1\,M_\odot$ and $k_{\rm BD}=4.46$ as being the best log-normal plus power-law representation of the canonical IMF. Alternatively, a reduced analysis based only on the trigonometric-parallax-based star counts has (table~1 in \citealt{Chabrier2003}) $m_c=0.079^{-0.016}_{+0.021}\,M_\odot$ and $\sigma_{lm}=0.69^{-0.01}_{+0.05}$ with $A=0.2791$ and $k_{\rm BD}=4.53$ \citep{Kroupa+2013}.

As explained in Sec.~\ref{chIMF:sec:origin},
the theoretical motivation for the log-normal form however fails below
$m\simless 0.1\,M_\odot$, and the analysis of Gaia data by
\cite{Kirkpatrick+2023} shows the log-normal form to be unviable below
about $0.3\,M_\odot$, confirming the results by \cite{Thies+2015}.
Given that quantities over a mass range $m_{\rm a}$ to $m_{\rm b}$ often need to
be computed such as the mass in stars,
$M_{m_{\rm a}}^{m_{\rm b}} = \int_{m_{\rm a}}^{m_{\rm b}}m\,\xi(m)\,{\rm d}m$, or the number of stars,
$N_{m_{\rm a}}^{m_{\rm b}} = \int_{m_{\rm a}}^{m_{\rm b}}\xi(m)\,{\rm d}m$, the advantage of the
two-part-power-law form becomes apparent in stellar sampling applications (Sec.~\ref{chIMF:sec:sampling}). 
Thus, according to the
canonical stellar IMF with $m_{{\rm s}0}=0.07\,M_\odot$, the average stellar mass 
\begin{equation}
m_{\rm av} = \frac 
{\int_{m_{{\rm s}0}}^{m_{\rm b}}\,m\,\xi(m)\,{\rm d}m}
{\int_{m_{{\rm s}0}}^{m_{\rm b}}\,\xi(m)\,{\rm d}m}  \, ,
\label{chIMF:eq:mav}    
\end{equation}
is $m_{\rm av}=0.5(0.3)\,M_\odot$ for $m_{\rm b}=150.0(1.0)\,M_\odot$. BDs contribute negligible mass (about 4~per cent) to a simple stellar population.
The above formulations count every single star in the stellar population. In the Solar neighbourhood about every second "star" is a binary system leading to the Solar neighbourhood IMF appearing flatter (\citealt{Kroupa+2013} and references therein).
This IMF of stellar systems, the "{\it system IMF}", has, however, no physical meaning.  Because most if not all stars are born as multiple systems (see Sec.~\ref{chIMF:sec:binaries}) a system-stellar IMF is meaningful if it is constructed from this initial or birth population \citep{Kroupa+2013}.
The analysis of star-forming regions and young stellar populations in
the Galaxy and the Large and Small Magellanic Clouds has been showing
the IMF to be similar to the canonical stellar IMF which has therewith
been deemed to be universal and invariant (fig.~1 in
\citealt{deMarchi+2010}).

\begin{figure}[t]
\centering
\includegraphics[width=.8\textwidth]{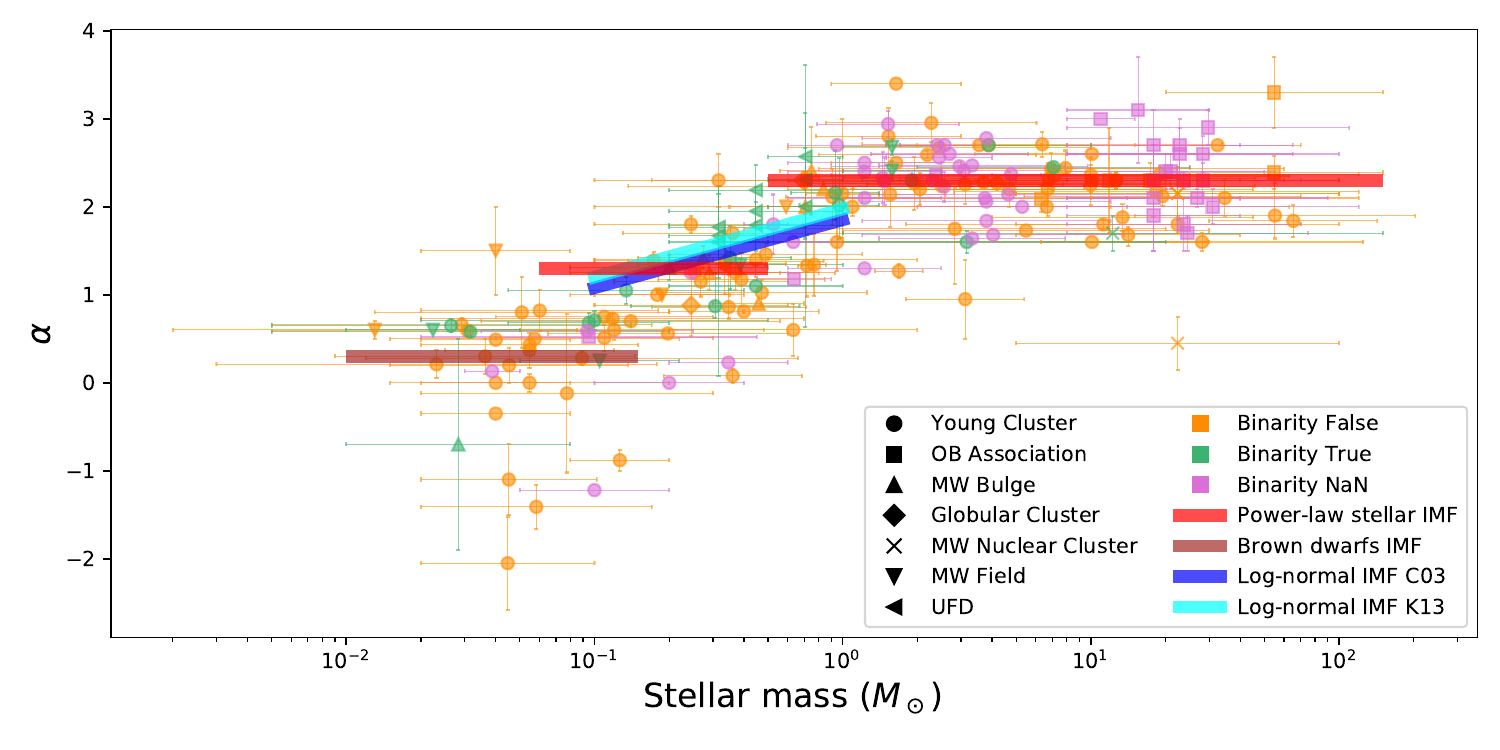}
\caption{
The alpha plot: The data points are the $\alpha$ indices for various young populations in dependence of the stellar mass range over which they are observationally constrained with the $\alpha$ index plotted at the mid-range value $\bar{m}$ (from \citealt{Kroupa2001, Kroupa2002} and \citealt{HennebelleGrudic2024}). The canonical stellar IMF is shown by the thick red lines while the brown dwarf regime ($\alpha_0$) is indicated by the thick brown line. Both overlap for reasons given in Sec.~\ref{chIMF:sec:BDs}. The canonical power-law form of the stellar IMF comes from a detailed star-count analysis using trigonometric {\it and} photometric surveys as described in the text. 
The systematic offset in the mass range $0.1-1\,M_\odot$ of the data compared to the canonical stellar IMF is due to unresolved multiple stars that hide faint companions in the orange data such that the observationally-deduced $\alpha$ values are smaller than the real ones defining the canonical stellar IMF. The green data points have been corrected for unresolved multiple systems. There is no universal correction for unresolved multiple systems because the details of a binary population depend on their past dynamical history. 
The light-blue line is the log-normal fit to the canonical stellar IMF (eq.~\ref{chIMF:eq:lognormal}, from \citealt{Kroupa+2013}, K13) while the dark blue line results from a reduced analysis of only the trigonometric parallax data (\citealt{Chabrier2003}, C03).  Adapted from fig.~26 in \cite{Kroupa+2013}, and fig.~1 in \cite{HennebelleGrudic2024} which contains a representation of the log-normal IMF based on the 8~pc sample of nearby stars and ignores the constraints on stellar number densities from the deep photometric-parallax-surveys thus leading to a flatter IMF (i.e. smaller $\alpha$ values).}
\label{chIMF:fig:alphaplot}
\end{figure}

Fig.~\ref{chIMF:fig:alphaplot} shows the canoncial stellar IMF in comparison to a compilation of power-law mass function indices of different stellar populations, $\alpha(\bar{m})$, in dependence of the stellar mass range with average value $\bar{m}$ over which they are constrained. The data are consistent with the canonical stellar IMF. 
At a given ${\bar m}$ the $\alpha$ values show a small and symmetric dispersion about the canonical index
despite different observing teams providing the data and the measurement uncertainties. 
Together with the existence of the $m_{\rm max}(M_{\rm ecl})$ relation (Fig.~\ref{chIMF:fig:mmax}),
this is an indication \citep{Kroupa+2013} 
for the stellar IMF being an optimally sampled distribution function (no dispersion of $\alpha$ values) rather than a probability distribution function that would lead to a larger dispersion of $\alpha$ indices than evident in the data (see Sec.~\ref{chIMF:sec:sampling}).

\section{The origin of the IMF, the binary-star population and the
  sub-stellar regime}
\label{chIMF:sec:origin}
The properties of the stellar IMF and its relation to a freshly born
stellar population in an embedded cluster depend on
the dynamical structures in which stars form and
on the feedback-self-regulated star-formation
process. Understanding these provides the context to the galaxy-wide population of newly formed stars (Sec.~\ref{chIMF:sec:gwIMF}).

\subsection{The dynamical structures in which stars form}
\label{chIMF:sec:dynstr}

Star formation occurs in spatially and
temporarily correlated structures, i.e. in \emph{molecular cloud
  clumps}, that developed to become gravitationally unstable and
collapse. A minimum mass, $M_{\rm J}$, for a gravitationally unstable
region in a molecular cloud can be estimated using the Jeans
instability criterion for characteristic densities in nearby molecular
clouds.  For a gas temperature of $T=10\,$K, the sound speed in
molecular hydrogen gas is

\begin{equation}
c_{\rm g} = \left(k_{\rm B}\,T /\left(\mu \, m_{\rm p} \right)
\right)^{1/2}, 
\end{equation}

\noindent where $k_{\rm B}$ is the Boltzmann constant, $\mu$ the
mean molecular weight and $m_{\rm p}$ the mass of a Hydrogen atom.
For Solar abundance $\mu \approx2.4$ and
$c_{\rm g0} \approx 0.063\, T^{1/2} \approx 0.2\,$km/s. According to
the Jeans instability criterion, a region of a hydrogen molecular
cloud with a density of $\rho_{\rm g0} \approx 10^3/$cm$^3$ becomes
gravitationally unstable when its mass surpasses

\begin{equation}
    M_{\rm J} / M_\odot \approx 2 \left(c_{\rm g}/c_{\rm
    g0}\right)^3\,\left(\rho_{\rm g}/\rho_{\rm g0}\right)^{-1/2}
\end{equation}
    
\noindent within a spherical radius
$r_{\rm J}/{\rm pc} \approx 0.4\, \left(c_{\rm g} / c_{\rm g0} \right)
\, \left(\rho_{\rm g}/\rho_{\rm g0} \right)^{-1/2} $. Overdense
regions, the molecular cloud clumps, collapse on a free-fall
time-scale,
$t_{\rm ff}/{\rm Myr} \approx 2\, \left( \rho_{\rm g0} / \rho_{\rm g}
\right)^{1/2}$. Given the empirically determined life-times of
molecular clouds of $\delta t \approx 10\,$Myr \citep{Murray2011}, ten
times less-dense regions will not collapse.
%
%
Less-massive cloud regions would need a higher density for collapse,
and thus the smallest observed star-forming events are indeed in
clumps of the above scale with a mass of a few~$M_\odot$ and diameters
of less than a~pc (e.g. the NESTs of \citealt{Joncour+2018} in the
nearest star-forming cloud Taurus-Auriga). 

Because $M_{\rm J} \gg m_{\rm av}$ an embedded cluster of stellar binaries
always forms within $r_{\rm J}$.  Surveys of nearby molecular clouds
show the youngest proto-stars to be highly clustered
(e.g. \citealt{Megeath+2016} for the Orion star-forming clouds). An
analysis of open star clusters based on the widest binary star orbits
in them suggests that the embedded-star-cluster forming cloud clumps that
are the progenitors of the open star clusters have half-mass radii at the stage of highest density of \citep{MarksKroupa2012},
\begin{equation}
r_{0.5} \approx 0.1^{+0.07}_{-0.04}\,\left(M_{\rm
    ecl}/M_\odot\right)^{0.13\pm0.04} \, , 
\label{chIMF:eq:r05}
\end{equation}
where $M_{\rm ecl}$ is the mass of the embedded cluster in stars.
It's relation to the residual gas
mass, $M_{\rm g}$, within $r_{0.5}$ is given by the canonical
star-formation efficiency (SFE) of about
$\epsilon\approx M_{\rm ecl}/\left(M_{\rm ecl}+M_{\rm g}\right)
\approx 1/3$ (\citealt{LadaLada2003}, e.g fig.~25 in
\citealt{Megeath+2016}). The deduced densities of the embedded clusters, $\rho_{\rm ecl}= \left(3\,M_{\rm ecl}\right)/\left(8\, \pi \, \epsilon \, r_{0.5}^3\right)$, are consistent with the observed molecular cloud clump densities. 
While this canonical value of $\epsilon$ has been useful in understanding observed open and globular star clusters and their relation to embedded clusters, it is likely to decrease with increasing metallicity and to increase with increasing $M_{\rm ecl}$ (e.g. \citealt{Dib+2011, Megeath+2016}). Also, $r_{0.5}$ is likely to be smaller at low than at high metallicity because the molecular cloud clump will have more time to collapse to a denser state before fragmenting. For a given $M_{\rm ecl}$, it is thus likely that the stellar IMF emerging from the molecular cloud clump and $r_{0.5}$ are correlated, but empirically this is not yet established.
Since only about 10~per cent of the mass of a
molecular cloud is in dense clumps \citep{BattistiHeyer2014}, the
overall star formation efficiency per molecular cloud is abour 3~per
cent.  

In galaxies similar to the Milky Way, individual molecular cloud clumps and their embedded star clusters are 
unlikely to merge to form more massive clusters \citep[][and references therein]{Mahani+2021}. 
The molecular clouds follow a mass--radius relation, $M_{\rm mcl}/M_\odot \approx 36.7\,\left(R_{\rm mcl}/{\rm pc} \right)^{2.2}$. The free-fall time, $t_{\rm ff}=\left( 3\,\pi / (32 \, G \, \rho_{\rm mcl}) \right)^{1/2}$ with the density $\rho_{\rm mcl} = 3\,M_{\rm mcl} / (4\, \pi\, R_{\rm mcl}^3)$ such that $ t_{\rm ff}/{\rm Myr} = 1.37 \, \left(M_{\rm mcl}/M_\odot \right)^{0.18}$. Under Milky Way conditions more massive clouds are therefore likely to form more, and more massive, individual embedded clusters. However, these take longer to fall towards each other the more massive the clouds are. Since the molecular cloud clumps turn into largely gas-free expanding very young star clusters within 1-2\,Myr these do not have time to fall towards each other to merge. That most open star clusters in the Galaxy form singly and monolithically is also suggested by the analysis of the self-similar cluster structures in massive star-forming molecular clouds by \cite{Zhou+2024}. In strongly interacting gas-rich galaxies the radii of molecular clouds are smaller for a given mass because the inter stellar medium is pressurised through the encounter, and this allows the massive embedded clusters to merge. If this occurs, the IMF of a merged cluster complex is the sum of the stellar IMFs in each pre-merger embedded cluster (\citealt{Mahani+2021}, see also Sec.~\ref{chIMF:sec:UCDs}). 

\begin{BoxTypeA}[chIMF:box:ECL]{Embedded clusters}

\noindent These space-and time-correlated star formation events within a
molecular cloud can practically and stellar-dynamically be described
as embedded star clusters \citep{LadaLada2003} that form in
gravitationally collapsing dense molecular cloud clumps ranging in
mass from $M_{\rm ecl, min}\approx 5\, M_\odot$ 
to $M_{\rm ecl}> {\rm many}\,10^6\, M_\odot$ in extreme
star-burst galaxies where convergent gas flows create large localised
gas densities. 

\end{BoxTypeA}

Across all stellar masses, a proto-star grows to
about 90~per cent of its final mass within about $10^5\,$yr
\citep{WuchterlTscharnuter2003,Duarte-Cabrall+2013}, with growth
tapering out through the competition between the proto-star's feedback
energy and outflow versus accretion \citep{AdamsFatuzzo1996, Dib+2011}. Within a
time $<10^5\,$yr the proto-star thus decouples from the hydrodynamical
motions in its filament and becomes a ballistic pre-main-sequence star subject to
strong gravitational forces in close encounters with other such
pre-main sequence stars. Because the embedded cluster assembles over
$t_{\rm ff}\approx\,$few\,Myr, its dynamical state at the time when
star formation ceases due to gas blow out can be approximated to be
virialised. These early processes on a time scale of a~Myr have
important implications on observationally deducing an IMF as
stellar-dynamical ejections, stellar mergers and binary-star disruptions
affect the young stellar population rendering the stellar IMF as not being observable (Sec.~\ref{chIMF:sec:existence}).
Stellar-feedback driven outflow of the residual gas
leads to stellar-dynamical expansion of the embedded clusters forming,
once the molecular cloud has dissolved, T-Tauri or OB associations
with remnants of the expanded embedded clusters surviving as open star
clusters \citep{Kroupa2008}. That young stellar populations are expanding has been confirmed by \citealt{Guilherme-Garcia+2023, Wright+2024, Jadhav+2024}.

Given that the stellar IMF is directly linked to the birth of stars in molecular cloud clumps, it is natural to seek for its origin in the hydrodynamics of the molecular cloud clump. The inter-stellar medium (ISM) is a turbulent fluid, so the shape of the stellar IMF is likely to be influenced by the cascade of kinetic energy at different scales in the ISM. Incidentally, the power spectrum, $P(k_{\rm p})\propto k_{\rm p}^{-\gamma_{\rm p}}$, of this cascade is described by a power law with an exponent of $\gamma_{\rm p}=-5/3$ for subsonic turbulence, and by a log-normal form ($\gamma_{\rm p}=-2$) for supersonic turbulence. Similarly then, under this premise the stellar IMF may be described by a combination of these two functions, a power law and a log-normal form. This is the rationale for the gravo-turbulent theory for the stellar IMF, and a detailed study incorporating the structure of molecular clouds is provided by \cite{Elmegreen2011}.
Under the gravo-turbulent theory, density peaks form in the shockwave fronts. These peaks, when dense enough, collapse to proto-stars. The stellar IMF can be deduced from the theoretical distribution of such peaks. But detailed simulations have shown that the low-mass density peaks  are destroyed by the next shock before they can collapse
to a proto-star. Therefore, it is implausible that the shape of the stellar IMF is driven by turbulence \citep{Bertelli+2016}. Furthermore, the gravo-turbulent theory of the stellar IMF is severely challenged by the observational result \citep[e.g.,][see also Sec.~\ref{chIMF:sec:emergenceLT}]{Andre+2014} that stars form in thin molecular cloud filaments with near-regularly spaced proto-stars. Such structures are not compatible with the stochastic appearance of density peaks in a turbulent medium.

\subsection{Emergence of the stellar IMF of late-type stars}
\label{chIMF:sec:emergenceLT}

Important progress in understanding the theoretical origin of the
stellar IMF came with the observation that late-type stars mostly form in certain regions of molecular clouds, specifically, 
$0.1\,$pc-thick and up to 1~pc long, cold, kinematically coherent filaments \citep{Larson1992, Myers2009, Andre+2010}. The detailed filamentary fine structure of the molecular cloud clump and its embedded proto-stars  has been documented in one of the nearest star-forming embedded clusters by \cite{Hacar+2017}.  
Assuming that pre-stellar cores form
in thermally virialised filaments and that the fragmentation of a
filament depends only on the line mass-density (mass per unit length)
of the parent filament the mass function of molecular cloud cores within which proto-stars form can be studied. 
A filament of total mass $M_{\rm tot, fil}$
and length $L$ has a line-mass $M_{\rm line}=M_{\rm tot, fil} / L$. It
evolves (or breaks up) into a set of overdensities, i.e.  pre-stellar cores if the line mass is larger than the critical mass, $M_{\rm line} > M_{\rm line,crit} = 2c_{\rm s}^2/G$, ignoring non-thermal motions and magnetic fields. Gas
falls along the filament into these cores feeding the proto-stellar multiple systems
forming there which self-regulates this accretion through feedback.
The mass function of pre-stellar cores emanating from a system of
molecular cloud filaments forming an embedded cluster is thus given by
integration over the filaments \citep{Andre+2019},
\begin{equation}
  \xi_{\rm psc}(m_{\rm psc}) = \int \, f_{M_{\rm line}}(m_{\rm psc})\,w(M_{\rm line}) \, g(M_{\rm
  line}) \, {\rm dlog}_{10}M_{\rm line} \, ,
  \label{chIMF:eq:filaments}
\end{equation}
where $f_{M_{\rm line}}(m_{\rm psc})$ is the differential mass function of pre-stellar cores in a filament of line-mass $M_{\rm line}$, $w(M_{\rm line})$ is the relative $M_{\rm line}$-dependent weight calculated from the pre-stellar formation efficiency and the line length, and $g(M_{\rm line})$ is the mass function of filaments, i.e. the differential number of filaments per unit length per unit log line mass.  This allows \cite{Andre+2019} to explore different toy models of filamentary structures and their relation to $\xi_{\rm psc}$, with further details being available in \cite{Zhang+2024}. The relation of the core-mass function, $\xi_{\rm psc}(m_{\rm psc})$, to the stellar IMF, $\xi(m)$, depends on the role the feedback from the forming proto-stars plays in throttling the mass growth of a proto-star (e.g. \citealt{AdamsFatuzzo1996, Dib+2011}, but the shape of the stellar IMF for late-type stars ($m<\,0.5-1.0\,M_\odot$) appears to be largely set through the primary fragmentation of the filaments.  A sufficiently long filament of uniform line-density will form similarly-spaced cores of similar mass. This theoretically suggests that the stellar IMF flattens at the low-mass end and may be a reason for $\alpha_1\approx\alpha_2-1$. In very low-mass embedded clusters as observed in the Taurus-Auriga molecular cloud, the flattening may be even more extreme ($\alpha_1\approx\alpha_2\approx 0$, \citealt{Luhman2004}. This star-forming region is spawning embedded clusters of very small mass \citep{Joncour+2018} in which the binary fraction remains high \citep{Kroupa1995a} such that the observationally-deduced IMF is still consistent with the canonical stellar mass (fig.7 in \citealt{ThiesKroupa2007}).

\subsection{Emergence of the stellar IMF of early-type stars}
\label{chIMF:sec:emergenceET}

At the intersection of filaments reside embedded clusters. Accretion rates along the filaments increase toward these intersections, reaching a peak in embedded clusters (e.g. \citealt{Kirk+2013}. Indeed, the most~massive~stars form in  concentrated regions in the centers of these embedded clusters \citep{Kirk+2011, Pavlik+2019, Zhou+2022}. More massive stars thus appear to be formed through more violent accretion and
dynamical processes in the densest regions of the embedded clusters and
may not be related directly to the hydrodynamical evolution of
individual filaments. 

In an embedded cluster with a mass $M_{\rm ecl}$, stars form with
masses ranging from the metallicity dependent hydrogen burning mass
limit $m_{\rm H}$ to a maximum stellar mass $m_{\rm max}(M_{\rm ecl}) \le m_{\rm max*}$ the empirically-determined upper limit of which is 
$m_{\rm max*} \approx 150\,M_\odot$
(\citealt{Kroupa+2013} and references therein). While the value of $m_{\rm max*}$ is not fully
understood, it is likely related to the feedback-self-regulation of
accretion in connection with stellar stability. A metallicity
dependence is expected but has not yet been confirmed.  While stars
with $m>m_{\rm max*}$ have been observed, these are likely to result
from stellar mergers that are common in initially binary-rich
populations in the massive and dense embedded clusters
\citep{BanerjeeKroupa2012}.  

\begin{figure}[t]
\centering
\includegraphics[width=.6\textwidth]{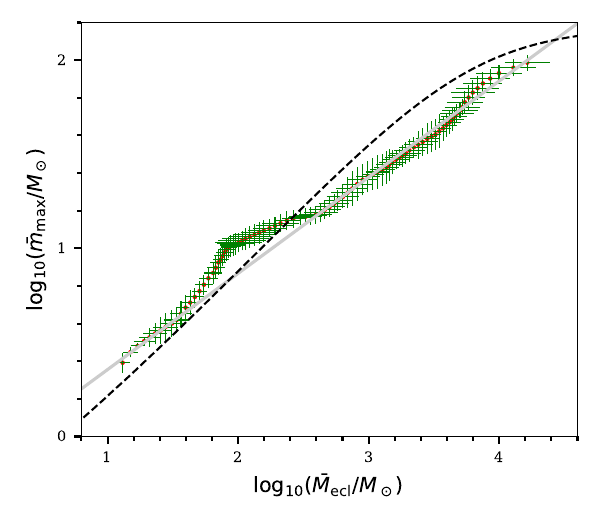}
\caption{
The observed correlation between the most-massive star, $m_{\rm max}$, and the stellar mass of the embedded cluster, $M_{\rm ecl}$, for nearby very young populations. The red data points constitute a running-mean of the observational data with associated uncertainties in green. The grey line is a linear relation to emphasise the flattening near ${\bar m}_{\rm max} \approx 10\,M_\odot$ which may be related to feedback processes affecting the formation of ionising stars. The solution (dashed black line) to eqs.~\ref{chIMF:eq:mmax_Mecl1} and~\ref{chIMF:eq:mmax_Mecl2} for the canonical stellar IMF  (eqs~\ref{chIMF:eq:canIMF1}--\ref{chIMF:eq:canIMF3}) is an adequate approximation to the data.  The observational data have a dispersion of $m_{\rm max}$ values at a given $M_{\rm ecl}$ which is consistent with the observational uncertainties leaving no evidence for an intrinsic dispersion \citep{Weidner+2013, Yan+2023} indicating a deterministic self-regulated star-formation process within a molecular cloud clump (i.e. an embedded star cluster) suggesting optimal sampling to be physically relevant (Sec.~\ref{chIMF:sec:sampling}). Adapted from fig.~6 in \cite{Yan+2023}.}
\label{chIMF:fig:mmax}
\end{figure}

Surveys of nearby molecular clouds and in the Large Magellanic Cloud 
uncover that the most massive star's mass,
$m_{\rm max}$, in an embedded cluster depends on the cluster's stellar
mass, $M_{\rm ecl}$. This $m_{\rm max}=m_{\rm max}(M_{\rm ecl})$ relation 
is shown in Fig.~\ref{chIMF:fig:mmax} and is
probably the result of feedback-modulation of the fragmentation and
accretion-growth of proto-stars, given the available mass in the
molecular cloud clump (\citealt{Yan+2023} and references therein). 
The bolometric luminosities of molecular cloud clumps in a volume-complete survey of Galactic molecular clouds are well represented by a model in which the clumps are spawning embedded clusters with accreting stars optimally sampled (Sec.~\ref{chIMF:sec:sampling}) from the canonical stellar IMF such that they obey this $m_{\rm max}=m_{\rm max}(M_{\rm ecl})$ relation \citep{Zhou+2024b}.
Thus, low-mass embedded
clusters ($M_{\rm ecl} < \, {\rm few}\,M_\odot$) cannot form stars
more massive than a few$\,M_\odot$ -- as particularly evident 
in the Orion star-forming region \citep{Hsu+2012}.
A good approximation to the empirically gauged
$m_{\rm max}=m_{\rm max}(M_{\rm ecl})$ relation (Fig.~\ref{chIMF:fig:mmax}) can be obtained by solving the two
equations,
\begin{equation}
1= \int_{m_{\rm max}}^{m_{\rm max*}}  \, \xi(m)\, \rm{d}m \, ,
\label{chIMF:eq:mmax_Mecl1}
\end{equation}
 \begin{equation}
M_{\rm ecl} =\int_{m_{\rm H}}^{m_{\rm max}} \,m\, \xi(m)\, \rm{d}m \, ,
\label{chIMF:eq:mmax_Mecl2}
\end{equation}
that link the one most massive star in an embedded cluster to the mass
in stars of the embedded cluster, respectively, for the normalisation
constant and $m_{\rm max}$.
The initial stellar population in an embedded cluster thus depends on
its mass through the $m_{\rm max}=m_{\rm max}(M_{\rm ecl})$ relation therewith 
constituting a variation of the stellar IMF without it changing its shape. 

\begin{BoxTypeA}[chIMF:box:mmaxMecl]{Star formation is feedback regulated}

\noindent The correlation
$m_{\rm max}=m_{\rm max}(M_{\rm ecl})$ (Fig.~\ref{chIMF:fig:mmax}) indicates that the
star-formation process within an embedded cluster may be highly
self-regulated, as is also implied by the small value of $\epsilon$ (see also Sec.~\ref{chIMF:sec:varIMF}).

\end{BoxTypeA}

That feedback-self-regulation in an embedded cluster may be a dominant
process rather than the Jeans mass in setting the stellar IMF has been
considered theoretically by \cite{AdamsFatuzzo1996, Dib+2011}. Recent
magneto-hydrodynamical simulations that take into account stellar
feedback (e.g. \citealt{Bate2019, Gonzalez+2020, Li+2021, Grudic+2022,
  Lewis+2023}) are approaching a high degree of realism but do not yet
reach reliable predictions concerning the shape of the stellar IMF and
its variation with physical conditions. Such numerical work needs to
demonstrate realism by reproducing the observed star-formation
activity in the nearby Taurus-Auriga molecular cloud in the form of
the many small embedded clusters, or ``NESTs''
\citep{Joncour+2018}.  Also, simulations have not yet been able to reproduce the Orion Nebula Cluster (ONC). The ONC  is the nearest star-forming site which formed and ejected ionizing stars (e.g. \citealt{Kroupa+2018, Jerabkova+2019}).

\subsection{The binary-star population}
\label{chIMF:sec:binaries}

\begin{figure}[t]
\centering
\includegraphics[width=.99\textwidth]{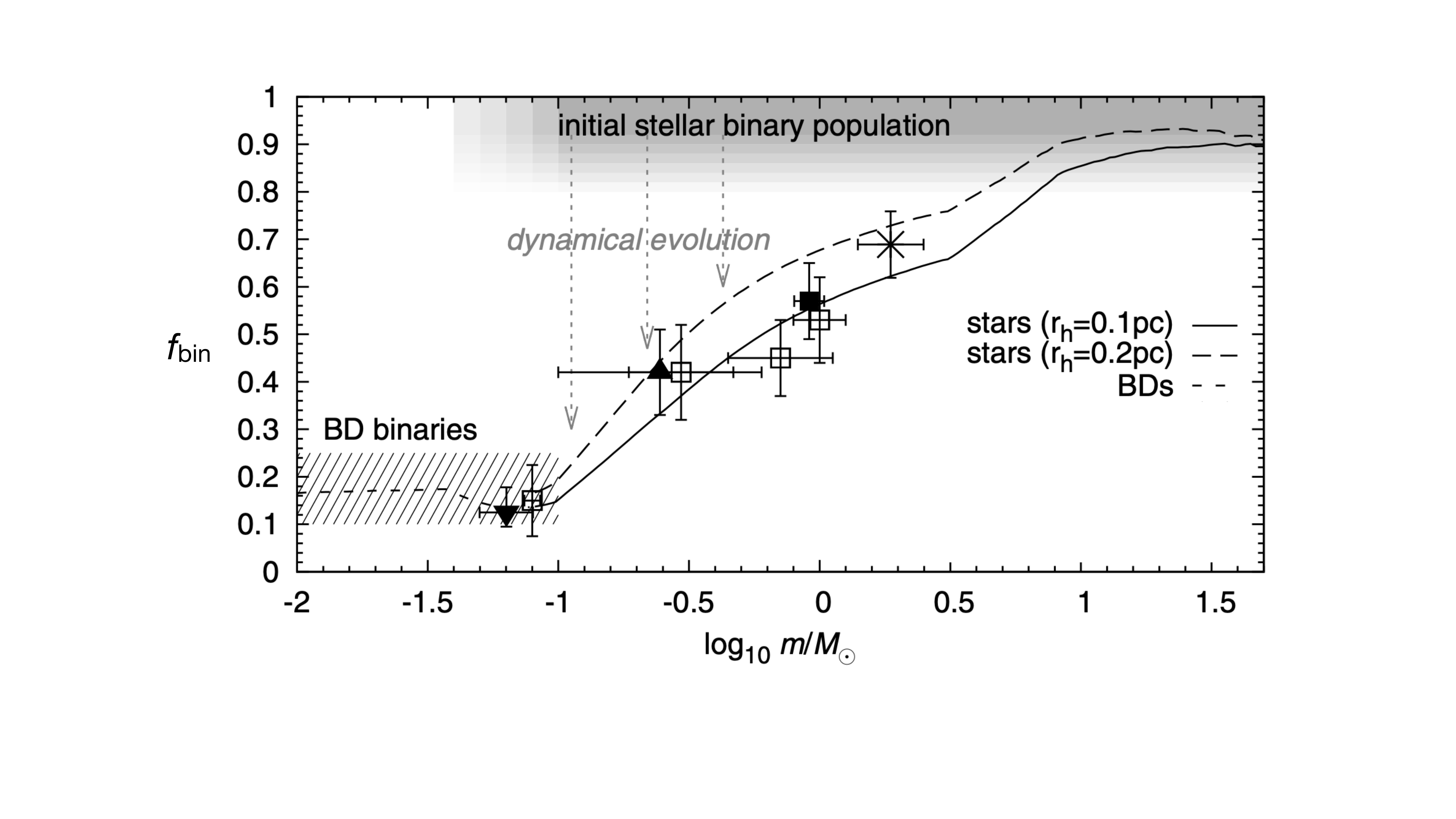}
\vspace{-15mm}
\caption{The binary fraction, $f_{\rm bin}$ (eqn.~\ref{chIMF:eq:fbin})
  as a function of the mass of the primary star. The extraction of the
  stellar IMF for any observed population of stars must carefully
  account for the different binary fractions in different environments
  and at different ages.  Thus, after the vast majority of stars form
  as binaries ($f_{\rm bin}\approx 1$), the binary fraction decreases
  through stellar-dynamical processing in the embedded clusters they
  were born in. The value of $f_{\rm bin}$ decreases and depends on
  the time when the stars are observed, on the initial density of the
  embedded cluster (here indicated by the half-mass radius,
  $r_{\rm h}$) and on the mass of the primary because with a
  less-massive primary the binary system has a smaller binding energy
  and is more easily disrupted in the embedded cluster. The Galactic
  field population shown by the data (for details see fig.~6 in
  \citealt{Thies+2015}) corresponds to the integrated population of about
  5~Gyr old stars that stem from embedded clusters, while the
  early-type primaries ($m>1\,M_\odot$) are young and have high
  binding energies and thus more directly indicate the high binary
  fraction at stellar birth. BDs (the shaded region indicates the
  uncertainty of their binary fraction) and very low mass stars ($m<0.2\,M_\odot$, VLMSs) appear to be a continuous
  extension of the stellar population but are in fact disjoint as  explained in Sec.~\ref{chIMF:sec:BDs}. From fig.~6 in \cite{Thies+2015}. Reproduced with permission. }
\label{chIMF:fig:binaries}
\end{figure}

Collapsing molecular cloud cores spin up and the contracting system's angular momentum is shared between at least two proto-stars. Star-counts in about Myr old low-density populations show these to be dominated by binary stars such that the stellar-dynamical decay of initial triple or quadruple systems on their crossing-time scale of $10^4-10^5\,$yr requires the vast majority of stars to form as binary or stable hierarchical multiple systems \citep{Goodwin+2007, MoeDiStefano2017}.  Defining the binary fraction
\begin{equation}
f_{\rm bin}= N_{\rm bin}/\left(N_{\rm sing} + N_{\rm bin}\right) \, ,
\label{chIMF:eq:fbin}
\end{equation}
where $N_{\rm sing}, N_{\rm bin}$ are, respectively, the number of single and binary stars in the sample, stars at birth have $f_{\rm bin}\approx 1$ (Fig.~\ref{chIMF:fig:binaries}).  The birth distribution functions of the semi-major axes or periods, $f_{\rm P}(P)$, mass-ratios, $f_{\rm q}(q)$, and eccentricities, $f_{\rm e}(e)$, of late-type binary systems, and their evolution with time are well understood \citep{Kroupa1995b, Belloni+2017}. Most initial binaries have wide obits (eqn.~46 in \citealt{Kroupa+2013}) and companion masses randomly distributed from the canonical stellar IMF.  This is qualitatively consistent with most late-type stars being born in fragmenting molecular cloud filaments (Sec.~\ref{chIMF:sec:emergenceLT}).  Systems with orbital periods shorter than about a 1000~days tend to have similar-mass companions which can be understood in terms of pre-main sequence eigenevolution (fig.~17 in \citealt{Kroupa2008}). The correction of star counts for unresolved stellar companions in any stellar population sensitively depends on how dynamically processed the observed population is (Fig.~\ref{chIMF:fig:binaries}). For example, a population of late-type stars that came from Taurus-Auriga type NESTs would retain a high late-type binary fraction, $f_{\rm bin}\approx 0.8$, while a population stemming from a dense embedded cluster similar or more massive than the ONC has $f_{\rm bin}<0.5$ \citep{Kroupa2008}. This leads to a deduced apparent variation of the stellar IMF if not accounted for in surveys of different young stellar populations.

Stars more massive than $\approx 1-5\,M_\odot$ follow different pairing
rules to late-type stars by having an initial distribution of periods
that is significantly shorter and an initial mass ratio distribution
with mass ratios $q=m_2/m_1 > 0.1$ ($m_1 > m_2$ being the primary and
secondary star masses, respectively). Also, while most late-type stars
form as star--star binaries, a large fraction of $m \simgreat 5\,M_\odot$ stars
appear to form in triple, duadruple or even higher-order systems
(e.g. \citealt{MoeDiStefano2017}, \citealt{KroupaJerabkova2021} and references therein).  This is probably related to early-type stars forming in the inner parts of more massive embedded clusters (Sec.~\ref{chIMF:sec:emergenceET}). 
The significantly larger
binding-energies of these massive star binaries and their preferred
formation in the innermost regions of their embedded clusters
 leads to them
being frequently violently stellar-dynamically ejected
\citep{OhKroupa2016}. The correction for unresolved mutliple
companions among the massive sellar binary population has a negligible
influence on the deduced IMF power-law slope. But near the transition mass
of about $5\,M_\odot$ where the initial multiplicity rules change, a
dip in the observationally deduced IMF is expected
\citep{KroupaJerabkova2018} even though the underlying stellar IMF is
given by the canonical two-part power law form of
Sec.~\ref{chIMF:sec:local}.

The massive ($m>5\,M_\odot$) part of the stellar IMF is thus subject
to complicated biases: for example, star counts in very young clusters
need to be corrected for the massive stars ejected from them. These
corrections are substantial and depend on the mass of the young star
cluster \citep{OhKroupa2016}. Also, wind mass loss can affect the
shape of the observationally deduced stellar IMF if not accounted for
correctly, and because most massive stars reside in binary systems,
rapid binary-stellar evolution affects the masses of the companions
through mutual mass overflow and accretion \citep{Schneider+2015}. The
detailed shape of the stellar IMF for massive stars remains thus
uncertain, although the canonical value, $\alpha_{3, \rm{ can}} = 2.3$,
serves as a benchmark.

The knowledge documented above is essential for uncovering the shape of the stellar IMF, but it is strictly only valid for near-Solar metallicity. The properties of the birth distribution functions of binaries are largely unknown for star formation at low metallicity. This will hopefully change in the future as large space-based telescopes may reach sufficient resolving power to allow an assessment of the stellar populations in star-forming regions in nearby low-metallicity dwarf galaxies (e.g. \citealt{Watts+2018}). Given observational data on a population of stars, the extraction of the birth functions, $f_{\rm P}(P), f_{\rm q}(q), f_{\rm e}(e)$, depend on the properties of the embedded clusters they stem from \citep{Kroupa1995a,Kroupa1995b}.  The relationship between these functions and the half-mass radii, $r_{0.5}$, of the embedded clusters at low metallicity need to be explored (e.g. \citealt{Marks+2022}). The observational constraints available to-date indicate that the old field population with $[Z]\approx -2$ has a similar binary fraction as the Solar neighbourhood ($[Z]\approx 0$, fig.~2 in \citealt{Carney+2005}). If $r_{0.5}([Z]\approx -2) < r_{0.5}([Z]\approx 0)$ (see Sec.~\ref{chIMF:sec:dynstr}) then this would indicate that $f_{\rm }(P, [Z]\approx -2)$ would be compressed to smaller $P$ if all stars also form as binaries at $[Z]\approx -2$ in order to allow a sufficient fraction of the initial binaries to survive the dynamical processing in the denser low-metallicity embedded clusters.

\subsection{Brown dwarfs}
\label{chIMF:sec:BDs}

It is often thought that BDs constitute the low-mass end of the stellar population.
But the probability that a density enhancement forms in a
filament that is sufficiently low-mass and dense to undergo
gravitational collapse without accreting gas to grow beyond
$m_{\rm H}$ is very small so that BDs are unlikely to arise by primary fragmentation of filaments as stars do. The IMF that emerges through primary
fragmentation of filaments thus falls off rapidly below about
$0.1\,M_\odot$, and in this context the observational finding that
$\alpha_{1, {\rm Kirk}} \approx 0.25 < \alpha_1 \approx 1.3$
(Sec.~\ref{chIMF:sec:local}) may be relevant. 

The origin of the large number of observed BDs -- about 1~BD per
4-5~stars -- therefore lies in the same embedded-cluster-forming
molecular cloud clumps but following different physical processes than
those of stars \citep{Thies+2015}. The hydrodynamical
simulations of accreting proto-stars by \cite{Thies+2010, Bate2012}
suggest BDs to mostly form through ``peripheral fragmentation'' at
distances of many dozens to hundreds of~AU from the proto-star.
For BD--BD binaries the simulations yield $f_{\rm bin, BD}\approx 0.15$
with a narrow distribution of semi-major axes around a few~AU
and a power-law mass function of BDs with $\alpha_{\rm BD}\approx 0.3$
therewith being well consistent with the observed population.  The
very small fraction amongst stars of observed star--BD systems (the brown dwarf desert) comes about from the wide BD orbits becoming unbound through
perturbations in the embedded cluster.

Understanding the properties of the initial binary population of stars
and of BDs and the stellar-dynamical evolution of the initial binary
population in embedded clusters is of essential importance for
deducing the canoncial IMF of stars and BDs. The high binary fraction
($f_{\rm bin}\approx 1$) in primary fragementation of a
filament (Sec.~\ref{chIMF:sec:binaries}) leads us to the {\it BD vs star
  origins problem}:

\begin{BoxTypeA}[chIMF:box:BD-starI]{The BD vs star origins problem I}
  
  \noindent BDs and very low mass stars (VLMSs) have significantly different
  binary properties than stars: {\bf (ia)}~most stars are born as star-star
  binaries {\bf (ib)} with periods ranging from days to millions of years,
  while {\bf (iia)}~only a small fraction of BDs are in binary systems with
  {\bf (iib)}~a tight semi-major axis distribution around~5AU,
  {\bf (iii)}~star--BD binaries are rare.

\end{BoxTypeA}

\noindent Therefore corrections of star and BD counts for unresolved companions
leads to the stellar IMF and BD IMF being distinct \citep{Kroupa+2013,
  Thies+2015} while overlapping (see Fig.~\ref{chIMF:fig:IMFshape} below). 
  Some VLMSs form by peripheral
fragmentation as most BDs do, while some massive BDs form as stars
from primary fragmentation of molecular cloud filaments. BDs thus form
with stars but following their own IMF, qualitatively similar to
planets forming along with stars but with their own mass
distribution.

\begin{BoxTypeA}[chIMF:box:BD-starII]{The BD vs star origins problem II}
  
  \noindent It is not possible to construct a population of
  stars and of BDs that fulfills the observed counts and at the same
  time the observed binary properties, unless stars and some of the massive BDs
  are treated separately from most BDs and some VLMSs in the initialisation. This is of critical
  importance for stellar and BD population synthesis.

\end{BoxTypeA}

\noindent While the ratio of BDs to stars probably varies with physical
conditions (e.g. on metallicity, in the presence or absence of ionising radiation, see
also \citealt{Bate2023}), such a variation has not yet been documented
in observed populations.

\section{The variation of the stellar IMF}
\label{chIMF:sec:varIMF}

\subsection{Theoretical considerations}
\label{chIMF:sec:varIMFtheory}

The cooling rate of metal-rich gas is significantly more rapid than that
of metal-free or metal-poor gas owing to the larger number of
electronic transitions in the former (e.g. \citealt{PloeckingerSchaye2020}). Also, the sound speed,
$c_{\rm g}$, is smaller in metal-rich than in metal-poor gas due to
the former's larger molecular weight, $\mu$, implying a smaller
$M_{\rm J}$.  A metal-rich gas filament in a cloud clump will thus fragment into
a larger number of less-massive proto-stars than an otherwise identical but
metal-free or metal-poor cloud clump, implying a stellar IMF that
contains a preponderance of low-mass stars (i.e. being bottom-heavy)
in the metal-rich case relative to the stellar IMF in the metal-poor
case.  The smaller sound speed in metal-rich than in metal-poor gas
limits the accretion rate onto proto-stars, slowing their mass
increase.  At the same time, metal-rich gas increases the
photon-to-matter coupling cross section such that accretion is also
more restrained through radiation feedback from the accreting
proto-star than in metal poor gas.  A metal-poor molecular cloud clump
is thus expected to produce an embedded cluster with a top-heavy
stellar IMF (containing relatively more massive stars) relative to a
metal-rich gas cloud. Additionally, magnetic fields (e.g. \citealt{KrumholzFederrath2019}) may play a role
inducing an additional metallicity dependence through the ionisation
fraction in the cloud  (removing an electron from a neutral H, C, Fe~atom requires, respectively, 13.6, 11.3, 7.9~eV), cosmic rays from nearby supernova explosions
lead to an increase in $T$ within the cloud clump
\citep{Papadopoulos2010}, and rotation of the molecular cloud clump
affects its fragmentation behaviour. In sufficiently dense molecular
cloud clumps, individual proto-stars may coalesce before they form
individual stars, thus pushing the emerging stellar IMF to be
top-heavy (e.g. \citealt{Dib2007} for simulations).  With the
formation times being about $10^5\,$yr for low- and massive stars (Sec.~\ref{chIMF:sec:dynstr}), the density
threshold of a molecular cloud clump above which the stellar IMF is
likely to become top-heavy through this coalescence process can be
estimated by equating the clump free fall time to $10^5\,$yr.  This
yields
$\rho_{\rm th, th} \approx 4\times 10^2\,\rho_{\rm g0} \approx 4\times
10^5/$cm$^3 \approx 10^4\,M_\odot/$pc$^3$
such that molecular cloud clumps with significantly larger density ought to be
forming embedded clusters with stellar IMFs that are noticeably
top-heavy.  

To attain a theoretical understanding of the change in the shape of
the IMF with physical conditions remains an active area of research (e.g. \citealt{HennebelleGrudic2024}).  
For example, 
\cite{Bate2023} points out that the metallicity
dependence through the cooling rate of gas of the stellar IMF may change with cosmic epoch given the evolving cosmological background
temperature.  The stellar IMF may have been bottom-light
and top-heavy in the early Universe due to the higher temperature of
the molecular gas being heated by the cosmic microwave background. This has
been studied by \cite{Jermyn+2018} in view of galaxies observed at a redshift $z>4$ appearing too massive in comparison to theoretically predicted ones. These authors suggest the stellar IMF to depend on the opacity, mean molecular weight, $\mu$, and ambient temperature, $T$, of the star-forming gas with the simplified form being given by
\begin{equation}
m_{{\rm s}0} = 0.08 \, \left( \frac{T}{T_0}  \right)^2 \, , \quad
m_{{\rm s}1} = 0.5\,M_\odot \, \left( \frac{T}{T_0}  \right)^2 \, , \quad
m_{{\rm s}2} = 1.0\,M_\odot \, \left( \frac{T}{T_0}  
\right)^2 \, .
\label{chIMF:eq:varIMF_T}
\end{equation}
where $m_{{\rm s}0}, m_{{\rm s}1}, m_{{\rm s}2}$, the power-law indices $\alpha_{1, 2, 3} = \alpha_{1, 2, 3, {\rm can}}$ are as in eqns~\ref{chIMF:eq:canIMF1}--\ref{chIMF:eq:canIMF3} and $T_0=20\,$K is the typical temperature of star-forming gas in the Galaxy.  This parametrisation (in comparison to eqn.~\ref{chIMF:eq:stIMF} below) thus allows the masses $m_{{\rm s}0}, m_{{\rm s}1}, m_{{\rm s}2}$ to change with $T$ but keeps the power-law indices invariant. It is thus possible to calculate the stellar IMF as a function of redshift, $z$: $\xi(m)=\xi(m : T(z))$.  This ansatz is used by \cite{Steinhardt+2022} to calculate $T(z)$ by fitting stellar population models generated with the above stellar IMF to observed galaxies to infer the evolution of $T(z)$ which can be compared to the background temperature of the cosmic microwave background. These authors thereby implicitly assume the galaxy-wide IMF (gwIMF) to be identical to the stellar IMF which need not be the case (Sec.~\ref{chIMF:sec:gwIMF}). This ansatz needs to be explored further by allowing for the time-evolving distribution of molecular cloud clumps within a galaxy and their fragmentation into stellar IMFs in dependence of their cooling rate as well as $T$.  Molecular clumps are expected to have different $T$ (e.g. a clump heated by cosmic rays from a nearby star-burst embedded cluster vs an isolated one in the outer galaxy) such that the gwIMF will in general not equal the stellar IMF.

The empirical approach to testing if the stellar IMF varies with physical conditions does not readily provide access to $T$ because when the stellar IMF of a given simple stellar population is constructed through star counts this information as been lost. Instead, the density of the molecular gas clump that formed the population can be estimated from the mass, $M_{\rm ecl}$, of this population and from its estimated half-mass radius at birth, $r_{0.5}$. These quantities can also be directly compared to observed clumps in molecular clouds to test for consistency (Sec.~\ref{chIMF:sec:dynstr}). Another quantity related to the ability of the molecular gas to fragment is the metallicity of the stars that formed from the gas, and this metallicity, $Z$, is also a measurable quantity. Observational data can thus provide information on $\xi(m)=\xi(m : M_{\rm ecl}, Z)$. Given such information, a composite or gwIMF can be calculated by integrating over $M_{\rm ecl}$ and $Z$. An additional dependency on $T$ (via the embedded-cluster-forming molecular cloud clump's pressure) may emerge by comparing, for example, galaxies observed at a high-redshift with models of the same galaxies obtained by the integration over all molecular cloud clumps in a galaxy since its formation. Discrepancies that arise between the model based on the stellar IMF as formulated by eqn~\ref{chIMF:eq:igimf_t} below may inform on this dependency.  In the next Sec.~\ref{chIMF:sec:varIMF_emp} evidence for a possible variation of the stellar IMF (i.e. on the cloud clump scale) is documented. The galaxy-wide problem is approached in Sec.~\ref{chIMF:sec:gwIMF}.

Thus, in general, the stellar IMF can be written in terms of a dependency on a set of $\{P_{\rm i}\}^{N_{\rm P}}$ physical parameters (e.g., $P_1=M_{\rm ecl}, P_2=Z, P_3=T, P_4=\omega$ with $\omega$ being the specific angular momentum of the molecular cloud clump), as

\begin{equation}
  \xi(m:\{P_{\rm i}\}^{N_{\rm {P}}}) = k_\xi \, k_{\rm i} \,
  m^{-\alpha_{\rm  i}(m : \{P_{\rm i}\}^{N_{\rm {P}}})}   \, ,
  \label{chIMF:eq:stIMFgen}
\end{equation}
where $k_{\rm i}$ assure continuity and $k_\xi$ is the normalisation constant ensuring solution of eqs~\ref{chIMF:eq:mmax_Mecl1}
and~\ref{chIMF:eq:mmax_Mecl2}, and the power-law indices, $\alpha_{\rm  i}(m : \{P_{\rm i}\}^{N_{\rm P}})$ are defined over some initial stellar mass ranges.

\subsection{Empirical evidence}
\label{chIMF:sec:varIMF_emp}

Do star-count data show evidence for a variation of the stellar IMF
with cloud clump density, temperature and metallicity as implied by the above? For
quantification purposes a few definitions are useful:

\begin{BoxTypeA}[chIMF:box:IMFshapes]{The relative shapes of IMFs}
  
  \noindent Relative to the canonical stellar IMF, a {\it top-heavy stellar IMF}
  is defined to have 
  \begin{equation}
      \alpha_3<2.3=\alpha_{\rm 3, can},
  \end{equation}
   while a {\it top-light stellar
    IMF} is defined to have 
  \begin{equation}
      \alpha_3>2.3=\alpha_{\rm 3, can}. 
  \end{equation} 
    A {\it bottom-heavy
    stellar IMF} is defined to have 
    \begin{eqnarray}
        \alpha_1  >1.3=\alpha_{\rm 1, can} \; , \; \alpha_2 >2.3=\alpha_{\rm 2, can},
    \end{eqnarray}
  while a {\it bottom-light stellar IMF} has
    \begin{eqnarray}
        \alpha_1   <1.3=\alpha_{\rm 1, can}  \; , \;\alpha_2  <2.3=\alpha_{\rm 2, can} .
    \end{eqnarray}
  A related definition is available through
  eqs~\ref{chIMF:eq:onenorm}--~\ref{chIMF:eq:shape2} below.

\end{BoxTypeA}

The stellar population forming in the observable part of the Galaxy is
limited to approximately Solar metallicity and embedded clusters with
$M_{\rm ecl} < \, {\rm few}\, \times 10^3\,M_\odot$ in stars such that any variation of the stellar IMF will be limited and observational surveys have been confirming this. 

\begin{figure}[t]
\centering
\includegraphics[width=.49\textwidth]{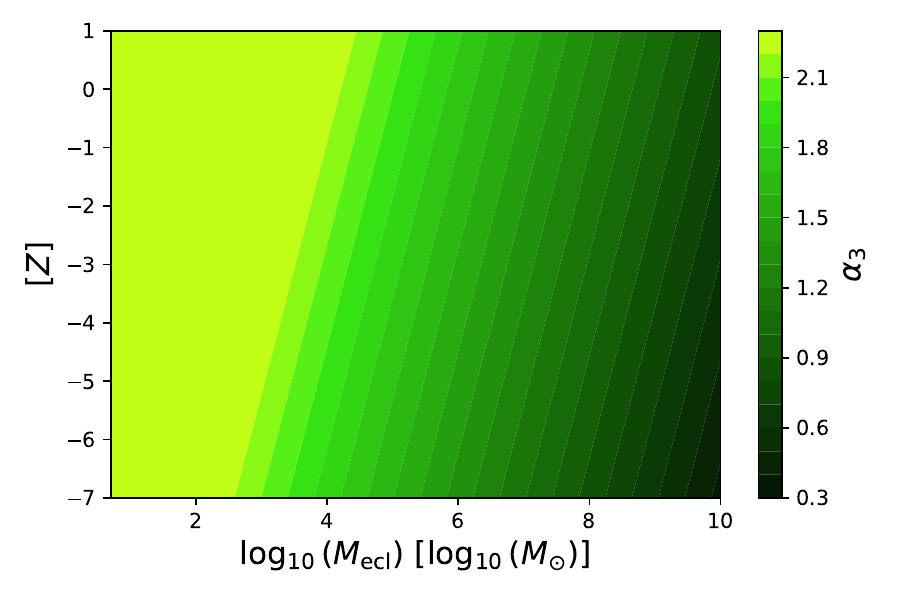}
\caption{The colour coding shows $\alpha_3$ values in dependence of [$Z$]
  and $M_{\rm ecl}$ (eqn.~\ref{chIMF:eq:alpha_3}). The canonical value, $\alpha_3=2.3$, is obtained in the lightest green area on the left. The
  stellar IMF is increasingly top-heavy for smaller metallicity,
  [$Z$], and larger stellar mass of the embedded cluster,
  $M_{\rm ecl}$, whereby this figure demonstrates that $\alpha_3$ mostly depends on the density of the molecular cloud clump, i.e. on $M_{\rm ecl}$. Reproducible with the {\tt pyIGIMF} software (Sec.~\ref{chIMF:sec:codes}). }
\label{chIMF:fig:Z_Mecl}
\end{figure}

The $1-4\,$Myr old and few$\,\times10^3\,M_\odot$ heavy Orion Nebula
Cluster is, with a distance of about $0.4\,$kpc, the closest
post-gas-expulsion embedded cluster that formed stars more massive
than $10\,M_\odot$.  It may have formed perfectly mass segregated
implying that the stellar IMF would be radially dependent within the
embedded cluster \citep{Pavlik+2019}. This case however also
demonstrates the complexity of the problem, because stellar-dynamical
activity has already shot out a substantial number of massive stars
such that the stellar IMF may have been top-heavy in the centermost
regions where the mass density surpassed $10^5\,M_\odot$/pc$^3$
\citep{Kroupa+2018}.  A top-heavy stellar IMF is implied for the
metal-poor star-burst R136 cluster
($M_{\rm ecl}\approx 10^5\,M_\odot$) once the star-counts are
corrected for the ejected massive stars \citep{BanerjeeKroupa2012b},
being supported by direct star counts in the 30\,Dor star-forming
region that harbours R136 \citep{Schneider+2018}.  Evidence for a
systematic variation of the stellar IMF has been uncovered from the
detailed analysis, taking into account the early gas-expulsion
process, of globular star clusters and ultra-compact dwarf galaxies (UCDs),
objects that formed more than a million stars (\citealt{Marks+2012},
see also \citealt{Dib2023}).  The power-law index for the high-mass
stellar IMF ($1.00 \leq m/M_{\odot} < m_{\rm max}$) is thus gauged by
empirical data to depend on the density and metallicity of the star
forming molecular gas cloud clump (\citealt{Haslbauer+2024} and referneces therein),
\begin{eqnarray}
     \centering
     \alpha_{3}=\alpha_3(M_{\rm ecl}, Z) = 
 \begin{cases}
 2.3  & \mathrm{for} \quad x < -0.87\, , \\
 -0.41 x + 1.94 & \mathrm{for} \quad x \geq -0.87\, , \\
 \end{cases}
     \label{chIMF:eq:alpha_3}
 \end{eqnarray}
 with
 \begin{eqnarray}
     \centering
     x~=~-0.14 [Z] + 0.6 \log_{10}\bigg(\frac{M_{\mathrm{ecl}}}{10^{6}M_{\odot}} \bigg)+2.82 \, ,
     \label{chIMF:eq:varIMF1}
 \end{eqnarray}
 where $M_{\rm ecl}$ is the initial stellar mass of the embedded star
 cluster with half-mass radius $r_{0.5}$ given by eqn.~\ref{chIMF:eq:r05},
 ${\rm [Z]}\equiv{\rm log}_{10}({\rm Z}/{\rm Z}_{\odot})$, $Z$ is its
 metallicity mass fraction (of elements beyond H and He) and
 $Z_\odot=0.0142$. This empirically-gauged stellar IMF becomes
 top-heavy in metal-poor and massive embedded clusters.
 Fig.~\ref{chIMF:fig:Z_Mecl} shows the dependency of $\alpha_3$ on  [$Z$] and $M_{\rm ecl}$.
Uncertainties remain in the exact form-variation of the stellar IMF
and it is possible that $\alpha_3>2.3$ for $x<-0.87$.  Given the range
of data available, the above formulation captures the variation of the
stellar IMF with density and metallicity for about $-4<[Z]<0.2$ and
$5< M_{\rm ecl}/M_\odot < 10^9$ beyond which the uncertainties are
very major. This can be seen that formally
$\alpha_3 \longrightarrow -\infty$ for $Z \longrightarrow 0$ (metal-free gas) which
is unphysical. Indeed, the stellar IMF of the first stars (commonly
referred to as the population~III IMF) remains largely unknown
(e.g. \citealt{KlessenGlover2023}).
That $M_{\rm ecl}$ and $Z$ may not be the only parameters that determine the shape of the stellar IMF is suggested by the metal-rich, high-density and strongly-magnetized star-forming region within about $0.1\,$pc of the Galactic center where \cite{Bartko+2010} report a significantly top-heavy stellar IMF. Only massive pre-stellar cores can collapse if their collapse time is shorter than the time the differential rotation around the Galactic centre shears the clouds apart \citep{NayakshinSunyaev2005, Hopkins+2024}.

 A tendency of the stellar IMF of late-type stars towards a bottom-light form for
 decreasing $Z$ was noted by \cite{Kroupa2002} from metal-poorer
 populations of stars in the Milky Way and its star clusters (see
 \citealt{Yan+2024} and references therein). A metallicity-dependency
 of the stellar IMF of late-type stars is supported by the
 spectroscopic analysis of 93000 Solar neighbourhood stars
 \citep{Li+2023}. In combination with data from recent work on
 early-type galaxies, the stellar IMF for late-type stars
 ($m<1\,M_\odot$) is gauged to depend on $Z$ \citep{Yan+2024},
\begin{equation}
\alpha_{1,2} = \alpha_{\rm 1,2, can} + 79.4\,\left(Z - Z_{\rm
    MW}\right) \, ,
\label{chIMF:eq:varIMF2}  
\end{equation}  
where ${\rm log}_{10} \left(Z_{\rm MW}/Z_\odot\right) = -0.10\pm 0.05$
is the average metallicity of the Solar neighbourhood stellar
ensemble. Metal-rich embedded star clusters ($Z>Z_\odot$) thus produce
a bottom-heavy stellar IMF in agreement with the constraints available
from elliptical galaxies \citep{Yan+2024}.  A major uncertainty
remains in this formulation for $Z \longrightarrow 0$, with
$\alpha_1 \longrightarrow 0.40, \alpha_2 \longrightarrow 1.40$. The
smallest mass that can form from direct collapse, the opacity limit
($\approx \, {\rm few}\, \times 10^{-3}\,M_\odot$), is given by the
ability of the gas to radiate thermal photons as it collapses and is
larger at low metallicity, but details remain unclear
(e.g. \citealt{HennebelleGrudic2024}). It is thus unknown how the IMF
of BDs depends on $Z$, but additional effects such as a strong
ionisation field from a top-heavy IMF and proto-stellar encounters are
likely to play an important role (e.g.,
\citealt{Kroupa+2003}). Searches for BDs in globular star clusters
will shed light on the variation of the shape of the sub-stellar IMF (e.g. \citealt{Marino+2024, Gerasimov+2024}).

\begin{BoxTypeA}[chIMF:box:stellarIMF_Mecl_Z]{The density and metallicity dependent stellar IMF}

\noindent Given the available observational knowledge ($N_{\rm P}=2$), the stellar IMF (eqn.~\ref{chIMF:eq:stIMFgen}) can be written in terms of a dependency on $M_{\rm ecl}$ (i.e. with $r_{0.5}$ and $\epsilon$ on the molecular cloud clump density) and $Z$ as
\begin{equation}
  \xi(m:M_{\rm ecl}, Z) = k_\xi \, k_{\rm i} \,
  m^{-\alpha_{\rm  i}(m : M_{\rm ecl}, Z)}   \, ,
  \label{chIMF:eq:stIMF}
\end{equation}
where $k_{\rm i}$ assure continuity, $k_\xi$ being the normalisation constant ensuring solution of eqs~\ref{chIMF:eq:mmax_Mecl1}
and~\ref{chIMF:eq:mmax_Mecl2} and the $\alpha_{\rm i}$ being given by eqns~\ref{chIMF:eq:alpha_3} and~\ref{chIMF:eq:varIMF2}. 

\end{BoxTypeA}

\begin{figure}[t]
\centering
\includegraphics[width=0.8\textwidth]{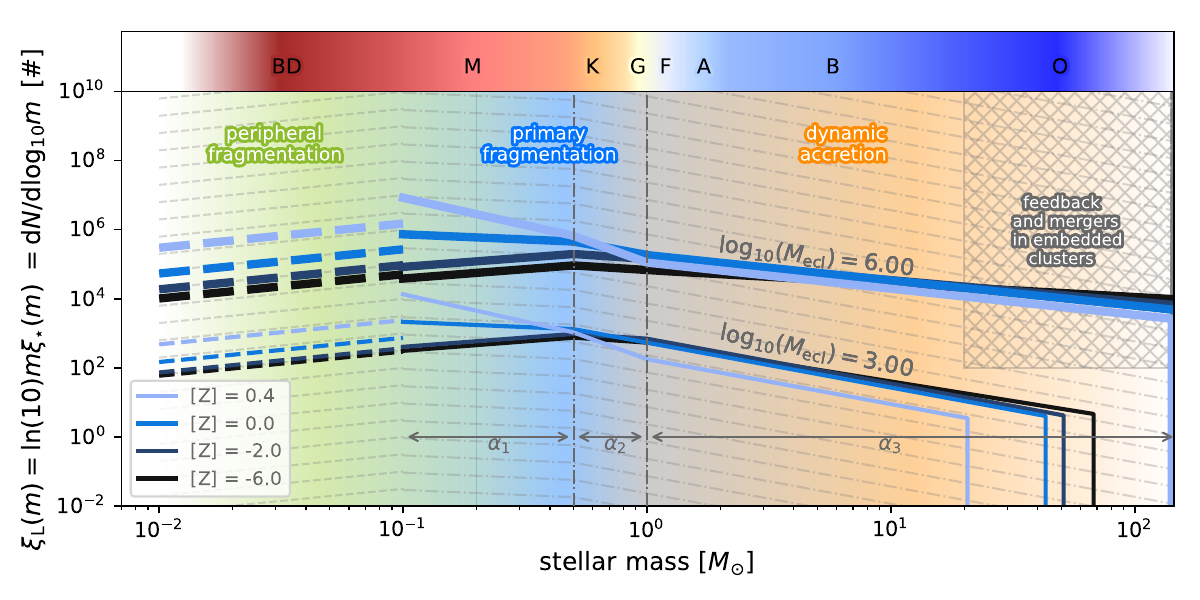}
\caption{The possible
  physical origin of the shape of the stellar IMF and its variation as
  encapsulated by eqn.~\ref{chIMF:eq:stIMF}. BDs and some VLMSs form through
  peripheral fragmentation in the outer regions of
  proto-stellar accretion  discs and late-type stars though primary fragmentation of
  molecular cloud filaments in the molecular cloud clump.
  Early-type stars increasingly (with larger $m$) self-regulate their accretion
  in the molecular cloud clump and are also increasingly
  (with larger $m$) subject to coalescence in the densest regions of
  the molecular cloud clump.  The resulting stellar IMF becomes
  increasingly top-heavy for larger
  stellar mass of the embedded cluster, $M_{\rm ecl}$, and for smaller metallicity, [$Z$], and 
increasingly bottom-heavy
  for increasingly larger [$Z$].  The stellar IMFs are normalised such
  that the area under the stellar IMF corresponds to the number of stars and have no stars for
  $m>m_{\rm max}(M_{\rm ecl})$. The dependency of $m_{\rm max}$ on $M_{\rm ecl}, Z$ comes from the conditions given by eqs.~\ref{chIMF:eq:mmax_Mecl1}
and~\ref{chIMF:eq:mmax_Mecl2}.
  The sub-stellar IMFs are shown as the dashed lines which overlap with the stellar IMFs and are normalised such that they contain $1/4.5$~the number of stars in their associated stellar population. For the late-type stars a dependency on the density of the gas cloud clump (via $M_{\rm ecl})$ is unknown, while for the BD IMF information on its possible dependency on $M_{\rm ecl}$ and $Z$ is unknown. 
  The thinner stellar and sub-stellar IMFs are computed for an embedded cluster with a stellar mass of $M_{\rm ecl}=10^3 M_{\odot}$, whereas the thicker lines represent computations for $10^6 M_{\odot}$. The color gradient of the IMFs from darker to lighter goes from lower to higher metallicity.   }
\label{chIMF:fig:IMFshape} 
\end{figure}

\begin{figure}[t]
\centering
\includegraphics[width=0.49\textwidth]
{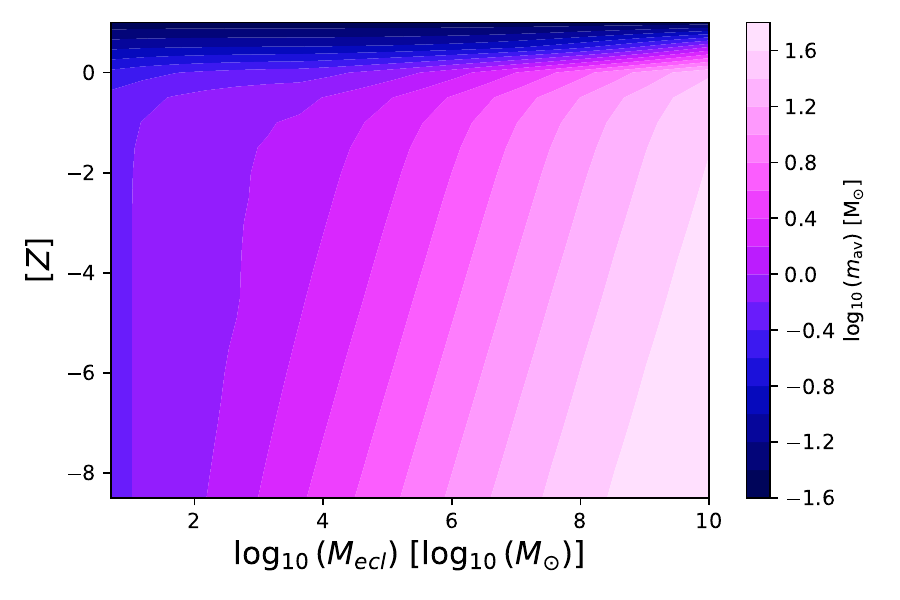}~\includegraphics[width=0.49\textwidth]{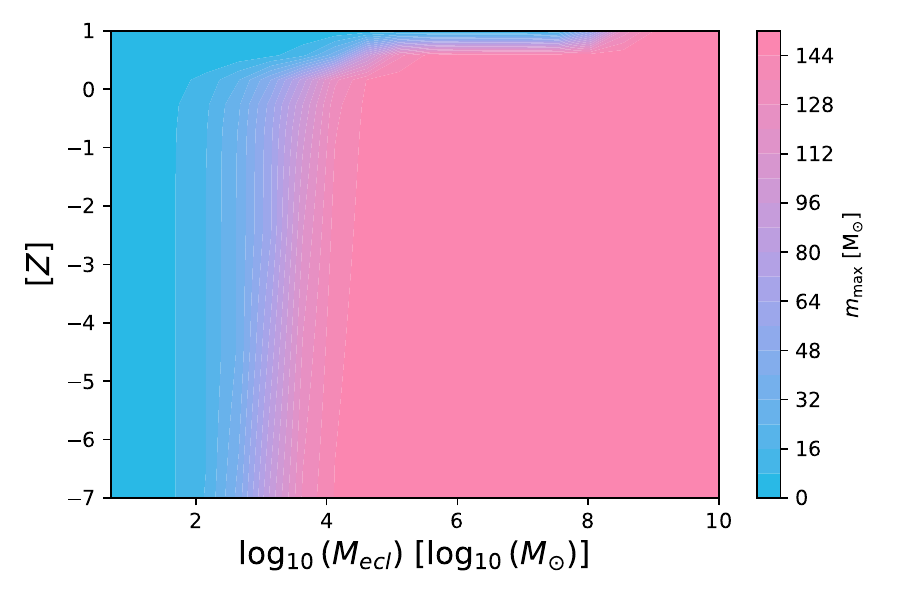}
\caption{The dependency of the average stellar mass, $m_{\rm av}$ (eqn~\ref{chIMF:eq:mav}, left panel), and of the maximum stellar mass, $m_{\rm max}$ (eqs.~\ref{chIMF:eq:mmax_Mecl1}
and~\ref{chIMF:eq:mmax_Mecl2}, right panel), on $M_{\rm ecl}$ and [$Z$].
 Reproducible with the {\tt pyIGIMF} software (Sec.~\ref{chIMF:sec:codes}). }
\label{chIMF:fig:mav_mmax1} 
\end{figure}

\noindent 
Note in comparison with eqn.~\ref{chIMF:eq:varIMF_T} that this parametrisation allows the 
power-law indices to change with $M_{\rm ecl}$ and $Z$ but keeps the 
masses $m_{{\rm s}0}, m_{{\rm s}1}, m_{{\rm s}2}$ 
invariant.
The variation of the stellar IMF, the mass-ranges over which dominant physical processes in shaping it probably play a role and its relation to the sub-stellar population is visualised in Fig.~\ref{chIMF:fig:IMFshape}. 
The dependency of the average stellar mass and of the most-massive star's mass on the mass in stars of the embedded cluster and on metallicity are displayed in Fig.~\ref{chIMF:fig:mav_mmax1}. In reality more massive stars can appear through mergers of binary components in the young binary-rich clusters \citep{BanerjeeKroupa2012}. The calculations show:

\begin{BoxTypeA}[chIMF:box:massiveclusters]{Massive star clusters as ionisation sources in the early Universe and as dark star clusters at super-Solar metallicity}

\noindent At low metallicity, massive ($M_{\rm ecl} \simgreat 10^7\,M_\odot$) clusters have a stellar population with an average stellar mass $m_{\rm av} \simgreat 10\,M_\odot$ and would be significant ionising sources in the early Universe reaching quasar luminosities while appearing as UCDs today \citep{Jerabkova+2017}. At high metallicity ($[Z]>0$), even massive clusters ($10^4 \simless M_{\rm ecl}/M_\odot \simless 10^8$) lack massive stars ($m_{\rm max} \simless 10\,M_\odot$) suggesting that very massive "dark clusters" can form in star-bursts at high metallicity. Such objects might be mistaken as being diffuse and only of low mass if an invariant canonical stellar IMF is assumed in the analysis of the observations. 

\end{BoxTypeA}

\begin{figure}[t]
\centering
\includegraphics[width=.49\textwidth]{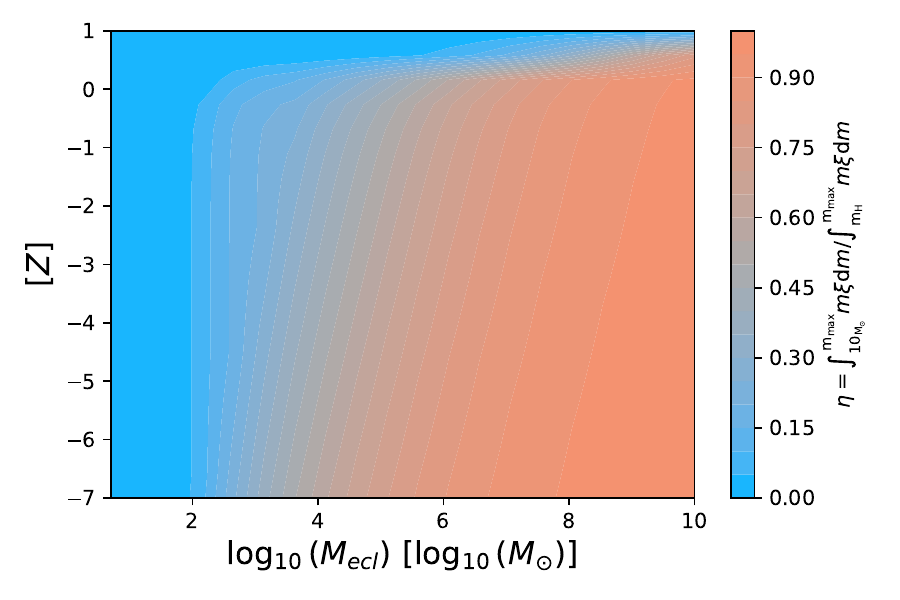}
\caption{The mass fraction, $\eta$ (eqn.~\ref{chIMF:eq:massfr}), in stars more massive than $10\,M_\odot$ in an initial
  stellar population  born in an embedded
  cluster as a function of its embedded cluster stellar mass, $M_{\rm ecl}$,  and its metallicity, [$Z$]. When $\eta>0.5$ the embedded cluster is likely to have become unbound after mass loss through stellar evolution. Values of $\eta=0$ arise through the $m_{\rm max}=m_{\rm max}(M_{\rm ecl})$ relation (Fig.~\ref{chIMF:fig:mmax}) which has an implicit metallicity dependence through eqs.~\ref{chIMF:eq:mmax_Mecl1} and~\ref{chIMF:eq:mmax_Mecl2}).  Reproducible with the {\tt pyIGIMF} software (Sec.~\ref{chIMF:sec:codes}).  }
\label{chIMF:fig:massfr}
\end{figure}

To help to assess how the stellar IMF dictates which embedded clusters are
likely to remain bound after loss of mass through stellar evolution,
the mass fraction in stars more massive than $10\,M_\odot$,
\begin{equation}
  \eta = \frac{
    \int_{10\,M_\odot}^{m_{\rm max}} \, \xi(m:M_{\rm ecl}, Z) \, m \, {\rm d}m
    }{
    \int_{m_{\rm H}}^{m_{\rm max}} \, \xi(m:M_{\rm ecl}, Z) \, m \, {\rm d}m
    } \, ,
  \label{chIMF:eq:massfr}
\end{equation}
is defined. When $\eta > 0.5$ the cluster is likely to have dissolved
when the last supernova has exploded (Fig.~\ref{chIMF:fig:massfr}).

\subsection{Discussion/caveats}

The temperature dependence discussed in Sec.~\ref{chIMF:sec:varIMFtheory} is at least partially implicitly contained in eqs.~\ref{chIMF:eq:alpha_3} and~\ref{chIMF:eq:varIMF2} through the explicit density and metallicity dependence.  Essentially, this formulation assumes that star formation always occurs at a similar low temperature near a dozen or two~dozen~K.  It is not likely to be a complete description though because the pressure in the molecular cloud clump does not appear explicitly -- a star-forming molecular cloud clump of a given density and thus $M_{\rm ecl}$-value can have different pressures depending on the temperature. For the time being no observational information exists on the dependency of this third parameter. It may be uncovered by comparing the gwIMFs calculated via the IGIMF theory (Sec.~\ref{chIMF:sec:gwIMF}) with the stellar populations of observed galaxies for which $T$ can be independently measured.

The dependency of the stellar IMF on the density and metallicity of the star-forming gas is thus still debated and not generally accepted.  For example, \cite{Baumgardt+2023} find 120~globular clusters (GCs) to have bottom-light IMFs but argue that a metallicity variation cannot be confirmed. As these are of low metallicity, this result does appear to be qualitatively consistent with eqn.~\ref{chIMF:eq:varIMF2}. \cite{Baumgardt+2023} also suggest GCs to have had more massive stars than provided by the canonical stellar IMF, again in qualitative agreement with eqn.~\ref{chIMF:eq:alpha_3}.  \cite{Dickson+2023} perform pure stellar-dynamical modelling of~37 Galactic GCs arguing that the data and modelling finds no evidence for a variation of the stellar IMF with density or metallicity in contradiction to eqn.~\ref{chIMF:eq:alpha_3}. The analysis of Galactic GCs is difficult because they have different Galactic orbits that determine the loss of low-mass stars and may depend on the metallicity as they reflect the formation process of the very young Galaxy.  The very early embedded phase and the violent emergence from it also play a role in shaping a present-day GC.

The stellar IMF dependency on density and metallicity as formulated by eqn.~\ref{chIMF:eq:alpha_3} rests on the analysis by \cite{Marks+2012} who used data for ultra-compact dwarf galaxies (UCDs), and the correlation between the concentration of 20~GCs and the power-law index of the stellar mass function at the half-mass radius of the GC where it represents the global stellar mass function of the GC (\citealt{deMarchi+2007} and references therein). The deficit of low-mass stars in GCs with lower concentration cannot be understood through stellar-dynamical evolution of GCs and is assumed instead to be due to a more top-heavy stellar IMF needed to violently expand the very young mass-segregated GC through feedback self-regulation in the formation process.  This approach leads to a strong correlation between $\alpha_3$, the GC-forming molecular cloud clump density and [$Z$] (eqn.~\ref{chIMF:eq:alpha_3}).  Data on the dynamical mass-to-light ratios and on the numbers of low-mass X-ray emitting sources in UCDs are consistent with this variation, as are the star-count data in the sub-solar metallicity starburst region of~30~Doradus in the Large Magellanic Cloud \citep{Schneider+2018}. The observed correlation of the dynamical mass-to-light ratios with metallicity of 163~GCs in the Andromeda galaxy is explained by the above density and metallicity dependency of the stellar IMF \citep{Haghi+2017}.

Thus, while more data analysis in conjunction with very detailed
stellar-dynamical and astrophysical understanding of the studied
systems is required, overall the available information suggests that
the stellar IMF may have a density and metallicity dependent form. 
Direct observational information on the stellar IMF's
dependency on other parameters such as the rotation of the embedded
cluster, magnetic field strengths and the temperature of the star-forming gas
has not emerged yet. 

\begin{BoxTypeA}[chIMF:box:consistencytest1]{Consistency Test on IMF
    variation}

  \noindent Any proposed formulation of the variation of the stellar IMF on the molecular cloud clump
scale (i.e. in an embedded star cluster), such as formulated by eqn.~\ref{chIMF:eq:stIMF}, needs to pass two conditions:

1) The galaxy-wide IMF calculated from it (Sec.~\ref{chIMF:sec:gwIMF})
must be consistent with the Solar neighbourhood constraint, i.e. the
canonical field-star IMF (eqns~\ref{chIMF:eq:canIMF1}--\ref{chIMF:eq:canIMF2} and~\ref{chIMF:eq:canIMF4}).

2) The galaxy-wide IMFs constructed from it need to come out
consistent with the observational constraints on the galaxy-wide IMFs
in star-forming dwarf, major disk galaxies and elliptical
galaxies (Fig.~\ref{chIMF:fig:extragalactic} below). This is addressed in Sec.~\ref{chIMF:sec:galobs}.
\end{BoxTypeA}

The variation of the stellar IMF formulated by
eqn.~\ref{chIMF:eq:stIMF} fulfills both conditions.

\section{Stochastic or optimal? The nature of the IMF}
\label{chIMF:sec:sampling}

The physical evolution of star clusters and of galaxies critically depends on the nature of the stellar IMF. When an embedded star cluster forms in a molecular cloud clump, is the sequence of stellar masses that appear in it (I)~purely random or (II)~deterministic? The former case~(I) means that 
forming stars do not influence each other and the stellar IMF must be interpreted as a probability density distribution function. 
The latter case~(II) comes about if the forming stars influence each other such that the sequence of stellar masses is strictly determined by the initial conditions (e.g. mass, temperature, rotation, magnetic field) defining the clump. In this case the stellar IMF can be described as being  an optimal distribution function \citep{Kroupa+2013}.
According to case~(I), low mass clumps can contain massive ionising stars, while according to~(II) this is never possible. Optimal sampling (case~II) thus implies the possibility of dim but significant star formation occurring in a dark system.
The mathematical procedures how to sample masses from the stellar IMF are described in detail in \cite{Kroupa+2013}, \cite{Schulz+2015} and \cite{Yan+2017}.  There are thus the 
two extreme interpretations (cases~I and~II) of the stellar IMF:

\begin{itemize}

\item[(Ia)] The stellar IMF is a probability density
distribution function. This means that a model stellar population 
consists of an ensemble of $N_{\rm s}$ stellar masses randomly drawn from the stellar IMF (e.g. \citealt{MaschbergerClarke2008}). After choosing $N_{\rm s}$, the stellar ensemble is constructed by randomly/stochastically drawing stellar masses from the stellar IMF until $N_{\rm s}$ stars are assembled.  Stochastic sampling has no constraints apart from the shape of the parent distribution function, i.e. of the stellar IMF, and the number of stars drawn from the stellar IMF, $N_{\rm s}$, defines the final mass of the population, $M_{\rm stars}$, through the sum of the drawn stellar masses.
Two populations of the same
mass, $M_{\rm stars}$, will thus contain different numbers and sets of stars.
Consider
the probability, $X_{m_1}^{m_2}$, of encountering a value for the
variable $m$ in the range $m_1$ to $m_2$,
\begin{equation}
X_{m_1}^{m_2}= \int_{m_1}^{m_2} p(m') \, {\rm d}m' \, ,
\label{chIMF:eq:randIMF}
\end{equation}
with $X_{m_{\rm H}}^{m_{\rm H}} = 0 \le X_{m_{\rm H}}^{m} \le X_{m_{\rm
    H}}^{m_{\rm max}} = 1$, and $p(m) \propto \xi(m)$ being the
probability distribution function. Inversion of the equation to the
form $m=m(X_{m_{\rm H}}^m)$ provides the {\it mass generating function} which allows efficient random sampling of the stellar IMF with
$X_{m_{\rm H}}^m$ being a random number distributed uniformly between~0 and~1.
This approach is unphysical when applied to embedded clusters that form in molecular cloud clumps because their masses play no role in the sampling. That is, a molecular cloud clump would produce a number of stars that can add up to any arbitrary mass.  \\

\item[(Ib)] Random sampling of stars from the stellar IMF is physically more realistic if the mass of an embedded cluster in stars, $M_{\rm ecl}$ (i.e., the star formation efficiency times the mass of the molecular cloud clump), is imposed as a constraint because $N_{\rm s}$ is not a physically meaningful quantity but the molecular cloud clump mass is.  That is, a molecular cloud clump of a given mass cannot spawn a population of stars that amounts to a larger mass than $M_{\rm ecl}$.  If $M_{\rm ecl}$ is applied as a mass-condition on the sampled population of stars, then stars are randomly drawn from the stellar IMF until 
$\sum_{{\rm i}=1}^{N_{\rm s}} m_{\rm i} = M_{\rm ecl} \pm \delta M_{\rm ecl}$,
where $\delta M_{\rm ecl} \ll M_{\rm ecl}$ is the tolerance within
which the sampling of the $N_{\rm s}$ stars is completed. This yields $N_{\rm s}$.  Case~Ib thus has the stellar IMF being a conditional probability distribution function.\\

\item[(II)] The alternative interpretation is that the stellar IMF is as an
optimally sampled distribution function. Optimal sampling from a defined distribution function leads to no statistical scatter of the sampled quantity. This means that every
embedded cluster of the same initial conditions (e.g. $M_{\rm ecl}$) spawns exactly the
same, i.e. deterministic, content of stars without Poisson uncertainties upon arbitrary binning of the sampled stellar masses.  The physical
interpretation of this is that two embedded clusters that form from
the same initial conditions $\{ P_{\rm i} \}^{N_{\rm P}}$ (Eqn.~\ref{chIMF:eq:stIMFgen}) also yield exactly the same sequence of
stellar masses and is related to the star-formation process being
strongly feedback regulated. 

To sample a stellar IMF optimally for a
given $M_{\rm ecl}$, first eqns~\ref{chIMF:eq:mmax_Mecl1}
and~\ref{chIMF:eq:mmax_Mecl2} are solved to obtain the
IMF-normalisation constant and $m_{\rm max} \equiv m_{\rm 1}$. The
remaining stellar masses, $m_{\rm i}$, in the embedded cluster are then
obtained by calculating 
\begin{eqnarray}
1 &= &\int_{m_{{\rm i}+1}}^{m_{\rm i}} \xi(m)\, {\rm d}m \, ,\\
m_{{\rm i}+1} &= &\int_{m_{{\rm i}+1}}^{m_{\rm i}} m\, \xi(m)\, {\rm d}m \, .
\label{chIMF:eq:optIMF}
\end{eqnarray}
An improved optimal sampling algorithm is provided by \cite{Schulz+2015}.
The sampling stops when $m_{\rm i+1} < m_{\rm H}$ yielding $N_{\rm s}$ stars.  \\

\end{itemize}

\noindent
Random sampling (case~Ia) has been the method of choice in most models of star formation and galaxy evolution because, historically, embedded star clusters were not considered to be relevant building blocks of galaxies.
Computer modelling of star formation in galaxies, for example, is simplest with this sampling method which applies no constraints. Statistically the same population of stars can be assumed to form anywhere where star formation can occur even if this means that fractions of supernova explosions need to be treated (see e.g. \citealt{SH2023} for a comparison of the implications of different sampling methods on galaxy evolution). Case~Ia has been the backbone for interpreting extragalactic star-formation activity (e.g. eqn.~\ref{chIMF:eq:Kennicutt} below). 
Evidence that the stellar IMF is a probability density distribution
function is for example fielded by 
\cite{Corbelli+2009} who find a randomly sampled IMF to provide the best fit to the young star-cluster sample of the galaxy~M33. 
\cite{Dib+2017} also concludes the IMF to be a probability density distribution function on the basis of an
analysis of a about 3200 star clusters from different surveys.
Systematic errors (through large distances, resolution, photometry)
and apparent variations caused by stellar ejections, stellar mergers and
different binary fractions as a result of different dynamical
histories of the individual stellar populations may contribute to this
interpretation needing further study of this problem.
Unconstrained stochastic sampling (case~Ia) leads to the most massive star in an embedded cluster with a stellar mass $M_{\rm ecl}$, $m_{\rm max}$, being independent of $N_{\rm s}$. But a weak $m_{\rm max}=m_{\rm max}(M_{\rm ecl})$ relation emerges because $m_{\rm max}$ cannot be larger than $M_{\rm ecl}$. The physical interpretation of this is that a molecular cloud clump fragments randomly into stars of arbitrary mass anywhere within it with the clump mass playing no role.

While unconstrained stochastic sampling is still much applied, 
an argument for embedded clusters being the
fundamental building blocks of galaxies is the need to account for the
binary fraction of $f_{\rm bin} \approx 0.5$ in the Solar
neighbourhood in view of $f_{\rm bin} \approx 1$ in nearby
star-forming regions: this difference is accounted for by the early
disruption of initial binaries in their birth embedded clusters (see
Fig.~\ref{chIMF:fig:binaries}, \citealt{MarksKroupa2011}). 
Direct infrared imaging surveys \citep{LadaLada2003} also increasingly led to the realisation that molecular clouds spawn embedded clusters such that embedded star clusters are the fundamental building block of galaxies \citep{Kroupa2005}. Extra-galactic surveys of the 
fraction of stars in young star clusters in disk galaxies 
also suggest that stars primarily form in embedded clusters \citep{Dinnbier+2022}. In constrained stochastic sampling (case~Ib) stars are sampled randomly from the stellar IMF until a value for the mass of the embedded cluster is reached. 
A $m_{\rm max}=m_{\rm max}(M_{\rm ecl})$ relation exists because of this constraint (e.g. \citealt{StanwayEldridge2023}). The scatter of $m_{\rm max}$ values produced with this sampling at a given $M_{\rm ecl}$ is however larger than allowed by the observational data (Fig.~\ref{chIMF:fig:mmax}, \citealt{Yan+2023, Weidner+2013}).

Extragalactic data of young star clusters can be used to test if the stellar IMF is an unconstrained (case~Ia) or a constrained (case~Ib) probability density distribution function or if it is an optimal distribution function (case~II) by 
comparing the values and the dispersion of the luminosities (e.g. in the H$\alpha$ band, see Sec.~\ref{chIMF:sec:caseC}, eqn.~\ref{chIMF:eq:Kennicutt}) of the models and data. 
In the past, an error occurred in this test if the random sampling from the stellar IMF imposes $m\le m_{\rm max}(M_{\rm ecl})$ as a condition: 
the pitfall of this approach is that randomly sampling stars from the stellar IMF up to the $m_{\rm max}$ value for a given $M_{\rm ecl}$ leads to a statistical underestimate of the average value of $m_{\rm max}$. This has been interpreted erroneously to mean that the 
$m_{\rm max}=m_{\rm max}(M_{\rm ecl})$ relation depicted in Fig.~\ref{chIMF:fig:mmax} is not generally valid (see \citealt{Weidner+2014} and references therein).

The (i) existence of a pronounced observed $m_{\rm max}=m_{\rm max}(M_{\rm ecl})$ relation (Fig.~\ref{chIMF:fig:mmax}), (ii) the negligible intrinsic scatter of the $m_{\rm max}$ values for a given $M_{\rm ecl}$ (Fig.~\ref{chIMF:fig:mmax}, \citealt{Yan+2023, Weidner+2013}), and (iii) the lack of a significant intrinsic dispersion in the power-law index $\alpha_3$ of the stellar IMF amongst many different young stellar ensembles (e.g. fig.~27 in \citealt{Kroupa+2013}) suggest that stochastic sampling (cases~Ia and~Ib) is not a physically relevant method of discretising star formation. All three (i--iii) pieces of evidence are naturally consistent with the star formation process being described by optimal sampling (case~II).  Optimal sampling requires the existence of a strict $m_{\rm max} = m_{\rm max}(M_{\rm ecl})$ relation without intrinsic scatter. The observed relation is consistent with this and shows features possibly related to the regulation of the growth of stars through feedback (Fig.~\ref{chIMF:fig:mmax}). Every ensemble of stars drawn from the stellar IMF will also yield exactly the same power-law indices such that any observed dispersion of, e.g., $\alpha_3$ values is only due to observational uncertainties, unrecognised multiple systems and stellar ejections and mergers. The small dispersion of observed $\alpha_3$ values is consistent with this interpretation (fig.~27 in \citealt{Kroupa+2013}).

As discussed in Sec.~\ref{chIMF:sec:origin} the stellar IMF of late-type ($m \simless 1.4 \,M_\odot$) stars appears to be given by the fragmentation of molecular cloud filaments which may be approximated as a random process (Fig.~\ref{chIMF:fig:IMFshape}).  This is consistent with these stars being born typically in wide binary systems with companion masses that are randomly drawn from the stellar IMF \citep{Kroupa1995b}. Feedback self-regulation of the star formation process and likely growth through coalescence of proto-stars in the dense central regions of sufficiently massive embedded clusters may make the stellar IMF an optimally sampled distribution function for larger stellar masses. This would suggest that the late-type stars of an embedded cluster may not be as mass segregated as the early-type ($m\simgreat 1.4 \,M_\odot$) stars are \citep{Pavlik+2019}.

It may thus be possible that the nature of the stellar IMF is intermediate between cases~(Ib) and~(II) for $m\simless 1.4\,M_\odot$ and more closely related to case~(II) for early-type stars. 
Magnetohydrodynamic simulations are being
conducted to shed light on whether star formation is a stochastic or a
physical process (e.g. \citealt{Grudic+2023}). In terms of comparing models with observed systems, the case~(II) interpretation leads to deterministic predictions that will appear to be stochastic through measurement uncertainties, multiplicity of stars, stellar ejections, mergers, and mass transfer in multiple systems.

\section{The composite stellar IMF in galaxies across cosmic time -- theoretical background}
\label{chIMF:sec:gwIMF}

The stellar population in a region in a galaxy (e.g. the Solar neighbourhood) or in a whole galaxy is the sum of the star formation events that contributed to this population. This concept, introduced by \cite{KW2003}, leads to powerful results concerning observable properties of galaxies when combined with the $m_{\rm max}=m_{\rm max}(M_{\rm ecl})$ relation (Fig.~\ref{chIMF:fig:mmax}), simply because galaxies or regions thereof that produce molecular cloud clumps of low mass only will have a deficit of massive stars. The consequences of this integration over star-forming regions strengthen if the physical parameters $P_{\rm i}$ (e.g. the gas density and temperature) upon which the stellar IMF depend vary with location in the system and with time.
How to calculate the composite, regional or galaxy-wide IMF is discussed in the following sections.

In general, the composite IMF (cIMF) of all stars ever formed with
masses in the range $m$ to $m+{\rm d}m$ in a region of volume $V$ can
be formulated to be the integral over this volume and over time
(cf. \citealt{Hopkins2018}),
\begin{equation}
\xi_{\rm cIMF}(m) = \int_V \, \int_P \, \int_t \, \xi_{\rm dens}(m:
{\vec r}, \{P_{\rm i}\}^{N_{\rm P}},t) \, {\rm
d}V \, {\rm d}^{N_{\rm P}}P_{\rm i} \, {\rm d}t \, ,
\label{chIMF:eq:IMFdens}
\end{equation}
where
${\rm d}N = \xi_{\rm dens}(m: \vec{r}, \{P_{\rm i}\}^{N_{\rm P}}, t)\,{\rm d}V \, {\rm d}^N P_{\rm i}
\,{\rm d}t$ is the number of stars with masses in the interval $m$ to
$m+{\rm d}m$ formed in the volume ${\rm d}V$ at position $\vec{r}$
(relative to some coordinate system defining the system) where the
$N_{\rm P}$ physical parameters are in the range $P_{\rm i}$ to $P_{\rm i}+{\rm d}P_{\rm i}$ and in the
time interval $t$ to $t+{\rm d}t$. $P_{\rm i}$ may be any or all of
chemical composition, temperature, pressure, specific angular momentum of the molecular cloud clump, cosmic ray flux etc., and depend
on the past history of the whole system, making this a non-linear and
non-trivial problem.  $\xi_{\rm dens}$ can then be interpreted as the
stellar-mass-dependent rate per unit volume with which a localised
region of volume ${\rm d}V$ turns gas into stars, given the physical
conditions described by the set $\{P_{\rm i}\}^{N_{\rm P}}$. 

Whether the stellar IMF is a probability density distribution function
or an optimally sampled distribution function (Sec.~\ref{chIMF:sec:sampling}) has fundamental
implications on the nature and variation of the galaxy-wide IMF of
stars (gwIMF) and on galaxy formation and evolution. In all cases, the 
shape of the stellar IMF which defines the mass-distribution of newly formed stars can depend on the physical properties of the star-forming gas (e.g. \citealt{Fumagalli+2011, Dib2022}). There are three broad formulations:

\subsection{Case A}
\label{chIMF:sec:caseA}
If the stellar IMF is an invariant probability density distribution
function (case~Ia in Sec.~\ref{chIMF:sec:sampling}) then the gwIMF will also be an invariant probability density
distribution function. This means stars form randomly throughout the
galaxy at a rate given by the galaxy-wide or local SFR, $\psi$ (herein always
in $M_\odot/$yr). With this assumption, each model galaxy needs to be
computed a large number of times to assess the range of variation of
the outcome (e.g. the metallicity enrichment and the photometric
properties differ between galaxies that have the same SFHs due to the
different stellar populations in each galaxy, \citealt{SLUG_II}).

\subsection{Case B}
\label{chIMF:sec:caseB}

Assuming all stars from in embedded clusters, the stellar IMF is a conditional probability density distribution function with the constraint being given by the mass of the embedded cluster, $M_{\rm ecl}$ (case~Ib in Sec.~\ref{chIMF:sec:sampling}). The gwIMF needs to be calculated by adding
all stellar IMFs in a galaxy of all embedded clusters
\citep{KW2003}. This implies the need to
describe where stars form in a galaxy because the mass an embedded cluster can reach depends on the local gas density \citep{PKr2008}. Strong evidence for this comes from the highly-significant dependency of the mass function of embedded clusters on the galactocentric distance in the galaxy~M33 \citep{Pflamm+2013}.
The calculation of the gwIMF thus
entails constrained probability distribution functions and for each
model of a galaxy a large number of renditions need to be computed to
assess that range of variability, as in Case~A above.

\subsection{Case C}
\label{chIMF:sec:caseC}

If the stellar IMF is an optimal distribution function (case~II in Sec.~\ref{chIMF:sec:sampling}), then the gwIMF needs to be calculated by adding all stellar IMFs contributed by all embedded clusters in a galaxy
\citep{KW2003, Yan+2017, Haslbauer+2024}.  Each galaxy needs to be
calculated only once because there is no statistical dispersion.  Thus
a large range of parameters can be explored (e.g. the implications of
using different stellar evolution models and/or element yield tables
can be explored readily).

\subsubsection{Case~C: the galaxy as a point mass object}
\label{chIMF:sec:caseCpoint}

Treating a galaxy as a point mass object,
eqn.~\ref{chIMF:eq:IMFdens} simplifies to an
integral over all embedded clusters \citep{KW2003}: for all embedded clusters that form within
a time span $t, t+\delta t$, the gwIMF contributed to the whole galaxy
in this time interval is given by the integrated initial mass function
of stars (IGIMF),
\begin{equation} 
\xi_{\rm IGIMF}(m;t) = 
\int_{M_{\rm ecl,min}}^{M_{\rm
ecl,max}(\psi(t))} \xi\left(m\le m_{\rm max}\left(M_{\rm
ecl}\right) : Z\right)~\xi_{\rm ecl}(M_{\rm ecl})~{\rm d}M_{\rm ecl},
\label{chIMF:eq:igimf_t}
\end{equation}
with the normalisation conditions eqs.~\ref{chIMF:eq:mmax_Mecl1}
and~\ref{chIMF:eq:mmax_Mecl2} which together yield the
$m_{\rm max}=m_{\rm max}(M_{\rm ecl})$ relation;
$\xi(m\le m_{\rm max}\left(M_{\rm ecl}\right) : Z)~\xi_{\rm ecl}(M_{\rm ecl})~dM_{\rm ecl}$ is the
stellar IMF contributed by $\xi_{\rm ecl}(M_{\rm ecl})~dM_{\rm ecl}$ embedded
clusters with stellar masses in the interval
$M_{\rm ecl}, M_{\rm ecl}+dM_{\rm ecl}$. 
$M_{\rm ecl,min} \approx 5\,M_{\odot}$ corresponds to the smallest
``embedded star-cluster units'' observed
(Sec.~\ref{chIMF:sec:origin}). 
Following
eqn.~\ref{chIMF:eq:IMFdens}, $\xi(m)$ can depend on physical
parameters such as $Z$ and $M_{\rm ecl}$ (eqn~\ref{chIMF:eq:stIMF}).  This variation is assumed to be valid in what follows. The embedded
cluster mass function (ECMF), $\xi_{\rm ecl}(M_{\rm ecl})$, is usually
taken to be a power-law probability distribution function, and in the
Case~C it is an optimally sampled function. Galactic and
extra-galactic data suggest $\xi_{\rm ecl}$ to be reasonably well
represented by a power-law function,
\begin{equation}
\xi_{\rm ecl}(M_{\rm ecl}) \propto M_{\rm ecl}^{-\beta} \, ,
\label{chIMF:eq:ECMF}
\end{equation}
with $\beta \approx 2$. Not much is known about the possible variation
of the ECMF power-law index $\beta$ with the physical properties of
the inter-stellar medium of a galaxy, but it may become top-heavy with
increasing SFR (e.g. \citealt{Yan+2017}),
\begin{equation}
\beta = -0.106\, {\rm log}_{10}\left(\psi / \left(M_\odot/{\rm yr} \right)\right) + 2.0   \, .
\label{chIMF:eq:beta}
\end{equation}
A metallicity dependence of $\beta$ would be naturally expected, but data on this are not yet available.
$M_{\rm ecl,max}$ follows from the maximum young star-cluster-mass {\it vs}
SFR ($M_{\rm ecl,max}=M_{\rm ecl, max}(\psi)$) relation. 
\cite{Weidner+2004, Randriamanakoto+2013} note that the  maximum
mass of a very young star cluster, $M_{\rm ecl,max}$, found in galaxies with different SFRs
correlates strongly with $\psi$, the scatter of the data being smaller than expected from randomly sampling embedded cluster masses from the ECMF. 
A fit to the data yields (\citealt{Kroupa+2013} and references therein)
\begin{equation}
\frac{M_{\rm ecl,max}}{M_\odot}  \approx 8.5 \times 10^4 \, 
\left(\frac{\psi}{M_\odot/{\rm yr}}\right)^{0.75} \, .
\label{chIMF:eq:Meclmax}
\end{equation}
A theoretical relation between
$M_{\rm ecl,max}$ and $\psi$, which is a good description of the
empirical data, is obtained as follows: If a galaxy has, at a time
$t$, a SFR of $\bar{\psi}(t)$ over a time span $\delta t$ over which
an optimally sampled embedded star cluster distribution builds up with
total stellar mass $M_{\rm tot}(t)$, then there is one most massive
embedded cluster,
\begin{equation}
1 = \int_{M_{\rm ecl, max}(\bar{\psi}(t))}^{M_{\rm U}} \xi_{\rm ecl}(M_{\rm
  ecl})\,{\rm d}M_{\rm ecl} \, ,
\label{chIMF:eq:MeclmaxSFR1}
\end{equation}
with $M_{\rm U}\approx 10^{10}\,M_\odot$ being the physical maximum
star cluster than can form, and
\begin{equation}
\bar{\psi}(t) = \frac{M_{\rm tot}(t) }{ \delta t} = \frac{1}{\delta t}
\int_{M_{\rm ecl, min}}^{M_{\rm ecl,max}(\bar{\psi}(t))} M_{\rm ecl}\,\xi_{\rm
  ecl}(M_{\rm ecl})\,{\rm d}M_{\rm ecl} \, .
\label{chIMF:eq:MeclmaxSFR2}
\end{equation}
The time span $\delta t$ is a
``star-formation epoch'', within which the ECMF is sampled optimally,
given a SFR. This formulation leads naturally to the observed
$M_{\rm ecl,max}(\bar{\psi})$ correlation if the ``epoch'' lasts about
$\delta t=10$~Myr.  This time-scale is consistent with the life-time
of molecular clouds in normal galactic disks
(Sec.~\ref{chIMF:sec:origin}). 
The power-law form for the ECMF (eq.~\ref{chIMF:eq:ECMF} is an approximation stemming from a census of nearby embedded star clusters \citep{LadaLada2003} and extragalactic data (\citealt{Weidner+2004}). The correct form, not yet implemented in IGIMF calculations, is to use a galaxy-wide ECMF which can be well represented by a Schechter function (Sec.~\ref{chIMF:sec:caseCextended}).

The above equations under Case~C formulate the ``IGIMF theory'' which allows
the freshly formed stellar populations to be calculated in dependence
of the galaxy's metallicity and SFR. The gwIMF depends on the
parametrisation of the dependency of the stellar IMF on the cloud-clump's density and metallicity which has been improved over time.  The galaxy-wide IMF
calculated according to the IGIMF theory cannot be represented by a
simple mathematical function, and grids of the gwIMF in
$Z - \bar{\psi}$ space are provided by \cite{Jerabkova+2018} and
\cite{Haslbauer+2024}.

\begin{BoxTypeA}[chIMF:box:consistencytest2]{Consistency Test of the IGIMF
    theory}

  \noindent An important result is that the IGIMF theory leads to the camonical field-star gwIMF (eqns~\ref{chIMF:eq:canIMF1}-\ref{chIMF:eq:canIMF2} and~\ref{chIMF:eq:canIMF4}) when
  the Milky Way conditions are reached ($Z\approx Z_\odot$,
  $\psi \approx 1-3\,M_\odot/$yr). This is a non-trivial but necessary
  outcome \citep{Guszejnov+2019} of the independently gauged stellar
  IMF (Fig.~\ref{chIMF:fig:Z_SFR1}, eqn.~\ref{chIMF:eq:stIMF}).
\end{BoxTypeA}

\begin{figure}
    \centering
\includegraphics[width=0.345\textwidth]{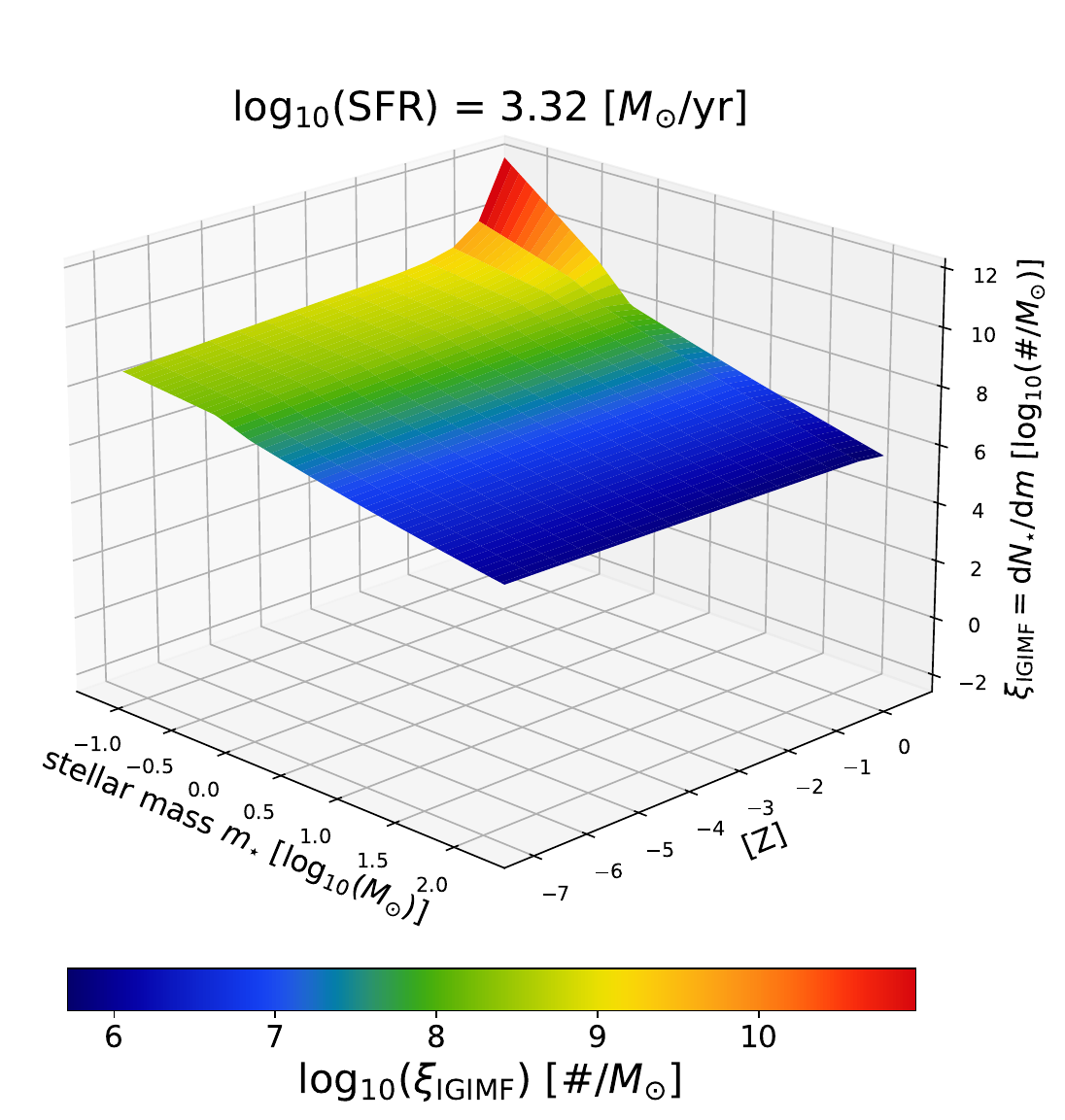}  \hspace{-.5cm}
\includegraphics[width=0.345\textwidth]{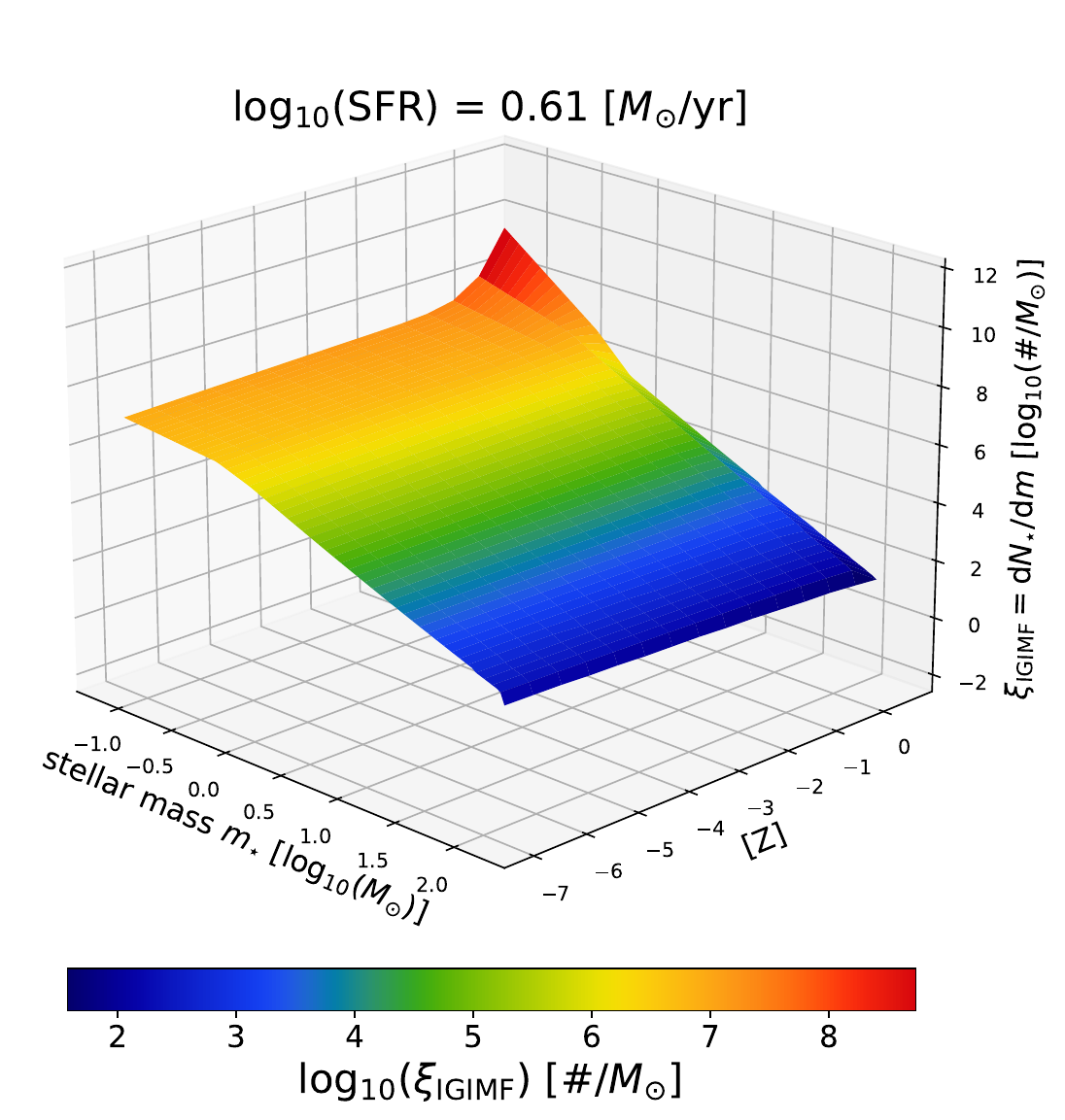}  \hspace{-.5cm}
\includegraphics[width=0.345\textwidth]{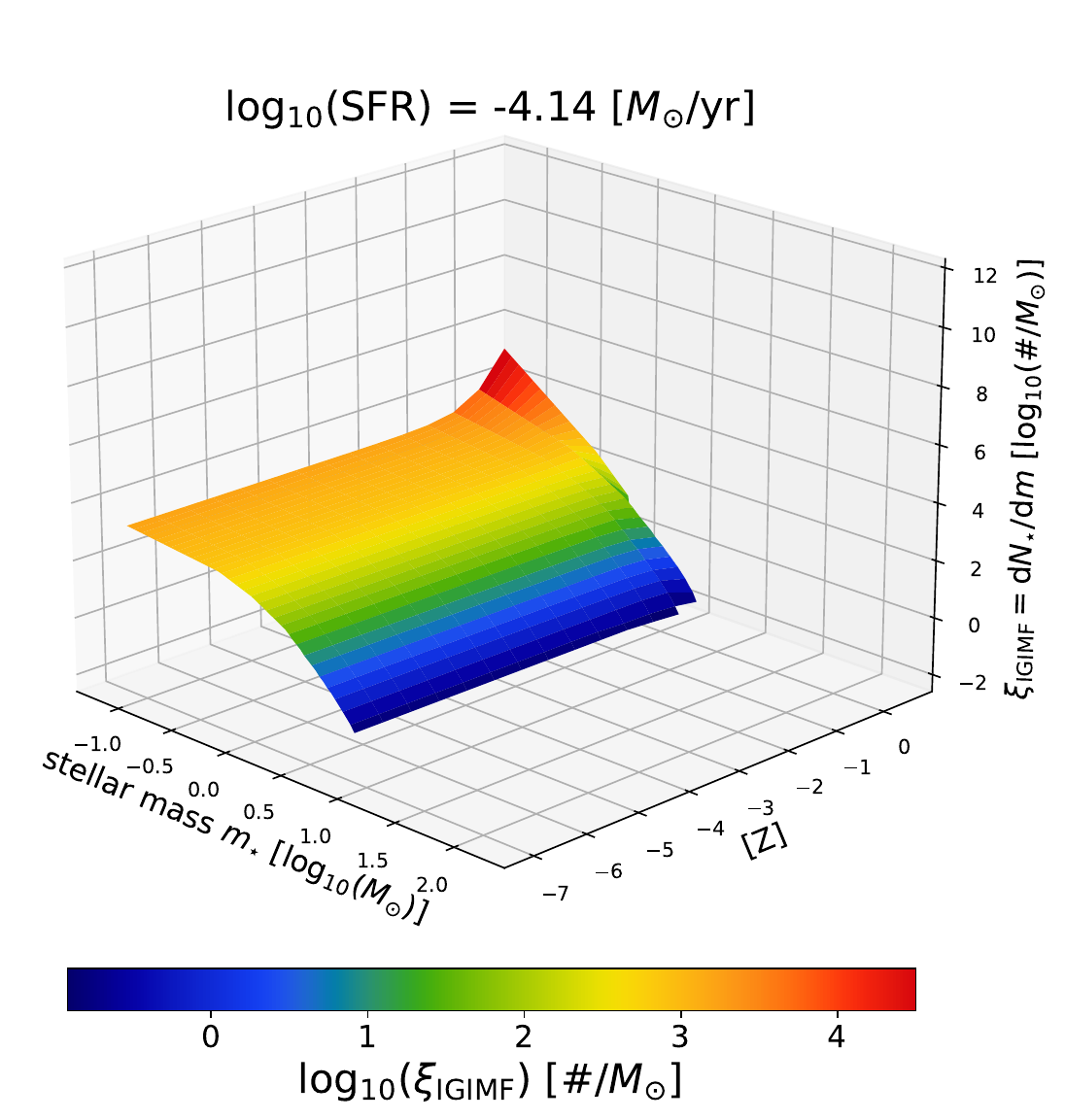}
    \caption{The galaxy-wide IMF, gwIMF (z-axis), calculated using the IGIMF theory (eq.~\ref{chIMF:eq:igimf_t})  as a 3D surface function of stellar mass and metallicity, for three SFRs: an extreme starburst (left panel), a typical Milky-Way-like spiral galaxy (middle panel) and an extremely low SFR,  as found in dwarf (irregular) galaxies (right panel). 2D slices of these surfaces are shown in Fig.~\ref{chIMF:fig:Z_SFR1}, where they appear, respectively and with similar SFRs, in the bottom right, bottom left and top-left panel. In each panel   
   the rainbow color-bar spans the highest and the lowest $\log_{10}(\xi_{\rm IGIMF})$ value in an isoluminant linear gradient. The red peak at high metallicity shows a strongly bottom-heavy gwIMF which is expected to be present in the inner regions of elliptical galaxies (Sec.~\ref{chIMF:sec:ETGs}).   
     Reproducible with the {\tt pyIGIMF} software (Sec.~\ref{chIMF:sec:codes}). }
    \label{chIMF:fig:3D}
\end{figure}

\begin{figure}[t]
\centering
\includegraphics[width=0.8\textwidth]{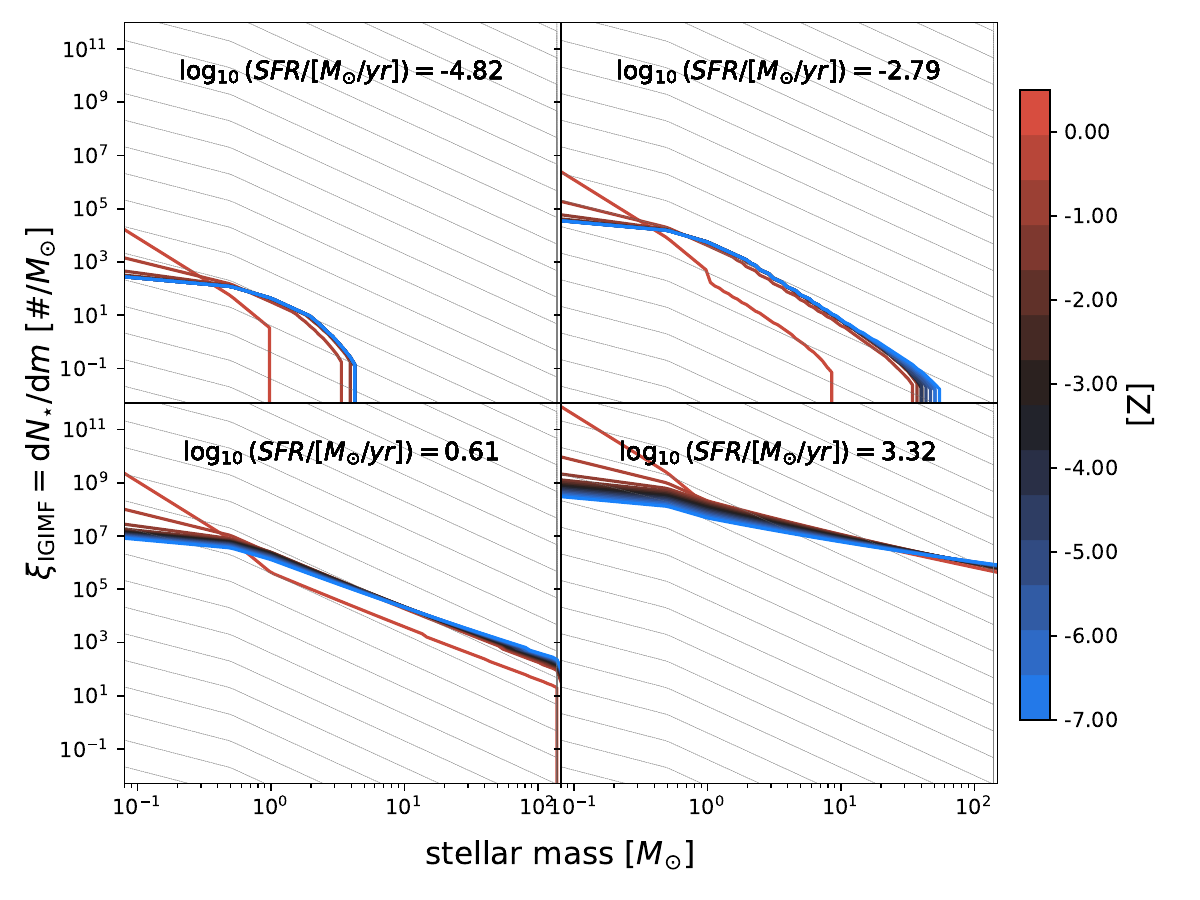}
\caption{The shape of the gwIMF as encapsulated by the IGIMF theory
  (eqn.~\ref{chIMF:eq:igimf_t}).  The area under each gwIMF curve shows the total number of stars formed in a galaxy with the given SFR and metallicity in the time span of $\delta t = 10\,$Myr.  The grey lines represent the canonical stellar IMF (eqs.~\ref{chIMF:eq:canIMF1}--\ref{chIMF:eq:canIMF3}). 
  The gwIMFs  
  become increasingly top-heavy
  for smaller metallicity, [$Z$], and larger SFR, and bottom-heavy for
  increasingly larger [$Z$]. For example, a massive elliptical galaxy begins to form with a low SFR and low metallicity (\emph{top left panel}: with a bottom-light and top light gwIMF), evolves through the \emph{top right} and \emph{lower left panel} with rising SFR and increasing [$Z$] and through to the \emph{bottom right panel} where the top-heavy gwIMF leads to rapid chemical enrichment such that the gwIMF also becomes bottom-heavy, ending again at the \emph{top left panel} with a decreasing SFR and a bottom-heavy gwIMF and large [$Z$]. 
 Note that the \emph{lower left  panel} matches closely the gwIMF of the Galaxy, except for supersolar metallicities. Compare with the full 3D gwIMF surfaces shown in Fig. \ref{chIMF:fig:3D}.  Reproducible with the {\tt pyIGIMF} software (Sec.~\ref{chIMF:sec:codes}). 
  }
\label{chIMF:fig:Z_SFR1}
\end{figure}

\begin{figure}[t]
\centering
\includegraphics[width=1.0\textwidth]{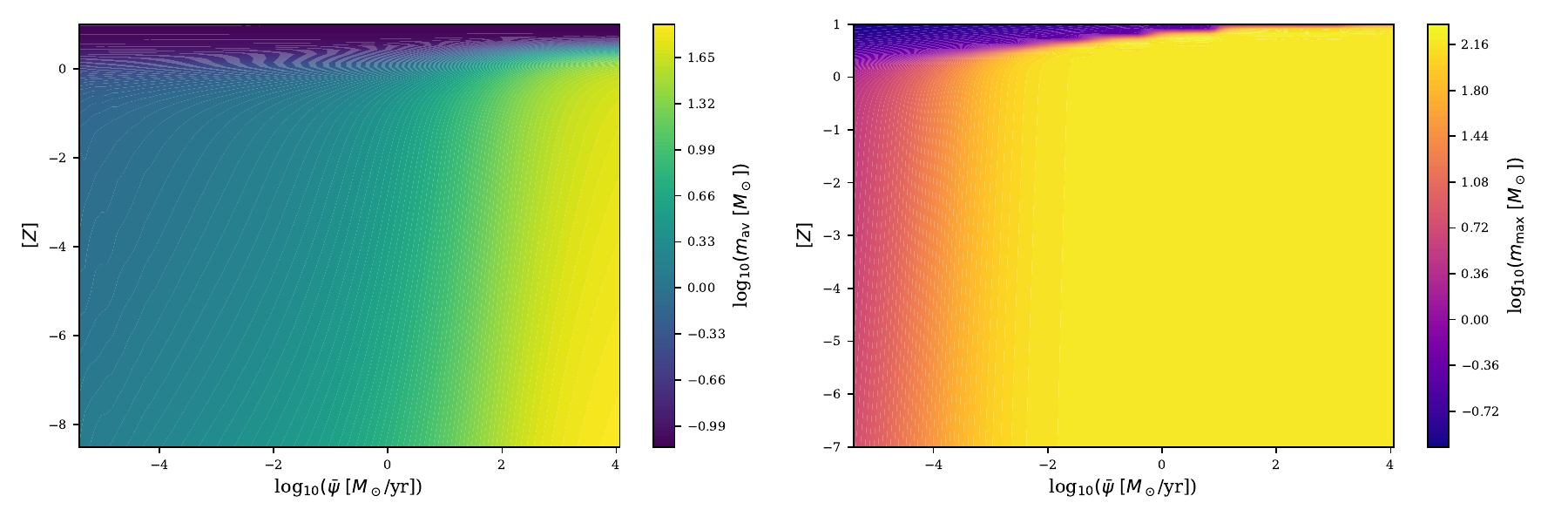}
\caption{The dependency of the average stellar mass, 
$m_{\rm av}$ (left panel), and of the maximum stellar mass, $m_{\rm max}$ (right panel), of the forming stellar population in a galaxy
on its SFR, $\bar{\psi}$, and $[Z]$, assuming the IGIMF theory and calculated using {\tt galIMF} (Sec.~\ref{chIMF:sec:codes}).}
\label{chIMF:fig:mav_mmax2} 
\end{figure}

The average stellar mass of all stars and the mass of the most massive star forming in a galaxy can be calculated by applying the IGIMF theory. The dependency of both quantities on the SFR and metallicity are displayed in Fig.~\ref{chIMF:fig:mav_mmax2}.  More massive stars can appear in galaxies through mergers of binary components in the young binary-rich populations \citep{BanerjeeKroupa2012}. From this figure follows:

\begin{BoxTypeA}[chIMF:box:SFRaveragemass]{Galaxies as early ionising sources and galaxies with dark star formation}

\noindent At high redshift and at low metallicity, galaxies with large SFRs ($\psi \simgreat 10^2\,M_\odot/$yr) have a stellar population with an average stellar mass $m_{\rm av}\simgreat 50\,M_\odot$ and would be significant ionising sources in the early Universe \citep{Schaerer+2024}. At high metallicity ($[Z]>\simgreat 0.9$), even massively star-bursting galaxies ($\psi \simgreat 10^2\,M_\odot/{\rm yr}$) lack massive stars ($m_{\rm max}<{\rm few}\,M_\odot$) suggesting that at high metallicity galaxies can have profuse dark star formation. Such galaxies might be misinterpreted as having only a minor, if any, SFR if an invariant canonical stellar IMF is assumed in the observational analysis. 

\end{BoxTypeA}

\begin{figure}[t]
\centering
\includegraphics[width=1.0\textwidth]{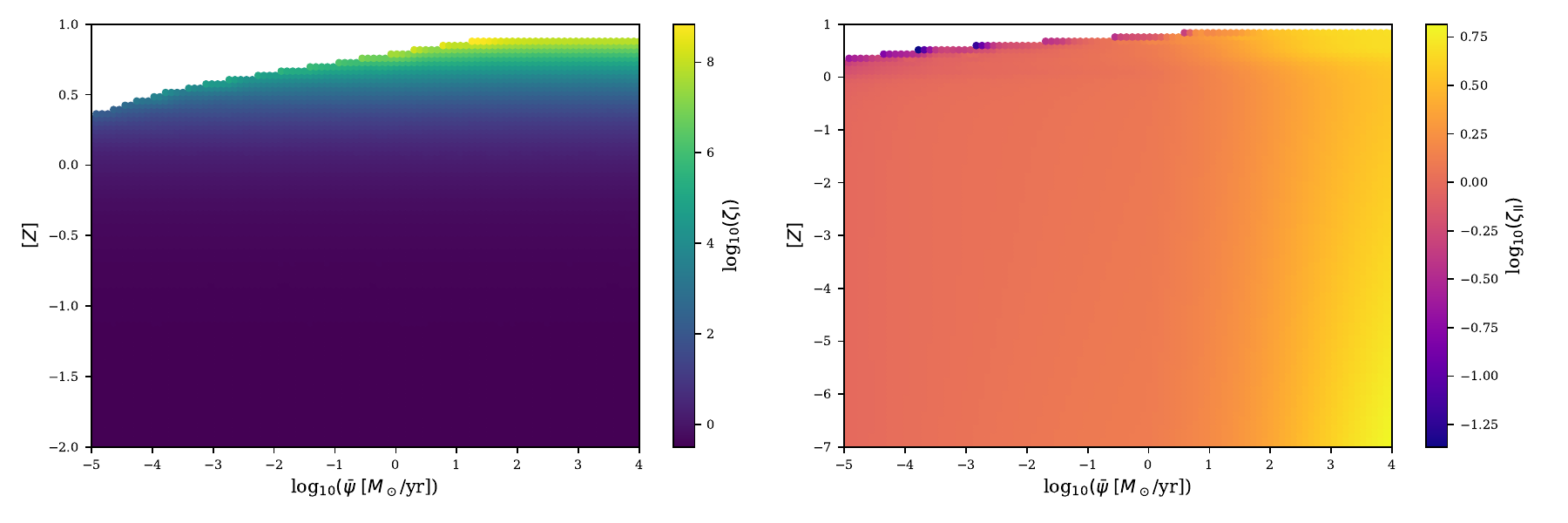}
\caption{The variation of the shape 
of the gwIMF calculated according to the IGIMF theory (eqn.~\ref{chIMF:eq:igimf_t}, calculated using {\tt galIMF}, Sec.~\ref{chIMF:sec:codes}) in terms of the 
parameters $\zeta_{\rm I}$ (eqn.~\ref{chIMF:eq:shape1}, left panel) and $\zeta_{\rm II}$ (eqn.~\ref{chIMF:eq:shape2}, right panel). According to the IGIMF theory, the gwIMF is 
increasingly bottom-heavy (larger $\zeta_{\rm I}$) for increasing metallicity, [$Z$], and 
increasingly top-heavy (larger $\zeta_{\rm II}$) for smaller [$Z$] and larger SFR, $\bar{\psi}$. 
The values of $\zeta_{I}, \zeta_{II}$, are cutoff at the top because their gwIMFs do not have stars with $m>1\,M_\odot$ such that the normalisation is undefined.
}
\label{chIMF:fig:Z_SFR2}
\end{figure}

The change in the shape of the gwIMF relative to the canonical stellar IMF (eqn.~\ref{chIMF:eq:canIMF1}--\ref{chIMF:eq:canIMF3}) for different values of $Z$ and $\psi(t)$ 
(Fig.~\ref{chIMF:fig:3D} and~\ref{chIMF:fig:Z_SFR1}) can be quantified by normalizing the functions to unity at a reference mass which is chosen to be $m=1\,M_\odot$. This then yields the normalized IGIMF, $\xi_{\rm nIGIMF}(m)$, and the normalised canonical stellar IMF, $\xi_{\rm ncan}(m)$,
\begin{equation}
\xi_{\rm nw}(m=1\,M_\odot) = 1 \, ,
\label{chIMF:eq:onenorm}
\end{equation}
where w$=\,$IGIMF or w$=\,$can (canonical). 
The shape of a gwIMF can thus be compared to that of the canonical stellar IMF
as defined in Sec.~\ref{chIMF:sec:local} by calculating the surplus vs
deficit of stars in different mass ranges. For example the ratio
$\zeta_{\rm I}$ informs whether a gwIMF is bottom-light
($\zeta_{\rm I} <1$) or bottom-heavy ($\zeta_{\rm I} >1$) relative to
the canonical stellar IMF,
\begin{equation}
\zeta_{\rm I} = \frac{ \int_{m_{\rm H}}^{1\,M_\odot} \, \xi_{\rm nIGIMF}(m) \, {\rm d}m  }{
  \int_{m_{\rm H}}^{1\,M_\odot} \, \xi_{\rm ncan}(m) \, {\rm d}m } \, .
\label{chIMF:eq:shape1}
\end{equation}
The following ratio informs if
the gwIMF is top-light ($\zeta_{\rm II} <1$) or top-heavy
($\zeta_{\rm II} >1$) relative to the canonical stellar IMF,
\begin{equation}
\zeta_{\rm II} = \frac{\int_{1\,M_\odot}^{150\,M_\odot} \, \xi_{\rm nIGIMF}(m) \, {\rm d}m  }{
  \int_{1\,M_\odot}^{150\,M_\odot}  \, \xi_{\rm ncan}(m) \, {\rm d}m } \, .
\label{chIMF:eq:shape2}
\end{equation}
Fig.~\ref{chIMF:fig:Z_SFR2} demonstrates how $\zeta_{\rm I}$ and~$\zeta_{\rm II}$
vary with metallicity and SFR.

Comparing different galaxies, the IGIMF theory leads to systematic differences in  the specific (i.e. per galaxy-wide mass in long-lived late-type stars) rate of the production of various stages of evolved stars, of
white dwarfs, core-collapse supernovae, and type~Ia explosions (SNIa, \citealt{WeidnerKroupa2005}). 
The number of objects in a particular tracer population per late-type star depends on the SFH of the galaxy, not only through the particular history of $\psi(t)$ but also through the gas density and metallicity dependent stellar IMF. This comes about
because the shape of the gwIMF systematically changes as a galaxy evolves though a changing metallicity and SFR such that the relative fractions of stars being born in different stellar mass ranges change. An important new insight obtained from detailed chemical abundance modelling of elliptical galaxies is the need for the specific rate of SNIa explosions to evolve beyond this rescaling of the specific SNIa rate \citep{Yan+2021}.
Galaxies with a higher SFR need an "overproduction" of Fe through an overproduction of SNIa explosions compared to galaxies with smaller SFRs. Galaxies that reach large SFR values need an additional source of SNIa progenitors to explain their metal abundances.
This is qualitatively consistent with the independent result obtained by \cite{SharaHurley2002} based on stellar-dynamical modelling of a massive star cluster: massive star clusters overproduce hard binaries of degenerate remnants as a result of stellar-dynamical encounters. A galaxy can only form massive star clusters ($M_{\rm ecl}>10^6\,M_\odot$) if its SFR is sufficiently large ($\psi \simgreat 27\,M_\odot/{\rm yr}$, eq.~\ref{chIMF:eq:Meclmax}) such that the specific rate of SNIa events is boosted when $\psi$ is large over galaxies with small SFRs. This problem needs further attention as it is important for understanding the detailed elemental abundances of elliptical galaxies.

\subsubsection{Case~C: the galaxy as an extended object}
\label{chIMF:sec:caseCextended}

The above discussion neglects the spatial extend of a galaxy. The
IGIMF theory can be formulated to calculate the {\it composite IMF} of
a region in a star-forming galaxy, i.e. the ``local IGIMF''.  The
theoretical development of this has not yet been much explored, but
the ansatz available in \cite{PKr2008} shows that the radial cutoff in
H$\alpha$ emission in galaxies that have extended-UV disks comes about
because the mass of the most-massive embedded cluster decreases with
the local gas density in the galactic disk. Outer regions of galactic disks can thus harbour dark (H$\alpha$-invisible) star formation. This follows because
embedded clusters with smaller $M_{\rm ecl, max}$ in the outer
galactic regions contain stars with masses
$m\le m_{\rm max}(M_{\rm ecl,max})$.

Introducing the {\it local ECMF} at position $x,y$ in an axisymmetric
rotationally-supported disk galaxy with the origin at its center,
\begin{equation}
  \xi_{\rm lecl}(M_{\rm ecl}, x, y) = \frac{{\rm d} N_{\rm ecl} }{ {\rm
      d}M_{\rm ecl}\, {\rm d}x \, {\rm d}y }
    \propto M_{\rm ecl}^{-\beta} \, ,
\label{chIMF:eq:lECMF}
\end{equation}
with the conjecture that the locally most massive embedded cluster
depends on the local gas density, $\Sigma_{\rm gas}(x,y)$,  according to
\begin{equation}
  M_{\rm lecl, max}(x, y) =M_{\rm ecl, max}(\bar{\psi}) \, \left( \frac{\Sigma_{\rm
      gas}(x,y) }{ \Sigma_{\rm gas, 0}    }\right)^\gamma \, ,
\label{chIMF:sec:lMecl}
\end{equation}
where $\Sigma_{\rm gas, 0}$ is the gas density at the origin (assuming
an axisymmetric galactic disk with an exponentially declining radial
gas disk surface density) and $M_{\rm ecl,max}$ is the globally
most-massive embedded cluster forming (in this particular application)
at the center of the galaxy with a SFR, $\bar{\psi}$
(eqs.~\ref{chIMF:eq:MeclmaxSFR1} and~\ref{chIMF:eq:MeclmaxSFR2}).  The
star-formation rate density,
$\Sigma_{\rm SFR}(x,y) \propto \Sigma_{\rm gas}^N(x,y)$ with
$N\approx 1.0$.  The local IGIMF can thus be calculated using the same
methodology as in Sec.~\ref{chIMF:sec:caseCpoint} in each radial bin. However,
eqn.~\ref{chIMF:sec:lMecl} and the required value of $\gamma=3/2$ are
merely an ansatz and this surface-density IGIMF approach needs deeper
elaboration.  That Case~C is applicable in this approach is supported
by the strong falsification of the random formation of star clusters
in the disk of the galaxy~M33 which shows a statistically highly
significant decrease of the most-massive young cluster with radial
distance \citep{Pflamm+2013}. By assuming that the local ECMF is a power-law function (eq.~\ref{chIMF:eq:lECMF}), \cite{LieberzKroupa2017} showed that the integration over the galaxy yields an integrated galaxy-wide ECMF which can be well approximated by a Schechter function. 

According to this approach, interarm regions in spiral galaxies have a
top-light local IGIMF relative to spiral arms in which the gas- and
star-formation rate density are higher. This comes about because (i)~the field is the
addition of many embedded clusters each with a stellar IMF that
(ii)~extends to a most-massive star, $m_{\rm max}(M_{\rm ecl})$, which
is a function of the stellar mass of the embedded cluster,
$M_{\rm ecl}$, and (iii)~massive embedded clusters are
underrepresented in interarm regions (eqn.~\ref{chIMF:sec:lMecl}). This
explains why $\alpha_{3, {\rm field}} > \alpha_{3, {\rm can}}$
(Sec.~\ref{chIMF:sec:local}).  A lack of massive embedded star
clusters may be an explanation for this difference because the Sun is
in a region between spiral arms lacking star-forming activity \citep{Vallee2018}.  The influence of the variation of the SFR on the deduced shape of the local composite (i.e. field) IMF is studied by \cite{ElmegreenScalo2006}. 

Concerning pressure-supported (i.e. elliptical) galaxies their formation through the monolithic collapse of a pre-galactic primordial gas cloud needs to be modelled \citep{Eappen+2022}. Every radially infalling volume element would be forming stars at a rate given by the gas density. A prescription related to the one used for axisymmetric disk galaxies above would be appropriate in terms of the locally most massive embedded cluster, $M_{\rm ecll, max}$, in an infalling radial shell for some assumed radial gas density law. Qualitatively this leads to the result that the time-integrated local IGIMF will be more bottom-heavy and more top-heavy in the innermost regions of the galaxy, and will be bottom light and top-light in the outermost regions where the metallicity will remain low. No explicit modelling exists on this however, and will require analytical models of collapsing gas clouds that form stars throughout to be developed. Alternatively, the IGIMF theory needs to be incorporated into galaxy formation codes (e.g. as suggested by \cite{Ploeckinger+2014}, Sec.~\ref{chIMF:sec:ETGs}).

\begin{BoxTypeA}[chIMF:box:buildingblocks]{Galactic building blocks}

\noindent In conclusion, all three cases~A, B~and~C link star-formation in
molecular cloud clumps to the galaxy scale. Case~A is the most
elementary according to which star-formation in molecular cloud clumps
is irrelevant. In Case~B the stellar IMF is also a probability density distribution function but subject to the condition that star-formation occurs in embedded clusters of given masses. Cases~A and~B lead to two galaxies with the same SFH having different intrinsic properties. 
In Case~C the stellar populations on the galaxy
scale are calculated from the properties of the molecular cloud
clumps assuming the stellar IMF to be an optimal distribution function, treating the embedded clusters as the fundamental building blocks
of galaxies. Case~C leads to two point-mass galaxies with the same SFH being identical. 

\end{BoxTypeA}

\section{Observational evidence}
\label{chIMF:sec:galobs}

Observations of stellar populations in galaxies provide important constraints on galaxy evolution models. These depend on cosmological theory and also on assumptions concerning the gwIMF.
The stellar IMF forms on the molecular clump scale (Sec.~\ref{chIMF:sec:sampling}), but observations relate to galaxy-wide scales. 

An important constraint is provided by the stellar population in the Milky Way and is discussed in Sec.~\ref{chIMF:sec:caseC}. An independent test of the Galaxy's gwIMF comes from microlensing events towards the inner Galaxy. This test relies on the phase-space distribution of all components of free-floating planets, brown dwarfs, stars, stellar remnants and on knowledge of the Galactic potential which is directly linked to this distribution. Available constraints on the brown dwarf and stellar gwIMFs are broadly consistent with the canonical form (eqn.~\ref{chIMF:eq:canIMF1}--\ref{chIMF:eq:canIMF3}, e.g. \citealt{Koshimoto+2021}). These constraints are subject to uncertainties from 
the varying stellar IMF which has not been accommodated in the data analysis, the formation history which remains largely unknown, the presence of stellar streams, and the Galactic potential in the inner Galaxy which depends on the mixture of unseen mass in various forms. 

Considering these factors, the available extragalactic information that may connect the molecular cloud clump and the galaxy-wide scales is explored next.

\subsection{Late type / star-forming galaxies}
\label{chIMF:sec:disks}

Massive stars die after a few Myr and because many are evident in galaxies these need to replenish their gas content to continue forming stars. In order to quantify the rate with which stars are being born from the gas various tracers of the star formation activity are employed (for a complete overview see \citealt{BuatZezas2021}).  
Massive stars ionise the gas in their wider vicinity which recombines with emission of H$\alpha$ photons. By relating the H$\alpha$ flux to the number of ionising massive stars and with an assumed shape of the gwIMF it is possible to calculate the total mass of stars formed per unit time within the life time of the massive stars. This is the basis of an often employed method to measure the SFRs of nearby galaxies.

Following \cite{Pflamm+2007}, the number of ionising photons a stellar population of a galaxy emits in the time interval $\delta t$ can be calculated from
\begin{equation}
N_{{\rm ion}, \delta t} = 
\int_{m_{\rm H}}^{m_{\rm max, g}}
\, \xi_x(m)\, N_{{\rm ion}, \delta t}(m)\,{\rm d}m \, ,  
\label{chIMF:eq:Nionising}
\end{equation}
where $\xi_x(m)$ can be, for example, the invariant canonical stellar IMF (eqn.~\ref{chIMF:eq:canIMF1}--\ref{chIMF:eq:canIMF3} ) or the IGIMF (eqn.~\ref{chIMF:eq:igimf_t}). $N_{{\rm ion}, \delta t}(m)$ is the number of ionising photons a star of mass $m$ emits in a time interval $\delta t$. The energy of one H$\alpha$ photon is $L_{{\rm H}\alpha} = 3.0207 \times 10^{-12}\,$erg and $\mu_{\rm r}$ is the fraction of ionising photons that lead to recombination (H$\alpha$) emission in the surrounding gas (typically $\mu_{\rm r}=1$ is adopted). It follows that
\begin{equation}
L_{{\rm H},\alpha} = \mu_{\rm r} \, 3.0207\times 10^{-12} \, {\rm erg} \, N_{{\rm ion}, \delta t} / \delta t \, .
\label{chIMF:eq:LHalpha}
\end{equation}
The $\psi=\psi(L_{{\rm H}\alpha})$ relation follows by calculating $L_{{\rm H}\alpha}$ via eqn.~\ref{chIMF:eq:Nionising} and~\ref{chIMF:eq:LHalpha} which links the IMF with the SFR via eqn.~\ref{chIMF:eq:MeclmaxSFR2} and Eq.~\ref{chIMF:eq:igimf_t}. 
 
By assuming the gwIMF to be an invariant probability distribution function (see Sec.~\ref{chIMF:sec:sampling}) and of \cite{MS1979} form (which is comparable to the canonical stellar IMF, see \citealt{Kroupa+2013}) such that it always retains the same fraction of ionising stars relative to all stars formed, \cite{Kennicutt1983} applied the linear relation 
\begin{equation}
\psi = L_{{\rm H}\alpha} / \left( 1.12 \times 10^{41}\,{\rm erg/s} \right) \, M_\odot/{\rm yr}
\label{chIMF:eq:Kennicutt}
\end{equation}
to quantify the SFRs of 170~nearby galaxies therewith quantifying, among other properties, gas consumption time scales and establishing this as the method of choice for subsequent research. 

An often practiced approach is thus to assume Case~A in Sec.~\ref{chIMF:sec:caseA} such that star-formation is a stochastic process in disk and dwarf galaxies and the gwIMF is an invariant probability distribution function (e.g. \citealt{SLUG_II}). According to this approach, all galaxies will be alike statistically in terms of their properties except for an increasingly large dispersion of these towards smaller galaxy masses.  However, a more recent survey of hundreds of nearby galaxies has shown that star-forming dwarf galaxies lack the emission of ionising radiation relative to that expected for an invariant canonical stellar IMF \citep{Lee+2009}. Similarly, massive disk galaxies have been found to have an overabundance of ionising stars \citep{Gunawardhana+2011}.  The authors of these observational studies discuss possible effects that may mimic such a variation with the underlying gwIMF being of Case~A. Dust obscuration and reddening is unlikely to be the culprit of an apparently increasing lack of massive stars with decreasing SFR because galaxies with small SFRs are dwarfs that are of lower-metallicity than their Milky-Way-type counterparts. A loss of ionising radiation in the dwarf galaxies (i.e. a small value of $\mu_{\rm r}$ in eq.~\ref{chIMF:eq:LHalpha}), and thus a decrease in the H$\alpha$ recombination flux, appears unlikely because dwarf galaxies have significantly larger gas fractions than their more massive counterparts (fig.~4 in \citealt{Lelli+2016}). It is also unlikely that the gwIMF is invariant with the expectation being that over a long-time-average dwarf galaxies have the same number of ionising stars relative to late type stars as their more massive counterparts. This would require the more than 100~galaxies with $\psi<10^{-2}\,M_\odot/$yr in the sample of \cite{Lee+2009} to currently have dropped their SFRs in-phase with each other despite not being in causal contact and being separated by a few~Mpc. An invariant, stochastically sampled gwIMF is favoured by \cite{KE2012}.

In contrast, the evidence based on isotopologue data
suggest galaxies with high SFRs at different redshifts to have top-heavy gwIMFs \citep{Zhang+2018}.
Further evidence for a top-heavy gwIMF through very early metal enrichment at high redshifts ($9.4<z<10.6$) is being studied based on JWST data
(e.g. \citealt{Curti+2024, BekkiTsujimoto2023, KobayashiFerrara2024}).

Thus, if the gwIMF varies according to the IGIMF theory (eqn.~\ref{chIMF:eq:igimf_t}) then the proportionality between $L_{{\rm H}\alpha}$ and $\psi$ (eqn~\ref{chIMF:eq:Kennicutt}) does not hold because dwarf galaxies with small SFRs lack ionising stars, i.e., for every H$\alpha$ photon they emit, many more low-mass stars are forming in them than in major disk galaxies that have gwIMFs that are comparable to the canonical form or are even top-heavy (e.g. Fig.~\ref{chIMF:fig:Z_SFR1}).  This leads to the SFRs deduced using eqn~\ref{chIMF:eq:Kennicutt} to be significantly too small for late-type dwarf galaxies and too large for late-type massive galaxies. 
In combination with the data by \cite{Lee+2009} and \cite{Gunawardhana+2011},
this suggests that the
gwIMF is increasingly top-light with decreasing SFR (for
$\psi < 1\,M_\odot/$yr), while it becomes increasingly top-heavy with
increasing SFR (for $\psi > 1\,M_\odot/$yr).  

Indeed, the calculation of the
SFR-dependent gwIMF using the IGIMF theory (Case~C in
Sec.~\ref{chIMF:sec:caseC}) predicts a dependency of the gwIMF on
the SFR with dwarf galaxies having a gwIMF with a deficit of ionising stars and disk galaxies with $\psi>1\,M_\odot/$yr having a top-heavy gwIMF consistent with the observational constraints (e.g. \citealt{Haslbauer+2024}). The radial cutoff  in H$\alpha$ emission despite star-forming galaxies having extended UV disks also follows from the IGIMF theory (Sec.~\ref{chIMF:sec:caseCextended}). 
Therewith the IGIMF theory passes an important observational test demonstrating the power of bridging the formation of stellar populations on the molecular clump scale to extragalactic data. 
A number of important theoretical implications arise from this if the true gwIMF is correctly represented by the IGIMF theory: 

\begin{itemize}

\item Dwarf galaxies with $\psi<10^{-3}\, M_\odot/$yr have a significant deficit of ionising stars relative to their UV fluxes that are a measure of the number of intermediate-mass stars (Fig.~\ref{chIMF:fig:Z_SFR2}, \citealt{Pflamm+2009}). Given observed H$\alpha$ recombination fluxes, the true SFRs of late-type dwarf galaxies are significantly larger such that their gas-consumption time scales become comparable to those of major disk galaxies being near $3\,$Gyr. This result implies that star-forming galaxies of all masses need to be accreting gas at a rate comparable to their SFRs to sustain their near-constant SFHs with implications for cosmological theory \citep{Haslbauer+2023, Haslbauer+2024}.

\item The systematically evolving gwIMF leads to a theoretical stellar-mass--metallicity relation of galaxies over the galaxy-mass range $10^4\,M_\odot$ to $10^{13}\,M_\odot$ matching the observed one of galaxies very well without the need of outflows of enriched gas from galaxies \citep{Haslbauer+2024}.

\item Traditionally, models with an invariant gwIMF have estimated, by eqn.~\ref{chIMF:eq:Kennicutt}, small SFRs for dwarf galaxies, suggesting it would take longer than the age of the universe (a Hubble time) for these galaxies to build up their observed stellar masses, as indicated by their B-band magnitudes. This has been a known long-standing problem. The application of the IGIMF theory, however, suggests that dwarf galaxies have higher true SFRs. This results in shorter timescales for these galaxies to accumulate their observed stellar masses, resolving this known problem \citep{PKr2009, Haslbauer+2024}. 

\item The significantly larger true SFRs of dwarf galaxies (assuming the gwIMF is given by the IGIMF theory) implies that dwarf and major disk galaxies have had approximately constant SFRs with implications for the interpretation of the Lilly-Madau plot of the redshift-dependent star-formation rate density in a comoving cosmological volume in terms of Gpc-scale cosmological structure \citep{Haslbauer+2023}.

\item With the implied significantly larger true SFRs of dwarf galaxies, and the implied smaller true SFRs of massive disk galaxies (the IGIMF theory predicting them to have top-heavy gwIMFs), the main sequence of galaxies flattens significantly \citep{Haslbauer+2024}.

\item Gas-rich dwarf galaxies can have H$\alpha$-invisible, i.e. dark, star formation appearing as intergalactic gas clouds with little traces of star formation \citep{Pflamm+2007}.
    
\end{itemize}

\begin{BoxTypeA}[chIMF:box:dwarfgalaxies]{Cosmological implications of the IGIMF theory}

\noindent
These implications demonstrate how an understanding of the gwIMF and its variation connects stellar populations to the cosmological arena.

\end{BoxTypeA}

\subsection{Early type galaxies (ETGs), high-redshift galaxies and ultra-faint dwarfs (UFDs)}
\label{chIMF:sec:ETGs}

ETGs, or more specifically elliptical galaxies, are often associated with cooling flows that
appear to deposit a large amount of gas in their inner regions the fate
of which remains unknown (\citealt{Fabian+2024} and references
therein). One way to assess if it turns into faint low-mass stars is
to use spectroscopic features that are sensitive to the gravitational
field strength in the photospheres of stars to assess the dwarf to
giant star ratio \citep{KroupaGilmore1994}. 
\cite{Matteucci1994} developed a chemical enrichment code to calculate models of elliptical galaxies and deduced, given observational constraints on the alpha-element over iron, [Mg/Fe], ratio that these formed within a time of about $3\times 10^8\,M_\odot$ and with a top-heavy gwIMF.  Less-massive elliptical galaxies formed over a longer time scale than more massive ones. \cite{Vazdekis+1996} also developed an elaborate model for calculating  the chemical enrichment and spectro-photometry and applied it to three well observed ETGs. The result of this work is that these ETGs formed in a short phase lasting less than a Gyr with a top-heavy gwIMF with some star formation continuing over a longer time with a bottom-heavy gwIMF.
Modern observational surveys using stellar surface-gravity-sensitive spectroscopic indices
confirm the result that elliptical galaxies
typically have significantly bottom-heavy gwIMFs (e.g. \citealt{Ferreras+2013} based on data for 40000 ETGs). This can be combined
with lensing and dynamical constraints (see \citealt{Smith2020} for a
review). These constraints however suffer from the degeneracy that faint
late-type stars and stellar remnants appear as dark matter that is
thought to dominate galaxy masses. 
It needs to be kept in mind that the inner regions of elliptical galaxies constitute high-density environments such that the long-lived late-type stars may have formed from and accreted over time heavily enriched gas such that the templates and stellar model libraries used in the above modelling that relies on surface-gravity-sensitive spectral indices may not represent the true stellar population. In the Solar neighbourhood the stars have been existing in a very different low density gas environment with significantly less-intense chemical enrichment.

The high alpha-element abundances
of elliptical galaxies imply these to have formed within typically
a~Gyr, with the more massive ones forming more rapidly (referred to as {\it downsizing})  and with star formation over the past few~Gyr adding negligible stellar mass \citep{Thomas+2005, SR+2020}.  Their solar
and super-solar metallicity implies the gwIMFs to have been top-heavy
\citep{Matteucci1994, Vazdekis+1996}. 
The constraints on the stellar and stellar-remnant content of elliptical galaxies,  
paired with the small mass-to-light ratios of high-redshift galaxies needed for consistency with cosmological structure formation theory (see \citealt{Haslbauer+2022}) suggest elliptical galaxies to have formed with bottom-heavy and top-heavy gwIMFs \citep{vanDokkumConroy2024}, thus confirming the  findings by \cite{Matteucci1994} and \cite{Vazdekis+1996}.

Interestingly, the application of the IGIMF theory 
(Sec.~\ref{chIMF:sec:caseC}) to the problem of forming and evolving elliptical galaxies leads to consistency with these
constraints including those from ultra-faint dwarf galaxies (UFDs, 
\citealt{Yan+2021, Yan+2024}). Elliptical galaxies thus appear to have formed monolithically on a gravitational collapse time which leads to the  downsizing relation of formation time {\it vs} mass \citep{Eappen+2022}. The observed increase of the dynamical
mass-to-light ratios of elliptical galaxies with increasing mass
follows from the top-heavy gwIMFs having not only produced the large
metallicity and thus a bottom-heavy gwIMF but also a very significant
mass in remnants \citep{DK2023}. Massive elliptical galaxies may in reality be a few to ten times more massive than implied when using an invariant canonical gwIMF. Given the downsizing times over which ETGs formed a correlation emerges naturally within the IGIMF theory: ETGs with a larger stellar velocity dispersion have a more top-heavy and a more bottom-heavy gwIMF than ETGs with a smaller velocity dispersion as these have a smaller mass formed over a longer time and thus with a smaller SFR.

The survey by \cite{LaBarbera+2019} of seven elliptical galaxies used stellar surface-gravity sensitive absorption lines in the empirical EMILES spectral library to model the radial dependency of the stellar population. The results show a pronounced radial gradient with the innermost regions of the galaxies by characterised by very bottom-heavy local gwIMFs while the outermost parts have a canonical IMF of late-type stars. This is  qualitatively consistent with the expectations of the IGIMF theory (Sec.~\ref{chIMF:sec:caseCextended}).

According to the IGIMF theory the gwIMF is very bottom-heavy at $[Z]>0$ such that even high SFRs remain dark (Fig.~\ref{chIMF:fig:mav_mmax2}). This may suggest a solution to the above mentioned cooling flow problem.

Empirical evidence for the gwIMF having been top-light and thus lacking massive stars in galaxies that had very low SFRs as calculated using the IGIMF theory is evident from detailed chemical evolution studies of the very old dwarf-spheroidal satellite galaxies (e.g. \citealt{Mucciarelli+2021, Tang+2023}).

Concerning the UFDs, these are resolved into individual stars such that the problem with using the correct stellar luminosity-mass relation is critically important. The published stellar mass functions for these need to be revisited in view of theoretical stellar luminosity-mass relations being incorrect (see Fig.~\ref{chIMF:fig:MLR}). The prediction from the IGIMF theory is that the gwIMF in UFDs is bottom-light due to the low metallicity and that it was top light due to the small SFR. (\citealt{Yan+2024} and references therein).

\subsection{Ultra compact dwarf galaxies (UCDs) and globular star clusters (GCs)}
\label{chIMF:sec:UCDs}

UCDs are typically many Gyr old and ten to a hundred times as massive as GCs and show a large range of dynamical mass-to-light ratios. These and their photometric properties can be readily understood \citep{Mahani+2021}: 

\begin{itemize}

\item UCDs with large dynamical mass-to-light ratios formed through a monolithic collapse as embedded clusters. These had very top-heavy IMFs (eqn.~\ref{chIMF:eq:stIMF}; see also \citealt{BekkiTsujimoto2023}) leaving a population of stellar remnants in the object that have a mass that is up to ten times larger than the mass in main sequence stars. When less than about 50~Myr old, the monolithically-formed UCDs attained luminosities comparable to high-redshift quasars \citep{Jerabkova+2017}.

\item Those UCDs that have GC-like dynamical mass-to-light ratios can be understood to have formed 
from the amalgamation of many star clusters in star-cluster complexes as are observed to be forming in massively interacting galaxies.
These latter UCDs have stellar populations as obtained though applying the IGIMF theory (Sec.~\ref{chIMF:sec:caseC}). 

\end{itemize}

\noindent
GCs are many Gyr old and have been thought to be simple stellar populations thus having formed monolithically in very massive molecular cloud clumps as massive embedded clusters ($M_{\rm ecl}>10^5\,M_\odot$). The observational finding that many GCs have sub-populations of stars that have different chemical abundances and also a small degree of spread in Fe suggests this understanding of their formation to be challenged with currently no consensus existing on their formation and the emergence of these multiple populations. The significant, and sometimes dominant, fraction of the enriched population poses a problem for nearly all chemical enrichment scenarios. This problem may however disappear if GCs do form monolithically with only one simple stellar population which, however, is binary-rich and with a top-heavy stellar IMF (eqn.~\ref{chIMF:eq:stIMF}). The stellar-dynamical encounters during the formation process of the stars lead to mergers between massive binary-star components and the winds from these expell synthesised elements into the interstellar medium in the young GC. Stars forming in the vicinity will thus be enriched to different degrees with light elements before core-collapse supernovae quench further star formation (\citealt{Wang+2020} and references therein). The quenching is not sudden and the first core-collapse supernovae enrich the gas from which stars are still forming with Fe accounting for the observed Fe spreads \citep{Wirth+2022, Wirth+2024}. The top-heavy stellar IMF accounts for the large fraction of enriched stars through the many enrichment events, and the large embedded cluster mass ensures that massive-star feedback alone cannot remove the gas with gas removal needing mutiple core-collapse supernova events (fig.~3 in \citealt{Baumgardt+2008}).

\subsection{Supermassive black holes}

The formation and early appearance of supermassive black holes (SMBHs)
remains a major area of research. The IMF of population III stars
appears to be relevant for this problem
\citep{KlessenGlover2023}. A possible solution
follows from elliptical galaxies and bulges forming on a free-fall
time from primordial pre-galactic gas clouds \citep{Eappen+2022}. The central
region of such a collapse undergoes a star burst as a UCD-type nuclear cluster.  Since elliptical galaxies and bulges more massive than a few $10^9\,M_\odot$ formed with $\psi \simgreat {\rm few} \, M_\odot/$yr this most massive cluster (eq.~\ref{chIMF:eq:Meclmax}) has a mass $M_{\rm ecl,max}\simgreat\; {\rm few} \;10^5\,M_\odot$. 
It has a quasar-like luminosity (Sec.~\ref{chIMF:sec:UCDs}) 
due to the top-heavy stellar IMF because of the
low initial metallicity and very high density
(eqn.~\ref{chIMF:eq:stIMF}). When the intense radiation field of
this cluster collapses after the massive stars die the cluster
contains more than $10^5$ stellar-mass black holes (BHs) and neutron
stars. Gas falling into this cluster compresses it until the BHs orbit
at relativistic speeds such that they radiate gravitational waves and
the BH cluster collapses. SMBHs weighing more than $10^5\,M_\odot$ can
thus form within a few hundred~Myr at the centres of forming elliptical galaxies and bulges and their masses correlate with the
masses of the hosting galaxies by the IGIMF theory
\citep{Kroupa+2020}.

\subsection{The combined extragalactic evidence}

\begin{figure}[t]
\centering
\includegraphics[width=.6\textwidth]{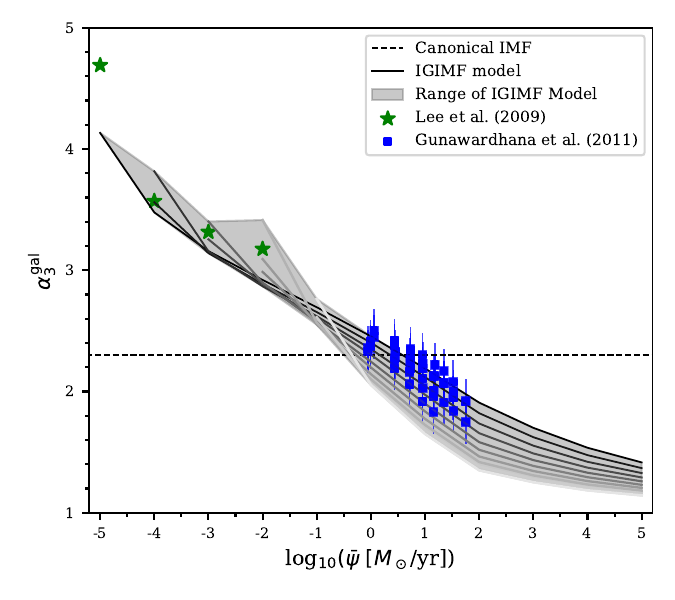}
\caption{
The variation of the shape of the IMF of stars forming in a galaxy with a SFR ${\bar \psi}$ (the average SFR in a time interval of $\delta t = 10\,$Myr). 
The green stars are from the observational survey of \cite{Lee+2009} of nearby dwarf 
galaxies, the blue squares are from the survey of major disk galaxies by 
\cite{Gunawardhana+2011}. 
The shaded/grey region gives the gwIMFs calculated with the IGIMF theory for different stellar mass ranges and metallicities. It shows 
the shape of the gwIMF in terms of its stellar-mass-dependent power-law index, $\alpha_3^{\rm gal}$, that represents the gwIMF for stars more
massive than $1\,M_\odot$ as a function of the galaxy-wide SFR, ${\bar \psi}$.
For a given SFR and at at each $m$ value, there exists a different $\alpha_3^{\rm gal}-\psi$
relation because the gwIMF cannot be represented by a small set of power laws (see Fig.~\ref{chIMF:fig:Z_SFR1}) but can be approximated by power-law sections with indices that change with the stellar mass, $m$. The solid lines are for mass ranges centred on $m/M_\odot=1.58, 2.51,..., 100$ from dark to light. 
Elliptical galaxies 
formed with ${\bar \psi} > 1000\,M_\odot/$yr with top-heavy gwIMFs that are 
needed to account for their Solar and super-Solar metallicities and high 
[$\alpha$/Fe] ratios \citep{Yan+2021}. The horizontal dashed line is the invariant canonical 
gwIMF (eq.~\ref{chIMF:eq:canIMF3})
which would be valid if the stellar IMF were an invariant probability 
density distribution function.  The survey data reproduce the IGIMF-theory quantitatively and do not support an invariant probability density distribution nature of the IMF. 
Adapted from fig.~6 in \cite{Yan+2017}. 
A depicton of the variation of the gwIMF 
 for low-mass stars in observational extragalactic data in comparison to the IGIMF-theoretical calculation is available in fig.~4 in \cite{Yan+2024}.}
\label{chIMF:fig:extragalactic}
\end{figure}

The various studies of star-forming dwarf and major (disk) galaxies, faint satellite galaxies and of massive elliptical galaxies suggest that the gwIMF changes as follows: with increasing SFR the gwIMF becomes increasingly top-heavy. At very low metallicity the gwIMF becomes bottom light which is evident in faint dwarf satellite galaxies.  This is shown graphically in Fig.~\ref{chIMF:fig:extragalactic}. The IGIMF theory accounts for this naturally without having been developed to do so as it primarily rests on the observations of nearby and resolved star forming regions and stellar populations. Other interpretations of the observational data insisting the gwIMF remains invariant (e.g. through a systematic current lull in the SFRs across all gas rich dwarf galaxies, a fine-tuned loss of ionising photons, particularly designed dust obscuration) have been suggested (Sec.~\ref{chIMF:sec:disks}) but do not naturally lead to the observed trend shown in Fig.~\ref{chIMF:fig:extragalactic}. 

\begin{BoxTypeA}[chIMF:box:extragalactic]{Extragalactic evidence for the variation of the gwIMF}

\noindent
From Fig.~\ref{chIMF:fig:extragalactic} it is clear that the extragalactic data {\it and} the IGIMF theory inform us that the gwIMF is top-light for $\psi\simless 1\,M_\odot/$yr and top-heavy for  $\psi\simgreat 1\,M_\odot/$yr.

\end{BoxTypeA}

\section{Existence statement}
\label{chIMF:sec:existence}

For modelling the evolution of star clusters and galaxies initial distribution functions are required for the initialisation of the stellar population. These are the stellar IMF, $\xi(m)$, and the birth distribution functions defining the binary population in terms of the primary-star-mass ($m_{\rm prim}$)  dependent fraction of stars born as binaries or multiple systems, $f_{\rm bin/mult}\approx 1$,  their distribution of periods, $f_{\rm P}(P:m_{\rm prim})$, their distribution of mass ratios, $f_{\rm q}(q:m_{\rm prim})$ and eccentricities, $f_{\rm e}(e:m_{\rm prim})$ as well as the birth half-mass radius, $r_{0.5}(M_{\rm ecl})$, of the embedded cluster this population with a stellar mass of $M_{\rm ecl}$ is born in.

It is useful to keep in mind that these functions do not exist in reality because stars form over a few free-fall times, $t_{\rm ff}$, of their proto-embedded cluster, they experience stellar-dynamical encounters within it and
are ejected from it within a fraction of $t_{\rm ff}$
such that an embedded cluster neither ever contains all
the stars born in it nor is there ever a time when an observer can, by direct star-counts, construct the above distribution functions.

By being mathematical descriptions of in reality non-existing functions, they are mathematical {\it hilfskonstrukts}, i.e., auxiliary functions needed for the mathematical description of idealised initial conditions (for example, models usually assume all stars to be born instantly for their initialisation in the computer). These hilfskonstrukts are needed because the construction of a realistic stellar population in an embedded cluster, let alone a whole galaxy, is not possible using magneto-hydrodynamical radiation-transfer computer simulations of sufficient resolution for star formation.  Although these functions cannot be constructed through direct observation, they and their possible dependency on the physical parameters, $\{ P_{\rm i}\}^N_{\rm P}$, of the star-forming molecular cloud clump, can be inferred from the observed properties of stars.

\section{Computer codes}
\label{chIMF:sec:codes}

As the shape of the IMF is essential information, estimators for the exponent and upper limit with goodness-of-fit tests for truncated power law distributions have been compared and published by \cite{MKr2009}. The estimators covered include binning, the cumulative distribution, Beg's estimator and the maximum likelihood estimator. The package {\sc statpl} is downloadable from the Astrophysics Source Code Library\footnote{\url{https://ascl.net/1206.006}}

Nearly all galaxy and cosmological structure formation simulation codes assume Case~A (Sec.~\ref{chIMF:sec:gwIMF}) such that fractions of supernovae can occur at small SFRs (e.g. one tenth of a supernova producing one tenth of the enrichment and one tenth of the explosion energy can occur if the IMF is undersampled such that it includes fractions of stars). An implementation into the adaptive mesh-refinement FLASH code of the IGIMF theory (Case~C) by \cite{Ploeckinger+2014} and \cite{SH2023} such that only full stars are allowed demonstrated that dwarf galaxies form significantly more stellar mass over time than when an invariant canonical gwIMF is assumed which produces more feedback through fractions of stellar winds and fractions of core-collapse supernovae than the top-light gwIMF computed with the IGIMF theory. These ideas are likely to have important implications for galaxies forming in a cosmological context (see Sec.~\ref{chIMF:sec:disks}).

Various publicly-available computer programs have been written to
model stellar populations under different assumptions Case~A,~B and ~C
(Sec.~\ref{chIMF:sec:gwIMF}). A few such codes are mentioned here:

For Case~A, the {\sc LIBIMF} C-library is available to set up a
stellar population according to a pre-chosen stellar IMF
\citep{LIBIMF} downloadable from the Astrophysics Source Code Library\footnote{\url{https://ascl.net/1206.009}}.
the {\sc Starburst99} code has been much used to model starbursts for
pre-chosen stellar IMFs \citep{Starburst99}\footnote{\url{http://ascl.net/1104.003}}. The {\sc PEGASE3} code is
available for a similar purpose \citep{PEGASE3}\footnote{\url{https://cdsarc.u-strasbg.fr/viz-bin/qcat?J/A+A/623/A143\#/browse}}. 

For Cases~A and~B, the binary population and spectral synthesis ({\sc BPASS}) programme\footnote{\url{https://bpass.auckland.ac.nz}} has been developed to calculate the properties and spectral energy distributions of synthetic stellar populations. The impact of binary star interactions and evolution processes are explicitly modelled. {\sc BPASS} assumes stochastic sampling of the stellar IMF subject to the $M_{\rm ecl}$ mass constraint to model a simple stellar population's properties and astrophysical (not star-cluster-dynamical) evolution.  The stochastically lighting up galaxies ({\sc SLUG}) software has been developed to model the photometric properties of star clusters and galaxies assuming the stellar IMF is invariant (e.g. \citealt{SLUGIII}\footnote{\url{https://bitbucket.org/krumholz/slug2/src/master/}}). To set up a realistic stellar plus binary star population for modelling star clusters the {\sc McLuster} code has been developed (e.g. \citealt{McLuster1}\footnote{\url{https://github.com/lwang-astro/mcluster}}). \cite{Garling+2024} published the open-source Julia package {\sc StarFormationHistories.jl} \footnote{\url{ https://github.com/cgarling/StarFormationHistories.jl}} for measuring resolved SFHs. It includes modelling unresolved photometric binaries and supports arbitrary IMFs.

For cases~A and~C, the initialisation and efficient dynamical
processing of a binary-star population in an embedded cluster and a
galactic field is coded in {\sc BiPos1}\footnote{\url{https://github.com/JDabringhausen/BiPoS1}}\citep{Dabringhausen+2022}.

For Case~C, the {\sc photGalIMF}\footnote{\url{https://github.com/juzikong/photGalIMF},\url{https://ascl.net/code/v/4066}} code is available as an extension of {\sc galIMF}\footnote{\url{https://github.com/Azeret/galIMF}}.  It allows the
calculation of the photometric evolution in various pass bands of  a one-zone chemical enrichment model of a galaxy by applying the IGIMF theory to track the changing gwIMF through an evolving SFR and 
due to the self-enrichment by metals (e.g. \citealt{Haslbauer+2024}). {\sc photGalIMF} is currently the only chemical enrichment code to self-consistently account for the evolution of the gwIMF and the luminosity of a forming galaxy. The {\sc GWIMF} FORTRAN code\footnote{\url{https://github.com/ahzonoozi/GWIMF}}, allowing the calculation of the gwIMF using the IGIMF theory following \cite{Jerabkova+2018}, has been written by A.~H.~Zonoozi, who has also developed the {\sc SPS-VarIMF} FORTRAN 
program to allow the calculation of spectral energy distributions of
galaxies evolved via the IGIMF theory (to be published by Zonoozi
and collaborators). The efficient {\sc GalCEM} code allows one-zone chemical
evolution models of galaxies to be calculated based on 86~elements and
451~isotopes for a pre-chosen invariant gwIMF \citep{Gjergo+2023}\footnote{\url{https://github.com/egjergo/GalCEM}}. It
is being extended as the {\sc pyIGIMF} module to include the IGIMF theoretical formalism\footnote{\url{https://github.com/egjergo/pyIGIMF}}.

\section{Conclusions and outlook}
\label{chIMF:sec:concs}

The understanding of the stellar IMF has advanced considerably over the past two decades such that a good understanding of stellar populations in the cosmological context is available (Fig.~\ref{chIMF:fig:extragalactic}).
But key issues remain: 
Relying only on theoretical stellar luminosity--mass relations is prone to lead to failure in the construction of a stellar IMF. 
Corrections for missed components of multiple stellar systems critically depend on the degree of dynamical processing of the stellar population under scrutiny -- there is no universal correction for unresolved multiple stellar systems that can be applied to star counts.
An important advance has been the realisation that the stellar IMF of a simple stellar population born in one molecular cloud clump as an embedded cluster is unlikely to be the same as the stellar IMF of all freshly formed stars in a galaxy, the gwIMF. 

Observational data are, however, often interpreted with the use of a stellar IMF that is assumed to be an invariant probability density distribution function and to be the same in star-forming molecular cloud clumps and on the galaxy scale. This provides a useful benchmark model. 

As the observations probe ever more extreme star-forming regions their interpretation increasingly points to the stellar IMF not being equal to the gwIMF. The observations of high redshift ($z>10$) star-formation activity also increasingly indicate the stellar IMF to depend on the physical properties of the star-forming gas, IMF$\,=$IMF$(\{P_{\rm i}\}^{N_{\rm P}})$, where $\{P_{\rm i}\}^{N_{\rm P}}$ is a set of $N_{\rm P}$ physical parameters such as the density of the molecular cloud clump, its metallicity, its temperature, specific angular momentum, magnetic field, the shear of the galaxy's rotational velocity field. This variation is expected to carry through to the corrections needed for calculating the stellar IMF from star-counts with unresolved multiple systems because the properties of their birth distribution functions as well as the properties of their birth embedded clusters are likely to also depend on the physical properties ($f_{\rm P} = f_{\rm P}(P,\{P_{\rm i}\}^{N_{\rm P}}:m_{\rm prim}), f_{\rm q} = f_{\rm q}(q,\{P_{\rm i}\}^{N_{\rm P}}:m_{\rm prim}), f_{\rm e} = f_{\rm e}(e,\{P_{\rm i}\}^{N_{\rm P}}:m_{\rm prim}), r_{0.5} = r_{0.5}(M_{\rm ecl}:\{P_{\rm i}\}^{N_{\rm P}})$). Currently available data are limited and do not indicate strong dependencies of these (Sec.~\ref{chIMF:sec:dynstr}, Sec.~\ref{chIMF:sec:binaries}).

One approach is to
condense the variation only to a background temperature dependency related to the cosmological model and to treat the gwIMF as being the same as the stellar IMF. This approach does not capture that the physical conditions vary within a galaxy and that information on the temperature under which a stellar population formed is lost with the gas such that a gauging of the stellar IMF/gwIMF dependence on temperature using star-counts is not easily possible. An alternative approach is to gauge the variation of the stellar IMF by the masses of the born simple stellar populations, the radii of the embedded clusters in which the populations were born, and the metallicities of the populations. The so-gauged IMF$\,=$IMF(molecular cloud clump density, metallicity) can then be used to construct the forming stellar population of a galaxy by integration over all new embedded clusters. 
Although previous studies had assumed it to be a probability density distribution function, 
inherent to the current IGIMF-theoretical approach is the interpretation that the stellar IMF emerging within a molecular cloud clump is an optimally sampled distribution function. This may be a result of significant feedback-self-regulation of the star-formation process on the molecular clump scale as is indicated to be the case by the small star-formation efficiency of $\epsilon < 0.4$ per clump. The existence of a tight $m_{\rm max}=m_{\rm max}(M_{\rm ecl})$ relation for embedded clusters supports significant feedback-self-regulation in the star-formation process. The correlation between the mass of the most massive young cluster, $M_{\rm ecl, max}=M_{\rm ecl,max}(\psi)$, and the SFR, $\psi$,  of its hosting galaxy suggests a self-regulation process on the galaxy scale, probably through the galaxy-wide potential determining the dynamics of the star-forming gas. 

According to the IGIMF theory, late-type dwarf galaxies would have a top-light gwIMF while massive disk galaxies would have a top-heavy gwIMF. Elliptical galaxies end up to have a bottom-heavy and had a top-heavy gwIMF. Ultra faint dwarf galaxies would have a bottom-light and had a top-light gwIMF. A correlation of the overall shape of the gwIMF with the velocity dispersion of an ETG arises naturally due to the radius-mass relation of ETGs and the mass--SFR relation via the downsizing time.  The properties of elliptical galaxies and the rapid formation of supermassive black holes, the stellar content of late type dwarf galaxies and the radial H$\alpha$ cutoff in UV-extended galactic disks would be explained naturally. Massive elliptical galaxies would be up to ten times as massive due to their content in faint red dwarf stars, neutron stars and black holes. According to the IGIMF theory, at super-Solar metallicity large SFRs remain dark because the gwIMF is bottom-heavy and top-light, suggesting a possible solution of the cooling flow problem.
Late-type dwarf disk galaxies would have similar gas-consumption time-scales as major disk galaxies and the galaxy main sequence would include late-type dwarf galaxies and would be flatter. The stellar IMF on the molecular cloud clump scale becomes connected to the cosmological scale. 

\emph{It is important to keep in mind} that any proposed model of the variation of the stellar IMF and of the gwIMF must account for their connection and the locally observed
  canonical shape of the stellar IMF and of the Milky Way's gwIMF. Given the available observational constraints spanning the local to the high redshift Universe, 
  the galaxy-wide stellar populations appear to require a varying stellar
  IMF on the molecular clump scale that needs to be essentially
  similar to the formulation provided by eqn.~\ref{chIMF:eq:stIMF}.

The future will inform us about the true nature and the variation of the
stellar IMF, the role of additional parameters, $P_{\rm i}$ beyond $M_{\rm ecl}$ and $Z$,  and the role of feedback self-regulation in the
star-formation process. 

\begin{ack}[Acknowledgments]

  We thank Jiadong Li at Heidelberg University for providing the data for
  Fig.~\ref{chIMF:fig:MLR}.  P.K. acknowledges support through
  the DAAD Bonn--Prague Eastern-Europe Exchange programme. 
  E.G. acknowledges the support of the National Natural Science Foundation of China (NSFC) under grants NOs. 12173016, 12041305.
  E.G. acknowledges the science research grants from the China Manned Space Project with NOs. CMS-CSST-2021-A08, CMS-CSST-2021-A07.
  E.G. acknowledges the Program for Innovative Talents, Entrepreneur in Jiangsu.

\end{ack}


\bibliographystyle{Harvard}
\bibliography{reference_IMF}

\end{document}